\documentclass[12pt]{article}

\usepackage{amsmath,amsfonts,amssymb,cite}

\usepackage[mathscr]{eucal}

\def\PCPqed{\hspace*{\fill}\ensuremath{\Box}}

\newcommand{\smin}{\,\raisebox{0.06em}{${\scriptstyle \in}$}\,}
\newcommand{\nsmin}{\,\raisebox{0.06em}{${\scriptstyle \notin}$}\,}
\newcommand{\ssmin}{\,\raisebox{0.06em}{${\scriptscriptstyle \in}$}\,}

\newcommand{\smcap}{\,\raisebox{0.06em}{${\scriptstyle \cap}$}\,}

\newcommand{\smwedge}{{\scriptstyle \wedge}}

\newcommand{\smcirc}{\,\raisebox{0.1ex}{${\scriptstyle \circ}$}\,}
\newcommand{\bwedge}{\raisebox{0.2ex}{${\textstyle \bigwedge}$}}

\newcommand{\upostar}{\ensuremath{%
 ^{\raisebox{0.05ex}{$\scriptscriptstyle \bigcirc$} \hspace{-0.5em} \star}}}

\newcommand{\mathslf}[1]{\ensuremath{\mbox{\slshape \textsf{#1}}}}

\newtheorem{prp}{Proposition}[section]

\newtheorem{thm}[prp]{Theorem}
\newtheorem{dfn}[prp]{Definition}
\newtheorem{cor}[prp]{Corollary}

\begin{document}

\title{Hamiltonian Multivector Fields and Poisson Forms \\
       in Multisymplectic Field Theory}
\author{
  Michael Forger$\,^1\,$,~
  Cornelius Pauf\/ler$\,^2\,$~
  and
  Hartmann R\"omer$\,^3\,$}
\date{\normalsize
      $^1\,$ Departamento de Matem\'atica Aplicada,~~\mbox{} \\
      Instituto de Matem\'atica e Estat\'{\i}stica, \\
      Universidade de S\~ao Paulo, \\
      Caixa Postal 66281, \\
      BR--05315-970~ S\~ao Paulo, S.P., Brazil \\[4mm]
      $^2\,$
      Matematisk Fysik / Alba Nova \\
      Kungliga Tekniska H\"ogskolan \\
      SE--10691~ Stockholm, Sweden \\[4mm]
      $^3\,$ Fakult\"at f\"ur Mathematik und Physik \quad \mbox{} \\
      Physikalisches Institut\\
      Albert-Ludwigs-Universit\"at Freiburg im Breisgau \\
      Hermann-Herder-Stra\ss e 3 \\
      D--79104~ Freiburg i.Br., Germany
}

\footnotetext[1]{\emph{E-mail address:}
 \textsf{forger@ime.usp.br}}
\footnotetext[2]{\emph{E-mail address:}
 \textsf{pcp@theophys.kth.se}}
\footnotetext[3]{\emph{E-mail address:}
 \textsf{hartmann.roemer@physik.uni-freiburg.de}}
\maketitle

\thispagestyle{empty}

\vspace*{-3mm}

\begin{abstract}
\noindent
We present a general classification of Hamiltonian multivector fields and of
Poisson forms on the extended multiphase space appearing in the geometric
formulation of first order classical field theories. This is a prerequisite
for \linebreak computing explicit expressions for the Poisson bracket between
two Poisson forms.
\end{abstract}
\vspace{-5mm}
\begin{flushright}
 \parbox{12em}
 {\begin{center}
   Universit\"at Freiburg \\
   THEP 04/13 \\
   Universidade de S\~ao Paulo \\
   RT-MAP-0402 \\
   July 2004
  \end{center}}
\end{flushright}

\newpage

%%%%%%%%%%%%%%%%%%%%%%%%%%%%%%%%%%%%%%%%%%%%%%%%%%%%%%%%%%%%%%%%%%%%%%%%%%%%%%%
\section{Introduction and General Setup}

The present paper is a continuation of previous work on Poisson brackets
of differential forms in the multiphase space approach to classical field
theory \cite{FR1,FPR}. Our aim is to specialize the general constructions
of Ref.~\cite{FPR} from abstract (exact) multi\-symplectic manifolds to
the extended multiphase spaces of field theory, which at present seem
to be the only known examples of multisymplectic manifolds, to clarify
the structure of Hamiltonian multivector fields, of Hamiltonian forms
and of Poisson forms on these spaces and to give explicit formulas for
the Poisson bracket between the latter introduced in Refs~\cite{FR1,FPR}.

The structure of the article is as follows. In the remainder of this
introduction, we briefly review the geometric constructions needed in
the paper.  We put particular emphasis on the consequences that arise
from the existence of a certain vector field, the scaling or Euler
vector field. Also, we fix the notation to be used in what follows.
In~Section~2, we present an explicit classification of locally
Hamiltonian multivector fields on extended multiphase space in
terms of adapted local coordinates and, following the logical
inclusion from locally Hamiltonian to (globally) Hamiltonian to
exact Hamiltonian multivector fields, show how the last two are
situated within the first. Section~3 is devoted to the study
of Hamiltonian forms and Poisson forms that are associated with
(globally) Hamiltonian multivector fields. In~Section~4, we use
the outcome of our previous analysis to derive expressions for
the Poisson bracket between two Poisson forms. In Section~5, we
summarize our main conclusions and comment on the relation of
our results to other approaches, as well as on perspectives
for future research. Finally, in order to make the article
self-contained, we include in an appendix a proposition that
is not new but is needed in some of the proofs.

We begin with a few comments on the construction of the extended multiphase
space of field theory \cite{Ki1,Ki2,CCI,Got,GIM}, which starts out from a
given general fiber bundle over space-time, with base space $M$ ($\dim M
= n$), total space $E$, bundle projection $\, \pi : E \longrightarrow M \,$
and typical fiber $Q$ ($\dim Q = N$). It is usually referred to as the
configuration bundle since its sections constitute the possible field
configurations of the system. (Of course, the manifold~$M$ represents
space-time, whereas the manifold~$Q$ plays the role of a configuration
space.) The extended multiphase space, which we shall simply denote by~$P$,
is then the total space of a larger fiber bundle over~$M$ and in fact the
total space of a vector bundle over~$E$ which can be defined in several
equivalent ways, e.g., by taking the twisted affine dual $J\upostar E$
of the first order jet bundle $JE$ of~$E$ or by taking the bundle
$\, \bwedge_{n-1}^n T^* E \,$ of $(n-1)$-horizontal $n$-forms on $E$;
see \cite{FPR,CCI,GIM} for details. Therefore, there is a natural class
of local coordinate systems on~$P$, namely those that arise from combining
fiber bundle charts of~$E$ over~$M$ with vector bundle charts of~$P$
over~$E$: these so-called adapted local coordinates $(x_{\vphantom{i}}^\mu,
q_{\vphantom{i}}^i,p\>\!_i^\mu,p)$ are completely fixed by specifying local
coordinates $x_{\vphantom{i}}^\mu$ for~$M$ (the space-time coordinates),
local coordinates $q_{\vphantom{i}}^i$ for~$Q$ (the position variables)
and a local trivialization of~$E$ over~$M$, and are such~that the induced
local coordinates $p\>\!_i^\mu$ (the multimomentum variables) and $p\>\!$
(the energy variable) are linear along the fibers of~$P$ over~$E$.
For details, we refer to Ref.~\cite{FPR}, where one can also find
the explicit transformation law for the multimomentum variables and
the energy variable induced by a change of the space-time coordinates,
of the position variables and of the local trivialization.

A first important feature of the extended multiphase space $P$ is that it
carries a naturally defined multicanonical form $\,\theta$ whose exterior
derivative is, up to a sign, the multisymplectic form $\,\omega\,$:
\begin{equation} \label{eq:EXMSM}
 \omega~= \; - \, d \>\! \theta~.
\end{equation}
The global construction can be found in Refs~\cite{FPR,CCI,GIM}, so we shall
just state their explicit form in adapted local coordinates:
\begin{equation} \label{eq:MCANF02}
 \theta~=~p\;\!_i^\mu \; dq^i \,\smwedge\; d^{\,n} x_\mu^{} \, + \;
          p \; d^{\,n} x~.
\end{equation}
\begin{equation} \label{eq:MSYMF02}
 \omega~=~dq^i \,\smwedge\; dp\;\!_i^\mu \,\smwedge\; d^{\,n} x_\mu^{} \, - \;
          dp \,\:\smwedge\; d^{\,n} x~.
\end{equation}
Here, we have already employed part of the following conventions concerning
local differential forms defined by a system of adapted local coordinates,
which will be used systematically throughout this paper:
\begin{equation} \label{eq:VOLDF1}
 d^{\,n} x~=~dx^1 \,\smwedge \ldots \smwedge\; dx^n~,
\end{equation}
\begin{equation} \label{eq:CMPDF1}
 d^{\,n} x_\mu^{}~=~i_{\partial_\mu}^{} \, d^{\,n} x~
 =~(-1)^{\mu-1} \, dx^1 \,\smwedge \ldots \smwedge\; dx^{\mu-1} \,\smwedge\;
                   dx^{\mu+1} \,\smwedge \ldots \smwedge\; dx^n~,
\end{equation}
\begin{equation} \label{eq:CMPDF2}
  d^{\,n} x_{\mu\nu}^{}~=~i_{\partial_\nu}^{} i_{\partial_\mu}^{} \, d^{\,n} x
  \quad \ldots \quad
  d^{\,n} x_{\mu_1 \ldots \mu_r}^{}~
  =~i_{\partial_{\mu_r}}^{} \ldots\, i_{\partial_{\mu_1}}^{} \, d^{\,n} x~.
\end{equation}
This implies
\begin{equation} \label{eq:CMPDF3}
 i_{\partial_\mu}^{} \, d^{\,n} x_{\mu_1 \ldots\, \mu_r}^{}~
 =~d^{\,n} x_{\mu_1 \ldots\, \mu_r \mu}^{}~,
\end{equation}
whereas
\vspace{2mm}
\begin{equation} \label{eq:CMPDF4}
 dx^\kappa \,\smwedge\; d^{\,n} x_\mu^{}~=~\delta_\mu^\kappa \, d^{\,n} x~,
\vspace{2mm}
\end{equation}
\begin{equation} \label{eq:CMPDF5}
 dx^\kappa \,\smwedge\; d^{\,n} x_{\mu\nu}^{}~
 =~\delta_\nu^\kappa \, d^{\,n} x_\mu^{} \, - \,
   \delta_\mu^\kappa \, d^{\,n} x_\nu^{}~,
\end{equation}
\begin{equation} \label{eq:CMPDF6}
 dx^\kappa \,\smwedge\; d^{\,n} x_{\mu_1 \ldots\, \mu_r}^{}~
 =~\sum_{s=1}^r \, (-1)^{r-s} \, \delta_{\mu_s}^\kappa \,
   d^{\,n} x_{\mu_1 \ldots\, \mu_{s-1} \mu_{s+1} \ldots\, \mu_r}^{}~.
\end{equation}
For later use, we also recall the definition of the Lie derivative of a
differential form $\alpha$ along an $r$-multivector field $X$,
\begin{equation} \label{eq:LDFMVF}
 L_X^{} \alpha~=~d \, i_X^{} \alpha \, - \,
                 (-1)^r \, i_X^{} \, d \>\! \alpha~,
\end{equation}
which leads to the following relations, valid for any differential form
$\alpha$ and any two multivector fields $X$ and $Y$ of tensor degrees~$r$
and~$s$, respectively,
\begin{eqnarray}
 &d \;\! L_X^{} \alpha~=~(-1)^{r-1} \, L_X^{} \;\! d \>\! \alpha~,&
 \label{eq:dLX-LXd} \\[1mm]
 &i_{[X,Y]}^{} \alpha~=~(-1)^{(r-1)s} \, L_X^{} i_Y^{} \alpha \, - \,
                       i_Y^{} L_X^{} \alpha~,&
 \label{eq:LXiY-iYLX} \\[1mm]
 &L_{[X,Y]}^{} \alpha~=~(-1)^{(r-1)(s-1)} \, L_X^{} L_Y^{} \alpha \, - \,
                       L_Y^{} L_X^{} \alpha~,&
 \label{eq:LXLY-LYLX} \\[1mm]
 &L_{X \wedge\>\! Y}^{} \alpha~=~(-1)^s \, i_Y^{} L_X^{} \alpha \, + \,
                                L_Y^{} i_X^{} \alpha~,&
 \label{eq:iYLX-LYiX}
\end{eqnarray}
where $[X,Y]$ denotes the Schouten bracket of $X$ and $Y$. For decomposable
multivector fields $\, X = X_1 \,\smwedge \ldots \smwedge\, X_r \,$ and
$\, Y = Y_1 \,\smwedge \ldots \smwedge\, Y_s$, it can be defined in terms
of the Lie bracket of vector fields according to the formula
\begin{equation}
 [X,Y]~=~\sum_{i=1}^r \sum_{j=1}^s (-1)^{i+j} \, [X_i,X_j] \,\smwedge\,
         X_1 \,\smwedge \ldots \widehat{X_i} \ldots \smwedge\, X_r \,\smwedge\,
         Y_1 \,\smwedge \ldots \widehat{\!\;Y_j\,\!} \ldots \smwedge\, Y_s~,
\end{equation}
where as usual the hat over a symbol denotes its omission. We shall also write
\begin{equation}
 L_X^{} Y~=~[X,Y]~,
\end{equation}
for any two multivector fields $X$ and $Y$. For properties of the Schouten
bracket, we refer to \cite{Tul}. A proof of the above identities relating
the Schouten bracket and the Lie derivative of forms along multivector
fields can be found in the appendix of Ref.~\cite{FPR}.

\vspace{\fill}

A second property of the extended multiphase space $P$ which provides
additional structures for tensor calculus on this manifold is that it
is the total space of a fiber bundle, which implies that we may speak
of vertical vectors and horizontal covectors. In fact, it is so in no
less than three different ways. Namely, $P$ is the total space of a
fiber bundle over~$M$ (with respect to the so-called source projection),
the total space of a vector bundle over~$E$ (with respect to the so-called
target projection) and the total space of an affine line bundle over the
ordinary multiphase space~$P_0$~\cite{FPR}. Therefore, the notions of
verticality for multivector fields and of horizontality for differential
forms on~$P$ admit different interpretations, depending on which projection
is used. In any~case, one starts by defining tangent vectors to the total
space of a fiber bundle to be vertical if they are annihilated by the
tangent map to the bundle projection, or what amounts to the same
thing, if they are tangent to the fibers. Dually, a $k$-form on the
total space of a fiber bundle is said to be $l$-horizontal if it
vanishes whenever one inserts at least $k-l+1$ vertical tangent
vectors; the standard horizontal forms are obtained by taking
$l\!=\!k$. Finally, an $r$-multivector on the total space of a
fiber bundle is said to be $s$-vertical if its contraction with any
$(r-s+1)$-horizontal form vanishes.  It~is not difficult to show that
these definitions are equivalent to requiring that, locally, an
$l$-horizontal $k$-form should be a sum of exterior products of $k$
one-forms, among which there are at least $l$ horizontal ones, and
that an $s$-vertical $r$-multivector field should be a sum of exterior
products of $r$ tangent vectors, among which there are at least $s$
vertical ones. Using this rule, properties of verticality for
multivectors or horizontality for forms are easily derived from the
corresponding properties for vectors or one-forms, respectively, which
in the case of the extended multiphase space~$P$ and in adapted local
coordinates $(x_{\vphantom{i}}^\mu,q_{\vphantom{i}}^i,p\>\!_i^\mu,p\>\!)$
are summarized in Tables~1 and~2 below.

\vspace{\fill}

In what follows, the terms ``vertical'' and ``horizontal'' will usually
refer to the source projection, except when explicitly stated otherwise.

\vspace{\fill}

\begin{center}
 \begin{tabular}{|c|c|c|c|} \hline
  \rule[-10mm]{0mm}{23mm} Tangent vectors \rule[-10mm]{0mm}{23mm}
  & \begin{minipage}{3.2cm}
     \begin{center}
      {\small vertical \\ with respect to the \\
              projection \\ onto $P_0$}
     \end{center}
    \end{minipage}
  & \begin{minipage}{3.2cm}
     \begin{center}
      {\small vertical \\ with respect to the \\
              target projection \\ onto $E$}
     \end{center}
    \end{minipage}
  & \begin{minipage}{3.2cm}
     \begin{center}
      {\small vertical \\ with respect to the \\
              source projection \\ onto $M$}
     \end{center}
    \end{minipage} \\ \hline
  \rule{0mm}{8mm}
  ${\displaystyle \frac{\partial}{\partial p}}$
  \rule{0mm}{8mm}
  & yes & yes & yes \\
  ${\displaystyle \frac{\partial}{\partial p\>\!_i^\mu}}$
  \rule{0mm}{8mm}
  & no  & yes & yes \\
  \rule{0mm}{8mm}
  ${\displaystyle \frac{\partial}{\partial q_{\vphantom{i}}^i}}$
  \rule{0mm}{8mm}
  & no  & no  & yes \\
  \rule[-5mm]{0mm}{14mm}
  ${\displaystyle \frac{\partial}{\partial x_{\vphantom{i}}^\mu}}$
  \rule[-5mm]{0mm}{14mm}
  & no  & no  & no  \\ \hline
 \end{tabular} \\[4mm]
 Table 1: Verticality of tangent vectors on extended multiphase space
\end{center}

\begin{center}
 \begin{tabular}{|c|c|c|c|} \hline
  \rule[-10mm]{0mm}{23mm} One-forms \rule[-10mm]{0mm}{23mm}
  & \begin{minipage}{3.2cm}
     \begin{center}
      {\small horizontal \\ with respect to the \\
              projection \\ onto $P_0$}
     \end{center}
    \end{minipage}
  & \begin{minipage}{3.2cm}
     \begin{center}
      {\small horizontal \\ with respect to the \\
              target projection \\ onto $E$}
     \end{center}
    \end{minipage}
  & \begin{minipage}{3.2cm}
     \begin{center}
      {\small horizontal \\ with respect to the \\
              source projection \\ onto $M$}
     \end{center}
    \end{minipage} \\ \hline
  \rule{0mm}{6mm} $dp$ \rule{0mm}{6mm}
  & no  & no  & no \\
  \rule{0mm}{6mm} $dp\>\!_i^\mu$ \rule{0mm}{6mm}
  & yes & no  & no \\
  \rule{0mm}{6mm}
  $dq_{\vphantom{i}}^i$
  \rule{0mm}{6mm}
  & yes & yes & no \\
  \rule[-3mm]{0mm}{9mm}
  $dx_{\vphantom{i}}^\mu$
  \rule[-3mm]{0mm}{9mm}
  & yes & yes & yes \\ \hline
 \end{tabular} \\[4mm]
 Table 2: Horizontality of cotangent vectors on extended multiphase space
\end{center}

A third important feature of the extended multiphase space $P$ is that it
carries a naturally defined vector field $\Sigma$, the scaling vector field
or Euler vector field, which exists on any manifold that is the total space
of a vector bundle. In adapted local coordinates,
\begin{equation} \label{eq:SVFEMPS}
 \Sigma~=~p\;\!_i^\mu \, \frac{\partial}{\partial p\;\!_i^\mu} \, + \,
          p \: \frac{\partial}{\partial p}~.
\end{equation}
It is then easy to verify the following relations (see Proposition~2.1 of
Ref.~\cite{FPR}):
\begin{eqnarray}
 &L_\Sigma^{} \theta~=~\theta~.&                     \label{eq:MCANF03} \\[1mm]
 &L_\Sigma^{} \omega~=~\omega~.&                     \label{eq:MSYMF03} \\[1mm]
 &i_\Sigma^{} \theta~=~0~.&                          \label{eq:MCANF04} \\[1mm]
 &i_\Sigma^{} \omega~= \; - \, \theta~.&             \label{eq:MSYMF04}
\end{eqnarray}
In particular, the last equation means that the scaling vector field
allows to reconstruct $\,\theta$ from~$\,\omega$. But the main utility of
$\Sigma$ is that taking the Lie derivative $L_\Sigma$ along $\Sigma$
provides a device for controlling the dependence of functions and, more
generally, of tensor fields on~$P$ on the multimomentum variables and the
energy variable, that is, along the fibers of~$P$ over~$E$: $L_\Sigma$ has
only integer eigenvalues, and eigenfunctions of $L_\Sigma$ with eigenvalue
$k$ are homogeneous polynomials of degree $k$ in these variables.

As we shall see soon, homogeneity under $L_\Sigma$ plays a central role in
the analysis of various classes of multivector fields and differential forms
on~$P$.

Let us recall a few definitions. An $r$-multivector field $X$ on~$P$
is called \emph{locally Hamiltonian\/} if $i_X^{} \omega$ is closed,
or equivalently, if
\begin{equation} \label{eq:LHAMMVF}
 L_X^{} \omega~=~0~.
\end{equation}
It is called \emph{globally Hamiltonian\/} if $i_X^{} \omega$ is exact,
that is, if there exists an $(n-r)$-form $f$ on~$P$ such that
\begin{equation} \label{eq:HAMMVFF}
 i_X^{} \omega~=~df~.
\end{equation}
In this case, $f$ is said to be a \emph{Hamiltonian form associated with $X$}.
Finally, it is called \emph{exact Hamiltonian\/} if
\begin{equation} \label{eq:EHAMMVF}
 L_X^{} \theta~=~0~.
\end{equation}
Of course, exact Hamiltonian multivector fields are globally Hamiltonian
(to show this, set $\, f = (-1)^{r-1} i_X^{} \theta \,$ and apply
eqs~(\ref{eq:LDFMVF}) and~(\ref{eq:EXMSM})), and globally Hamiltonian
multivector fields are obviously locally Hamiltonian. Conversely, an
$(n-r)$-form $f$ on~$P$ is called a \emph{Hamiltonian form\/} if there
exists an $r$-multivector field $X$ on~$P$ such that eq.~(\ref{eq:HAMMVFF})
holds; in this case, $X$ is said to be a \emph{Hamiltonian multivector field
associated with $f$}. Moreover, $f$ is called a \emph{Poisson form\/} if
in addition, it vanishes on the kernel of~$\,\omega$, that is, if for any
multivector field~$Z$, we have
\begin{equation} \label{eq:KERN1}
 i_Z^{} \, \omega~=~0~~\Longrightarrow~~i_Z^{} f~=~0~.
\end{equation}
A trivial example of a Poisson form is the multisymplectic form $\,\omega$
itself. Another example is provided by the multicanonical form $\,\theta$,
since it follows trivially from eq.~(\ref{eq:MSYMF04}) that $\,\theta$
vanishes on the kernel of $\,\omega$.

Concerning stability under the Lie derivative along the scaling vector field
$\Sigma$, we have the following
\begin{prp} \label{prp:INVLIESIGMVF}~~
 The space $\mathfrak{X}_{LH}^\wedge(P)$ of locally Hamiltonian multivector
 fields, the space $\mathfrak{X}_H^\wedge(P)$ of globally Hamiltonian
 multivector fields, the space $\mathfrak{X}_{EH}^\wedge(P)$ of exact
 Hamiltonian multivector fields and the space $\mathfrak{X}_0^\wedge(P)$
 of multivector fields taking values in the kernel of $\,\omega$ are all
 invariant under the Lie derivative along the scaling vector field $\Sigma$:
 \begin{equation} \label{eq:INVLIESIGLH}
  L_X^{} \omega~=~0 \quad \Longrightarrow \quad L_{[\Sigma,X]}^{} \omega~=~0~,
 \end{equation}
 \begin{equation} \label{eq:INVLIESIGGH}
  i_X^{} \omega~=~df \quad \Longrightarrow \quad
  i_{[\Sigma,X]}^{} \omega~=~d \left( L_\Sigma^{} f - f \right)\!~,
 \end{equation}
 %\pagebreak
 \begin{equation} \label{eq:INVLIESIGEH}
  L_X^{} \theta~=~0 \quad \Longrightarrow \quad L_{[\Sigma,X]}^{} \theta~=~0~,
 \end{equation}
 \begin{equation} \label{eq:INVLIESIGKO}
  i_\xi^{} \omega~=~0 \quad \Longrightarrow \quad
  i_{[\Sigma,\xi]}^{} \omega~=~0~.
 \end{equation}
\end{prp}
\textbf{Proof.}~~All these relations can be shown by direct calculation.
For example, eqs~(\ref{eq:INVLIESIGLH}) and~(\ref{eq:INVLIESIGEH}) follow
directly from combining eq.~(\ref{eq:LXLY-LYLX}) with eqs~(\ref{eq:MSYMF03})
and~(\ref{eq:MCANF03}), respectively. Similarly, eq.~(\ref{eq:INVLIESIGGH})
follows directly from combining eq.~(\ref{eq:LXiY-iYLX}) with
eqs~(\ref{eq:MSYMF03}) and~(\ref{eq:dLX-LXd}).  Finally,
eq.~(\ref{eq:INVLIESIGKO}) is a special case of
eq.~(\ref{eq:INVLIESIGGH}), obtained by putting $f=0$. \\
\PCPqed \\
Dually, we have
\begin{prp} \label{prp:INVLIESIGFOR}~~
 The space $\Omega_H^{}(P)$ of Hamiltonian forms, the space $\Omega_0^{}(P)$
 of forms that vanish on the kernel of $\,\omega$ and the space
 $\Omega_P^{}(P)$ of Poisson forms are all invariant under the
 Lie derivative along the scaling vector field $\Sigma$:
 \begin{equation} \label{eq:INVLIESIGHF}
  df~=~i_X^{} \omega \quad \Longrightarrow \quad
  d \left( L_\Sigma^{} f \right)\!~=~i_{X+[\Sigma,X]}^{} \omega~.
 \end{equation}
\end{prp}
\textbf{Proof.}~~The first statement is a consequence of
eq.~(\ref{eq:INVLIESIGHF}), which follows directly from combining
eqs~(\ref{eq:dLX-LXd}) and~(\ref{eq:LXiY-iYLX}) with
eq.~(\ref{eq:MSYMF03}).  For the second statement, assume that $f$
vanishes on the kernel of~$\,\omega$. Then if $\xi$ is any multivector
field $\xi$ taking values in the kernel of~$\,\omega$, the multivector
field $[\Sigma,\xi]$ takes values in the kernel of~$\,\omega$ as well
(cf. eq.~(\ref{eq:INVLIESIGKO})), so that according to
eq.~(\ref{eq:LXiY-iYLX}),
\[
 i_\xi^{} \left( L_\Sigma^{} f \right)\!~
 =~L_\Sigma^{} i_\xi^{} f \, - \, i_{[\Sigma,\xi]}^{} f~=~0~.
\]
But this means that $L_\Sigma^{} f$ vanishes on the kernel of~$\,\omega$.
Finally, the third statement follows by combining the first two. \\
\PCPqed

A special class of multivector fields and of differential forms on~$P$ which
will be of particular importance in what follows is that of \emph{fiberwise
polynomial multivector fields} and of \emph{fiberwise polynomial differential
forms} on~$P$: their coefficients are polynomials along the fibers of~$P$
over~$E$, or in other words, polynomials in the multimomentum \linebreak
variables and the energy variable. The main advantage of working with
tensor fields on the total space of a vector bundle which are fiberwise
polynomial is that they allow a unique and globally defined (or in other
words, coordinate independent) decomposition into homogeneous components,
according to the different eigenspaces of the Lie derivative $L_\Sigma^{}$
along $\Sigma$; the corresponding eigenvalue will in what follows be called
the \emph{scaling degree}  (to distinguish it from the ordinary tensor degree).
In doing so, it must be borne in mind that, in an expansion with respect to an
adapted local coordinate system, the scaling degree receives contributions not
only from the coefficient functions but also from some of the coordinate
vector fields and differentials since the vector fields $\, \partial/
\partial x_{\vphantom{i}}^\mu$, $\partial/\partial q_{\vphantom{i}}^i$,
$\partial/\partial p\>\!_i^\mu$ and $\partial/\partial p \,$ carry scaling
degree $0$, $0$, $-1$ and $-1$, respectively, while the differentials
$\, dx_{\vphantom{i}}^\mu$, $dq_{\vphantom{i}}^i$, $dp\;\!_i^\mu$ and
$dp \,$ carry scaling degree $0$, $0$, $+1$ and $+1$, respectively;
moreover, the scaling degree is additive under the exterior product,
since $L_\Sigma^{}$ is a derivation. Therefore, a fiberwise polynomial
$r$-multivector field on~$P$ admits a globally defined decomposition
into a finite sum
\begin{equation} \label{eq:LIESIGD1}
 X~=~\sum_{s {\scriptscriptstyle \geqslant} -r} X_s~,
\end{equation}
where $X_s$ is its homogeneous component of scaling degree $s$:
\begin{equation} \label{eq:LIESIGD2}
 L_\Sigma^{} X_s~=~s \, X_s~.
\end{equation}
Each $X_s$ can be obtained from $X$ by applying a projector which is
itself a polynomial in~$L_\Sigma^{}$:
\begin{equation} \label{eq:LIESIGD3}
 X_s~=~\prod_{\stackrel{s' {\scriptscriptstyle \geqslant} -r}
             {s' {\scriptscriptstyle \neq} s}} \,
       \frac{1}{s\!-\!s'} \left( L_\Sigma^{} - s' \right) X~.
\end{equation}
Similarly, a fiberwise polynomial $(n-r)$-form $f$ on~$P$ admits a globally
defined decomposition into a finite sum
\begin{equation} \label{eq:LIESIGD4}
 f~=~\sum_{s {\scriptscriptstyle \geqslant} 0} f_s~,
\end{equation}
where $f_s$ is its homogeneous component of scaling degree $s$:
\begin{equation} \label{eq:LIESIGD5}
 L_\Sigma^{} f_s~=~s \, f_s~.
\end{equation}
Each $f_s$ can be obtained from $f$ by applying a projector which is
itself a polynomial in~$L_\Sigma^{}$:
\begin{equation} \label{eq:LIESIGD6}
 f_s~=~\prod_{\stackrel{s' {\scriptscriptstyle \geqslant} 0}
                       {s' {\scriptscriptstyle \neq} s}} \,
       \frac{1}{s\!-\!s'} \left( L_\Sigma^{} - s' \right) f~.
\end{equation}

The relevance of these decompositions for locally Hamiltonian multivector
fields and for Hamiltonian forms on the extended multiphase space~$P$ stems
from the following theorems, whose proof will follow from statements to be
derived in the course of the next two sections, by means of explicit
calculations in adapted local coordinates.
\begin{thm} \label{thm:FIBPOL}~~
 Except for trivial contributions, locally Hamiltonian multivector fields
 and Hamiltonian forms on~$P$ are fiberwise polynomial. More precisely, we
 have:
 \begin{itemize}
  \item Any locally Hamiltonian $r$-multivector field on~$P$, with
        $\, 0 < r < n$, can be decomposed into the sum of a fiberwise
        polynomial locally Hamiltonian $r$-multivector field and an
        $r$-multivector field taking values in the kernel of~$\,\omega$.
        Such a decomposition is unique up to fiberwise polynomial
        $r$-multivector fields taking values in the kernel of~$\,\omega$.
        (Note that for $r\!=\!1$, this decomposition is trivial.)
  \item Any Hamiltonian form (Poisson form) of degree $n-r$ on~$P$,
        with $\, 0 < r < n$, can be decomposed into the sum of a
        fiberwise polynomial Hamiltonian form (fiberwise polynomial
        Poisson form) of degree $n-r$ and a closed form (closed form
        vanishing on the kernel of $\,\omega$) of degree $n-r$. Such
        a decomposition is unique up to fiberwise polynomial closed
        forms (up to fiberwise polynomial closed forms vanishing on
        the kernel of~$\,\omega$) of degree $n-r$.
 \end{itemize}
\end{thm}
More specifically, we have:
\begin{thm} \label{thm:DECSCD}~~
 Fiberwise polynomial locally Hamiltonian $r$-multivector fields and fiberwise
 polynomial Hamiltonian forms of degree $n-r$ have non-trivial homogeneous
 components of scaling degree $s$ only for $\, s = -1,0,\ldots,r-1 \,$ and
 for $\, s = 0,1,\ldots,r \,$, respectively. More precisely, we have:
 \begin{itemize}
  \item Every fiberwise polynomial locally Hamiltonian (Hamiltonian, exact
        Hamiltonian) $r$-multivector field~$X$ on~$P$, with $\, 0 < r < n$,
        admits a unique, globally defined decomposition into homogeneous
        components with respect to scaling degree, which can be written
        in the form\footnote{We abbreviate $X_{-1}^{}$ as $X_-^{}$.}
        \begin{equation} \label{eq:HAMMVFSCD1}
         X~=~X_-^{} + X_+^{} + \, \xi \qquad \mbox{with} \qquad
         X_+^{}~=~\sum_{s=0}^{r-1} X_s^{}~,
        \end{equation}
        where each $X_s^{}$ is locally Hamiltonian (Hamiltonian, exact
        Hamiltonian) and
        \begin{equation} \label{eq:HAMMVFSCD2}
         \xi~=~\sum_{-r {\scriptscriptstyle \leqslant} s
                        {\scriptscriptstyle \leqslant} -2} \xi_s \, + \,
               \sum_{s {\scriptscriptstyle \geqslant}\, r} \xi_s
        \end{equation}
        is a fiberwise polynomial $r$-multivector field on~$P$ taking
        values in the kernel of~$\,\omega$.
  \item Every fiberwise polynomial Hamiltonian form (Poisson form) $f$
        of degree $n-r$ on~$P$, with $\, 0 < r < n$, admits a unique,
        globally defined decomposition into homogeneous components with
        respect to scaling degree, which can be written in the form
        \begin{equation} \label{eq:HAMFORSCD1}
         f~=~f_0^{} + f_+^{} + f_c^{} \qquad \mbox{with} \qquad
         f_+^{}~=~\sum_{s=1}^r f_s^{}~,
        \end{equation}
        where each $f_s^{}$ is Hamiltonian (Poisson) and
        \begin{equation} \label{eq:HAMFORSCD2}
         f_c^{}~=~\sum_{s {\scriptscriptstyle \geqslant}\, r+1} (f_c^{})_s^{}
        \end{equation}
        is a fiberwise polynomial closed $(n-r)$-form on~$P$.
 \end{itemize}
\end{thm}

The cases $r\!=\!0$ and $r\!=\!n$ are exceptional and must be dealt with
separately; see Propositions~\ref{prp:HAMFUN} and~\ref{prp:POISFN} for
$r\!=\!0$ and Propositions~\ref{prp:HAMNVF} and~\ref{prp:POISF0} for
$r\!=\!n$.

In view of these theorems, it is sufficient to study locally
Hamiltonian multivector fields and Hamiltonian forms which are
homogeneous under the Lie derivative along the scaling vector field
$\Sigma$. This condition of homogeneity is also compatible with the
correspondence between globally Hamiltonian multivector fields~$X$
and Hamiltonian forms~$f$ established by the fundamental
relation~(\ref{eq:HAMMVFF}), because $\,\omega$ itself is homogeneous:
according to eq.~(\ref{eq:MSYMF03}), $\omega$ has scaling degree $1$.
Indeed, except for the ambiguity inherent in this correspondence
($f$ determines $X$ only up to a multivector field taking values in
the kernel of~$\,\omega$ and $X$ determines $f$ only up to a closed
form), eq.~(\ref{eq:HAMMVFF}) preserves the scaling degree, up to a
shift by $1$: $X$ is homogeneous with scaling degree $s-1$ if and
only if $f$ is homogeneous with scaling degree $s$:
\begin{equation} \label{eq:HOMHAM1}
 \begin{array}{c}
  L_\Sigma^{} X~=~(s-1) X \\[1mm]
  \mbox{modulo multivector fields} \\
  \mbox{taking values in the kernel of $\,\omega$}
 \end{array}
 \quad \Longleftrightarrow \quad
 \begin{array}{c}
  L_\Sigma^{} f~=~s f \\[1mm]
 \mbox{modulo closed forms}
 \end{array}
\end{equation}
For a proof, note that the condition on the lhs amounts to requiring that
$\, i_{[\Sigma,X]}^{} \omega = (s-1) \, i_X^{} \omega$, while the condition
on the rhs amounts to requiring that $\, d \, L_\Sigma^{} f = s \, df$,
so the equivalence stated in eq.~(\ref{eq:HOMHAM1}) is an immediate
consequence of eq.~(\ref{eq:INVLIESIGHF}). A particular case occurs
when $\, s=1$, since the locally Hamiltonian multi\-vector fields which
are homogeneous of scaling degree $0$ are precisely the exact Hamiltonian
multivector fields: for $\, L_X^{} \omega = 0$,
\begin{equation} \label{eq:HOMHAM2}
 \begin{array}{c}
  L_\Sigma^{} X~=~0 \\[1mm]
  \mbox{modulo multivector fields} \\
  \mbox{taking values in the kernel of $\,\omega$}
 \end{array}
 \quad \Longleftrightarrow \quad
 L_X^{} \theta~=~0~.
\end{equation}
Indeed, combining eqs~(\ref{eq:MSYMF04}) and~(\ref{eq:LXiY-iYLX}) gives
\begin{equation} \label{eq:LTIO}
 L_X^{} \theta~= \; - \, L_X^{} i_\Sigma^{} \omega~
 =~(-1)^r \left( i_{[X,\Sigma]}^{} \omega \, - \,
                 i_\Sigma^{} L_X^{} \omega \right)\!~
 =~(-1)^{r-1} \, i_{[\Sigma,X]}^{} \omega~.
\end{equation}
More generally, the fundamental relation~(\ref{eq:HAMMVFF}) preserves the
property of being fiberwise polynomial, in the following sense: If~$X$ is
a fiberwise polynomial Hamiltonian $r$-multivector field and $f$ is a
Hamiltonian $(n-r)$-form associated with~$X$, then modifying $f$ by
addition of an appropriate closed $(n-r)$-form if necessary, we may
always assume, without loss of generality, that $f$ is fiberwise
polynomial as well. Conversely, if~$f$ is a fiberwise polynomial
Hamiltonian $(n-r)$-form and $X$ is a Hamiltonian $r$-multivector
field associated with~$f$, then modifying $X$ by addition of an
appropriate $r$-multivector field taking values in the kernel
of~$\,\omega$ if necessary, we may always assume, without loss
of generality, that $X$ is fiberwise polynomial as well.

%%%%%%%%%%%%%%%%%%%%%%%%%%%%%%%%%%%%%%%%%%%%%%%%%%%%%%%%%%%%%%%%%%%%%%%%%%%%%%%
\section{Hamiltonian multivector fields}

Our aim in this section is to determine the explicit form, in adapted local
coordinates, of locally Hamiltonian $r$-multivector fields on the extended
multiphase space~$P$, where $\, 0 \leqslant r \leqslant n+1$. (Multivector
fields of tensor degree $> n+1$ are uninteresting since they always take
their values in the kernel of $\,\omega$.)

As a first step towards this goal, we shall determine the explicit form,
in adapted local coordinates, of the multivector fields on~$P$ taking
values in the kernel of $\omega$; this will also serve to identify, in the
next section, the content of the kernel condition~(\ref{eq:KERN1}) that
characterizes Poisson forms. To this end, note first that $\,\omega$ being
a homogeneous differential form (of degree $n+1$), its kernel is graded,
that is, if an inhomogeneous multivector field takes values in the kernel
of $\,\omega$, so do all its homogeneous components.
\begin{prp} \label{prp:KERNOMEGA}~~
 An $r$-multivector field on~$P$, with $\, r > 1$, takes values in the
 kernel of~$\,\omega$ if and only if, in adapted local coordinates, it
 can be written as a linear combination of\/ $3$-vertical terms, of the
 $2$-vertical terms
 \begin{equation} \label{eq:KERN4}
  \begin{array}{cccc}
   {\displaystyle
    \frac{\partial}{\partial q^i} \;\smwedge\;
    \frac{\partial}{\partial q^j} \;\smwedge\;
    \frac{\partial}{\partial x_{\phantom{i}}^{\mu_3}}
    \;\smwedge \ldots \smwedge\;
    \frac{\partial}{\partial x_{\phantom{i}}^{\mu_r}}} &,&
   {\displaystyle
    \frac{\partial}{\partial p\>\!_k^\kappa} \;\smwedge\;
    \frac{\partial}{\partial p\>\!_l^\lambda} \;\smwedge\;
    \frac{\partial}{\partial x_{\phantom{i}}^{\mu_3}}
    \;\smwedge \ldots \smwedge\;
    \frac{\partial}{\partial x_{\phantom{i}}^{\mu_r}}} & ,              \\[4mm]
   {\displaystyle
    \frac{\partial}{\partial q^i} \;\smwedge\;
    \frac{\partial}{\partial p} \;\smwedge\;
    \frac{\partial}{\partial x_{\phantom{i}}^{\mu_3}}
    \;\smwedge \ldots \smwedge\;
    \frac{\partial}{\partial x_{\phantom{i}}^{\mu_r}}} &,&
   {\displaystyle
    \frac{\partial}{\partial p\>\!_k^\kappa} \;\smwedge\;
    \frac{\partial}{\partial p} \;\smwedge\;
    \frac{\partial}{\partial x_{\phantom{i}}^{\mu_3}}
    \;\smwedge \ldots \smwedge\;
    \frac{\partial}{\partial x_{\phantom{i}}^{\mu_r}}} &,
  \end{array}
 \end{equation}
 and of the $1$-vertical terms
 \begin{equation} \label{eq:KERN5}
  \left( \frac{\partial}{\partial q^i} \;\smwedge\;
         \frac{\partial}{\partial p\>\!_k^\kappa} \; + \; \delta_i^k \;
         \frac{\partial}{\partial p} \;\smwedge\;
         \frac{\partial}{\partial x_{\phantom{i}}^\kappa} \right) \smwedge\;
         \frac{\partial}{\partial x_{\phantom{i}}^{\mu_3}}
         \;\smwedge \ldots \smwedge\;
         \frac{\partial}{\partial x_{\phantom{i}}^{\mu_r}}~,
 \end{equation}
 \begin{equation} \label{eq:KERN6}
  \left( \frac{\partial}{\partial p\>\!_i^{\mu_1}} \;\smwedge\;
         \frac{\partial}{\partial x_{\phantom{i}}^{\mu_2}} \; + \;
         \frac{\partial}{\partial p\>\!_i^{\mu_2}} \;\smwedge\;
         \frac{\partial}{\partial x_{\phantom{i}}^{\mu_1}} \right) \smwedge\;
  \frac{\partial}{\partial x_{\phantom{i}}^{\mu_3}}
  \;\smwedge \ldots \smwedge\,
  \frac{\partial}{\partial x_{\phantom{i}}^{\mu_r}}~.
 \vspace{4mm}
 \end{equation}
 Thus every $r$-multivector field $X$ on~$P$ admits, in adapted local
 coordinates, a unique decomposition of the form
 \begin{eqnarray} \label{eq:HAMMVF1}
  X \!\!
  &=&\!\! \frac{1}{r!} \; X_{\phantom{i}}^{\mu_1 \ldots\, \mu_r} \;
          \frac{\partial}{\partial x_{\phantom{i}}^{\mu_1}}
          \;\smwedge \ldots \smwedge\;
          \frac{\partial}{\partial x_{\phantom{i}}^{\mu_r}}   \nonumber \\[1mm]
  & &\!   \mbox{} + \; \frac{1}{(r\!-\!1)!} \;
          X_{\phantom{i}}^{i,\mu_2 \ldots\, \mu_r} \;
          \frac{\partial}{\partial q^i} \;\smwedge\;
          \frac{\partial}{\partial x_{\phantom{i}}^{\mu_2}}
          \;\smwedge \ldots \smwedge\;
          \frac{\partial}{\partial x_{\phantom{i}}^{\mu_r}}   \nonumber \\
  & &\!   \mbox{} +~~~~\frac{1}{r!}~~~~
          X_i^{\mu_1 \ldots\, \mu_r} \;
          \frac{\partial}{\partial p\>\!_i^{\mu_1}} \;\smwedge\;
          \frac{\partial}{\partial x_{\phantom{i}}^{\mu_2}}
          \;\smwedge \ldots \smwedge\;
          \frac{\partial}{\partial x_{\phantom{i}}^{\mu_r}}             \\
  & &\!   \mbox{} + \; \frac{1}{(r\!-\!1)!} \;
          \tilde{X}_{\phantom{i}}^{\mu_2 \ldots\, \mu_r} \;
          \frac{\partial}{\partial p} \;\smwedge\;
          \frac{\partial}{\partial x_{\phantom{i}}^{\mu_2}}
          \;\smwedge \ldots \smwedge\;
          \frac{\partial}{\partial x_{\phantom{i}}^{\mu_r}}   \nonumber \\[4mm]
  & &\!   \mbox{} + \; \xi~,                                  \nonumber
 \end{eqnarray}
 where all coefficients are totally antisymmetric in their space-time indices
 and $\xi$ takes values in the kernel of $\,\omega$; then
 \begin{equation} \label{eq:HAMMVF2}
  \begin{array}{rcl}
   i_X^{} \omega \!\!
   &=&\!\! {\displaystyle \mbox{} - ~
            \frac{1}{(r\!-\!1)!}~ \;
            \tilde{X}_{\phantom{i}}^{\mu_2 \ldots\, \mu_r} \;
            d^{\,n} x_{\mu_2 \ldots\, \mu_r}^{}}                        \\[4mm]
   & &\!\! {\displaystyle \mbox{} +~~
            \frac{(-1)^r}{r!}~~X_i^{\mu_1 \ldots\, \mu_r} \;
            dq^i \,\smwedge\; d^{\,n} x_{\mu_1 \ldots\, \mu_r}^{}}      \\[4mm]
   & &\!\! {\displaystyle \mbox{} + \;
            \frac{(-1)^{r-1}}{(r\!-\!1)!} \;
            X_{\phantom{i}}^{i,\mu_2 \ldots\, \mu_r} \;
            dp\>\!_i^\mu \,\smwedge\;
            d^{\,n} x_{\mu \mu_2 \ldots\, \mu_r}^{}}                    \\[4mm]
   & &\!\! {\displaystyle \mbox{} + \;
            \frac{1}{r!} \; X_{\phantom{i}}^{\mu_1 \ldots\, \mu_r} \;
            dq^i \,\smwedge\; dp\>\!_i^\mu \,\smwedge\;
            d^{\,n} x_{\mu \mu_1 \ldots\, \mu_r}^{} \; - \;
            \frac{(-1)^r}{r!} \; X_{\phantom{i}}^{\mu_1 \ldots\, \mu_r} \;
            dp \;\smwedge\; d^{\,n} x_{\mu_1 \ldots\, \mu_r}^{}}~,
  \end{array}
 \end{equation}
 and similarly,
 \begin{equation} \label{eq:HAMMVF3}
  \begin{array}{rcl}
   i_X^{} \theta \!\!
   &=&\!\! {\displaystyle
            \frac{1}{(r\!-\!1)!} \;
            p\>\!_i^\mu X_{\phantom{i}}^{i,\mu_2 \ldots\, \mu_r} \;
            d^{\,n} x_{\mu \mu_2 \ldots\, \mu_r}^{} \; + \;
            \frac{1}{r!} \; p \, X_{\phantom{i}}^{\mu_1 \ldots\, \mu_r} \;
            d^{\,n} x_{\mu_1 \ldots\, \mu_r}^{}}                        \\[4mm]
   & &\!   {\displaystyle
            \mbox{} + \;
            \frac{(-1)^r}{r!} \;
            p\>\!_i^\mu X_{\phantom{i}}^{\mu_1 \ldots\, \mu_r} \;
            dq^i \,\smwedge\; d^{\,n} x_{\mu \mu_1 \ldots\, \mu_r}^{}}~,
  \end{array}
 \end{equation}
 where, in each of the last two equations, the first term is to be omitted
 if $\, r=n$, whereas only the last term in the first equation remains and
 $i_X^{} \theta$ vanishes identically if $\, r=n+1$.
\end{prp}
\textbf{Proof.}~~First of all, the fact that $\,\omega$ vanishes
on $3$-vertical multivector fields and on the $2$-vertical and
$1$-vertical local multivector fields written down in
eqs~(\ref{eq:KERN4})--(\ref{eq:KERN6}) follows directly from the local
coordinate expression for $\,\omega$, eq.~(\ref{eq:MSYMF02}). To prove
the converse, we write down the local coordinate expression for a
general $r$-multivector field $X$,
\begin{eqnarray*}
 X \!\!
 &=&\!\! \frac{1}{r!} \; X_{\phantom{i}}^{\mu_1 \ldots\, \mu_r} \;
         \frac{\partial}{\partial x^{\mu_1}} \;\smwedge \ldots \smwedge\;
         \frac{\partial}{\partial x^{\mu_r}}                            \\[1mm]
 & &\!   \mbox{} + \; \frac{1}{(r\!-\!1)!} \;
         X_{\phantom{i}}^{i,\mu_2 \ldots\, \mu_r} \;
         \frac{\partial}{\partial q^i} \;\smwedge\;
         \frac{\partial}{\partial x^{\mu_2}} \;\smwedge \ldots \smwedge\;
         \frac{\partial}{\partial x^{\mu_r}}                            \\
 & &\!   \mbox{} + \; \frac{1}{(r\!-\!1)!} \;
         X_{\;\,k}^{\prime \, \kappa,\mu_2 \ldots\, \mu_r} \;
         \frac{\partial}{\partial p\>\!_k^\kappa} \;\smwedge\;
         \frac{\partial}{\partial x^{\mu_2}} \;\smwedge \ldots \smwedge\;
         \frac{\partial}{\partial x^{\mu_r}}                            \\
 & &\!   \mbox{} + \; \frac{1}{(r\!-\!1)!} \;
         X^{\prime \, \mu_2 \ldots\, \mu_r} \;
         \frac{\partial}{\partial p} \;\smwedge\;
         \frac{\partial}{\partial x^{\mu_2}} \;\smwedge\;
         \frac{\partial}{\partial x^{\mu_3}} \;\smwedge \ldots \smwedge\;
         \frac{\partial}{\partial x^{\mu_r}}                            \\
 & &\!   \mbox{} + \; \frac{1}{(r\!-\!2)!} \;
         X_{\phantom{\prime \, i,}\>\!k}
          ^{\prime \, i,\kappa,\mu_3 \ldots\, \mu_r} \;
         \frac{\partial}{\partial q^i} \;\smwedge\;
         \frac{\partial}{\partial p\>\!_k^\kappa} \;\smwedge\;
         \frac{\partial}{\partial x^{\mu_3}} \;\smwedge \ldots \smwedge\;
         \frac{\partial}{\partial x^{\mu_r}}                            \\[3mm]
 & &\!   \mbox{} + \; \xi^\prime~,
\end{eqnarray*}
where $\xi^\prime$ contains the $3$-vertical terms as well as the
$2$-vertical terms listed in eq.~(\ref{eq:KERN4}) that occur in $X$ and
hence are annihilated under contraction with $\,\omega$. This leads to

%\pagebreak

\begin{eqnarray*}
 i_X^{} \omega \!\!
 &=&\!\! \frac{1}{r!} \; X_{}^{\mu_1 \ldots\, \mu_r} \;
         dq^i \,\smwedge\; dp\>\!_i^\mu \,\smwedge\;
         d^{\,n} x_{\mu \mu_1 \ldots\, \mu_r}^{} \; - \;
         \frac{(-1)^r}{r!} \; X_{}^{\mu_1 \ldots\, \mu_r} \;
         dp \;\smwedge\; d^{\,n} x_{\mu_1 \ldots\, \mu_r}^{}            \\[1mm]
 & &\!   \mbox{} + \;
         \frac{(-1)^{r-1}}{(r\!-\!1)!} \; X_{}^{i,\mu_2 \ldots\, \mu_r} \;
         dp\>\!_i^\mu \,\smwedge\;
         d^{\,n} x_{\mu \mu_2 \ldots\, \mu_r}^{}                        \\
 & &\!   \mbox{} - \;
         \frac{(-1)^{r-1}}{(r\!-\!1)!} \;
         X_{\;\;i}^{\prime \, \mu,\mu_2 \ldots\, \mu_r} \;
         dq^i \,\smwedge\; d^{\,n} x_{\mu \mu_2 \ldots\, \mu_r}^{}      \\
 & &\!   \mbox{} - \;
         \frac{1}{(r\!-\!1)!} \; X_{}^{\prime \, \mu_2 \ldots\, \mu_r} \;
         d^{\,n} x_{\mu_2 \ldots\, \mu_r}^{} \; + \;
         \frac{1}{(r\!-\!2)!} \;
         X_{\phantom{\prime \, i,}\,i}
          ^{\prime \, i,\mu,\mu_3 \ldots\, \mu_r} \;
         d^{\,n} x_{\mu \mu_3 \ldots\, \mu_r}^{}~.
\end{eqnarray*}
These two equations can conveniently be rewritten in the form
(\ref{eq:HAMMVF1}) and (\ref{eq:HAMMVF2}), respectively, by setting
\[
 X_i^{\mu_1 \ldots\, \mu_r}~
 =~\frac{1}{r} \, \sum_{s=1}^r \, (-1)^{s-1} \,
   X_{\;\;i}^{\prime \, \mu_s,\mu_1 \ldots\,
              \mu_{s-1} \mu_{s+1} \ldots\, \mu_r}~,
\vspace{-2mm}
\]
\[
 \tilde{X}_{}^{\mu_2 \ldots\, \mu_r}~
 =~X_{}^{\prime \, \mu_2 \ldots\, \mu_r} \, - \,
   \sum_{s=2}^r \, (-1)^s
   X_{\phantom{\prime \, i,}\,i}
    ^{\prime \, i,\mu_s,\mu_2 \ldots\,
      \mu_{s-1} \mu_{s+1} \ldots\, \mu_r}~,
\]
and
\begin{eqnarray*}
 \xi \!\!
 &=&\!\! \frac{1}{(r\!-\!1)!}
         \left( X_{\;\;i}^{\prime \, \mu,\mu_2 \ldots\, \mu_r} \, - \,
                X_i^{\mu \mu_2 \ldots\, \mu_r} \right)
                \frac{\partial}{\partial p\>\!_i^\mu} \;\smwedge\;
                \frac{\partial}{\partial x_{\phantom{i}}^{\mu_2}}
                \;\smwedge \ldots \smwedge\;
                \frac{\partial}{\partial x_{\phantom{i}}^{\mu_r}}            \\
 & &\!   + \; \frac{1}{(r\!-\!2)!} \;
         X_{\phantom{\prime \, i,}\>\!k}
          ^{\prime \, i,\kappa,\mu_3 \ldots\, \mu_r}
         \left( \frac{\partial}{\partial q^i} \;\smwedge\;
                \frac{\partial}{\partial p\>\!_k^\kappa} \; + \; \delta_i^k \;
                \frac{\partial}{\partial p} \;\smwedge\;
                \frac{\partial}{\partial x^\kappa} \right) \smwedge\;
         \frac{\partial}{\partial x^{\mu_3}} \;\smwedge \ldots \smwedge\;
         \frac{\partial}{\partial x^{\mu_r}} \\[2mm]
 & &\!   + \; \xi^\prime~,
\end{eqnarray*}
which is the general local coordinate expression for an $r$-multivector field
taking values in the kernel of $\,\omega$. \\
\PCPqed

\vspace{2mm}

With the standard local coordinate representation (\ref{eq:HAMMVF1})
for $r$-multivector fields $X$ at hand, we are now in a position to
analyze the restrictions imposed on the coefficients
$X_{\phantom{i}}^{\mu_1 \ldots\, \mu_r}$,
$X_{\phantom{i}}^{i,\mu_2 \ldots\, \mu_r}$,
$X_i^{\mu_1 \ldots\, \mu_r}$ and
$\tilde{X}_{\phantom{i}}^{\mu_2 \ldots\,\mu_r}$
by requiring $X$ to be locally Hamiltonian.\footnote{Of course, it
  makes no sense to discuss the question which locally Hamiltonian
  multivector fields are also globally Hamiltonian when working in
  local coordinates.} \linebreak
As a warm-up exercise, we shall settle the extreme cases of tensor degree
$0$ and $n+1$.
\begin{prp} \label{prp:HAMFUN}~~
 A function on~$P$, regarded as a $0$-multivector field, is locally Hamiltonian
 if and only if it is constant; it is then also exact Hamiltonian. Similarly,
 an $(n+1)$-multivector field on~$P$, with standard local coordinate
 representation
 \begin{equation} \label{eq:HAMMVF4}
  X~=~\tilde{X}~\frac{\partial}{\partial p} \,\:\smwedge\,\,
      \frac{\partial}{\partial x^1} \,\:\smwedge \ldots \smwedge\,\,
      \frac{\partial}{\partial x_{\phantom{i}}^n} \;
      + \; \xi~,
 \end{equation}
 where $\xi$ takes values in the kernel of $\,\omega$, is locally Hamiltonian
 if and only if the coefficient function $\tilde{X}$ is constant and is exact
 Hamiltonian if and only if it vanishes.
\end{prp}

%\pagebreak

\noindent
\textbf{Proof.}~~For functions, we use the fact that the operator $i_1^{}$
corresponding to the constant function $1$ on a manifold is defined to be
the identity, so that the operator $i_f^{}$ corresponding to an arbitrary
function $f$ on a manifold is simply multiplication by $f$. Therefore, we
have for any differential form $\alpha$
\[
 L_f^{} \>\! \alpha~
 =~d \left( i_f^{} \>\! \alpha \right) \, - \, i_f^{} \, d \alpha~
               =~d \left( f \alpha \right) \, - \, f \, d \alpha~
               =~df \,\smwedge\, \alpha~,
\]
implying that if $f$ is constant, $L_f^{} \alpha = 0 \,$ no matter what
$\alpha$ one chooses. On the other hand, we compute in adapted local
coordinates
\begin{eqnarray*}
 L_f^{} \>\! \omega \!\!
 &=&\!\! \left( \frac{\partial f}{\partial x^\nu} \, dx^\nu \, + \,
                \frac{\partial f}{\partial q^j} \, dq^j \, + \,
                \frac{\partial f}{\partial p\>\!_j^\nu} \, dp\>\!_j^\nu \, + \,
                \frac{\partial f}{\partial p} \, dp \right) \smwedge
         \left( dq^i \,\smwedge\; dp\;\!_i^\mu \,\smwedge\; d^{\,n} x_\mu^{} \,
                - \; dp \,\:\smwedge\; d^{\,n} x \right)                \\[3mm]
 &=&\!\! \frac{\partial f}{\partial x^\mu} \;
         dq^i \,\smwedge\; dp\;\!_i^\mu \,\smwedge\; d^{\,n} x \; - \;
         \frac{\partial f}{\partial q^j} \;
         dq^i \,\smwedge\; dq^j \,\smwedge\;
         dp\;\!_i^\mu \,\smwedge\; d^{\,n} x_\mu^{} \; - \;
         \frac{\partial f}{\partial q^i} \;
         dq^i \,\smwedge\; dp \,\:\smwedge\; d^{\,n} x                  \\
 & &\!   \mbox{} + \; \frac{\partial f}{\partial p\>\!_j^\nu} \;
         dq^i \,\smwedge\; dp\;\!_i^\mu \,\smwedge\;
         dp\>\!_j^\nu \,\smwedge\; d^{\,n} x_\mu^{} \; - \;
         \frac{\partial f}{\partial p\>\!_i^\mu} \;
         dp\;\!_i^\mu \,\smwedge\; dp \,\:\smwedge\; d^{\,n} x          \\
 & &\!   \mbox{} + \; \frac{\partial f}{\partial p} \;
         dq^i \,\smwedge\; dp\;\!_i^\mu \,\smwedge\;
         dp \,\:\smwedge\; d^{\,n} x_\mu^{}~.
\end{eqnarray*}
Inspecting the various terms, we see that this expression can only vanish
if all partial derivatives of $f$ are identically zero. Similarly, for
multivector fields of degree \mbox{$n+1$}, it is clear that when $r$
equals $n+1$, the last four terms in eq.~(\ref{eq:HAMMVF2}) vanish by
antisymmetry, so that~-- in contrast to what happens in the general case~--
the first three terms in eq.~(\ref{eq:HAMMVF1}) also take values in the
kernel of $\omega$ and can thus be incorporated into $\xi$. Therefore,
by putting $\, \tilde{X}_{\phantom{i}}^{\mu_1 \ldots\, \mu_n} =
\epsilon_{\phantom{i}}^{\mu_1 \ldots\, \mu_n} \tilde{X}$, we can
reduce the standard local coordinate representation of $X$ to the form
given in eq.~(\ref{eq:HAMMVF4}) and the expression (\ref{eq:HAMMVF2})
to $\, i_X^{} \omega = - \tilde{X}$. But $\, L_X^{} \omega \,
= \, d ( i_X^{} \omega ) - (-1)^n \, i_X^{} d \>\! \omega \, = \,
d ( i_X^{} \omega ) \,$ and $\, L_X^{} \theta \, = \, d ( i_X^{} \theta )
- (-1)^{n+1} \, i_X^{} d \>\! \theta \, = \, (-1)^{n+1} \, i_X^{} \omega$,
so the proposition follows. \\
\PCPqed \\

The intermediate cases ($0 < r \leqslant n$) are much more interesting.
However, the situation for tensor degree $n$ is substantially different
from that for tensor degree $< n$, mainly due to the fact that when $r$
equals $n$, the penultimate term in eq.~(\ref{eq:HAMMVF2}) and the last
term in eq.~(\ref{eq:HAMMVF3}) still vanish by antisymmetry; this case
will therefore be dealt with first. To this end, we begin by simplifying
the notation, writing
\begin{equation} \label{eq:COEFNVF}
 \begin{array}{cccc}
  X_{\phantom{i}}^{\mu_1 \ldots\, \mu_n}~
  =~\epsilon_{\phantom{i}}^{\mu_1 \ldots\, \mu_n} \, \tilde{X} & , &
  X_{\phantom{i}}^{i,\mu_2 \ldots\, \mu_n}~
  =~\epsilon_{\phantom{i}}^{\mu_2 \ldots\, \mu_n \mu} \, X_\mu^i &, \\[3mm]
  X_i^{\mu_1 \ldots\, \mu_n}~
  =~\epsilon_{\phantom{i}}^{\mu_1 \ldots\, \mu_n} \, X_i^{} & , &
  \tilde{X}_{\phantom{i}}^{\mu_2 \ldots\, \mu_n}~
  =~\epsilon_{\phantom{i}}^{\mu_2 \ldots\, \mu_n \mu} \, X_\mu^{} &,
 \end{array}
\end{equation}
so that the standard local coordinate representation (\ref{eq:HAMMVF1}) of~$X$
takes the form

%\pagebreak

\begin{eqnarray} \label{eq:HAMNVF1}
 X \!\!
 &=&\!\! \tilde{X} \;\, \frac{1}{n!} \;
         \epsilon_{\phantom{i}}^{\mu_1 \ldots\, \mu_n} \,
         \frac{\partial}{\partial x_{\phantom{i}}^{\mu_1}}
         \;\smwedge \ldots \smwedge\;
         \frac{\partial}{\partial x_{\phantom{i}}^{\mu_n}}    \nonumber \\[1mm]
 & &\!   \mbox{} + \; X_\mu^i \;\, \frac{1}{(n\!-\!1)!} \;
         \epsilon_{\phantom{i}}^{\mu_2 \ldots\, \mu_n \mu} \,
         \frac{\partial}{\partial q^i} \;\smwedge\;
         \frac{\partial}{\partial x_{\phantom{i}}^{\mu_2}}
         \;\smwedge \ldots \smwedge\;
         \frac{\partial}{\partial x_{\phantom{i}}^{\mu_n}}    \nonumber \\
 & &\!   \mbox{} + \; X_i^{} \;\; \frac{1}{n!} \;
         \epsilon_{\phantom{i}}^{\mu_1 \ldots\, \mu_n} \,
         \frac{\partial}{\partial p\>\!_i^{\mu_1}} \;\smwedge\;
         \frac{\partial}{\partial x_{\phantom{i}}^{\mu_2}}
         \;\smwedge \ldots \smwedge\;
         \frac{\partial}{\partial x_{\phantom{i}}^{\mu_n}}              \\
 & &\!   \mbox{} + \;\; X_\mu^{} \;\, \frac{1}{(n\!-\!1)!} \;
         \epsilon_{\phantom{i}}^{\mu_2 \ldots\, \mu_n \mu} \,
         \frac{\partial}{\partial p} \;\smwedge\;
         \frac{\partial}{\partial x_{\phantom{i}}^{\mu_2}}
         \;\smwedge \ldots \smwedge\;
         \frac{\partial}{\partial x_{\phantom{i}}^{\mu_n}}    \nonumber \\[3mm]
 & &\!   \mbox{} + \; \xi~,                                   \nonumber
\end{eqnarray}
where $\xi$ takes values in the kernel of $\,\omega$, while
eqs.~(\ref{eq:HAMMVF2}) and~(\ref{eq:HAMMVF3}) take the form
\begin{equation} \label{eq:HAMNVF2}
 i_X^{} \omega~
 =~(-1)^{n-1} \, \tilde{X} \, dp \; + \, X_\mu^i \, dp\>\!_i^\mu \, - \,
   (-1)^{n-1} \, X_i^{} \, dq^i \, - \, X_\mu^{} \, dx^\mu~,
\end{equation}
and
\begin{equation} \label{eq:HAMNVF3}
 i_X^{} \theta~=~p \, \tilde{X} \, + \,
                 (-1)^{n-1} \, p\>\!_i^\mu X_\mu^i~,
\end{equation}
respectively.
\begin{prp} \label{prp:HAMNVF}
 An $n$-multivector field $X$ on $P$ is locally Hamiltonian if and only if,
 locally and modulo terms taking values in the kernel of $\,\omega$, it can
 be written in terms of a single function $f$, as follows:
 \begin{equation} \label{eq:HAMNVF4}
  \begin{array}{rcl}
   X &=& {\displaystyle - \;
          \frac{1}{(n\!-\!1)!} \; \epsilon^{\mu_2 \ldots\, \mu_n \mu} \;
          \left( \frac{\partial f}{\partial x_{\phantom{i}}^\mu} \,
                 \frac{\partial}{\partial p} \; - \;
                 \frac{1}{n} \, \frac{\partial f}{\partial p} \,
                 \frac{\partial}{\partial x_{\phantom{i}}^\mu} \right)
          \,\smwedge\;
          \frac{\partial}{\partial x_{\phantom{i}}^{\mu_2}}
          \;\smwedge \ldots \smwedge\;
          \frac{\partial}{\partial x_{\phantom{i}}^{\mu_n}}} \\[5mm]
     & & {\displaystyle + \;
          \frac{1}{(n\!-\!1)!} \; \epsilon^{\mu_2 \ldots\, \mu_n \mu} \;
          \left( \frac{\partial f}{\partial p\>\!_i^\mu} \,
                 \frac{\partial}{\partial q^i} \; - \;
                 \frac{1}{n} \, \frac{\partial f}{\partial q^i} \,
                 \frac{\partial}{\partial p\>\!_i^\mu} \right)
          \,\smwedge\;
          \frac{\partial}{\partial x_{\phantom{i}}^{\mu_2}}
          \;\smwedge \ldots \smwedge\;
          \frac{\partial}{\partial x_{\phantom{i}}^{\mu_n}}}~.
  \end{array}
 \end{equation}
 Moreover, $X$ is exact Hamiltonian if and only if $f$ is a linear function
 of the multi\-momentum variables $p\>\!_r^\rho$ and of the energy variable
 $p\>\!$.
\end{prp}
\textbf{Proof.}~~Obviously, $X$ is locally Hamiltonian if and only if,
locally, $i_X^{} \omega = df \,$ for some function $f$, which in view
of eq.~(\ref{eq:HAMNVF2}) leads to the following system of equations for
the coefficients $\tilde{X}$, $X_\mu^i$, $X_i^{}$ and $X_\mu^{}$ of~$X$
in its standard local coordinate representation~(\ref{eq:HAMNVF1}):
\begin{equation} \label{eq:HAMNVF5}
 \tilde{X}~=~(-1)^{n-1} \, \frac{\partial f}{\partial p}~~,~~
 X_\mu^i~=~\frac{\partial f}{\partial p\>\!_i^\mu}~~,~~
 X_i^{}~=~(-1)^n \, \frac{\partial f}{\partial q^i}~~,~~
 X_\mu^{}~= \; - \, \frac{\partial f}{\partial x^\mu}~~.~~
\end{equation}
Inserting this back into eq.~({\ref{eq:HAMNVF1}) and rearranging the terms,
we arrive at eq.~(\ref{eq:HAMNVF4}). Note also that then,
\[
 i_X^{} \theta~
 =~(-1)^{n-1} \, p \; \frac{\partial f}{\partial p} \, + \,
   (-1)^{n-1} \, p\>\!_i^\mu \, \frac{\partial f}{\partial p\>\!_i^\mu}~,
\]
that is,
\begin{equation} \label{eq:HAMNVF6}
 i_X^{} \theta~=~(-1)^{n-1} \, L_\Sigma^{} f~.
\end{equation}
Next, $X$ will be exact Hamiltonian if and only if, in addition,
\begin{equation} \label{eq:HAMNVF7}
 f~=~(-1)^{n-1} \, i_X^{} \theta~,
\end{equation}
which in view of the previous equation means that $f$ must be an eigenfunction
of the scaling operator $L_\Sigma^{}$ with eigenvalue $1$: this is well known
to be the case if and \linebreak only if $f$ is linear in the multimomentum
variables $p\>\!_r^\rho$ and the energy variable $p\>\!$. \\
\PCPqed

\vspace{3mm}

Now we turn to multivector fields of tensor degree $\, <n$.
Here, the main result is
\begin{thm} \label{thm:HAMMVF}~~
 An $r$-multivector field $X$ on~$P$, with $\, 0<r<n$, is locally Hamiltonian
 if and only if the coefficients $X_{\phantom{i}}^{\mu_1 \ldots\, \mu_r}$,
 $X_{\phantom{i}}^{i,\mu_2 \ldots\, \mu_r}$, $X_i^{\mu_1 \ldots\, \mu_r}$
 and $\tilde{X}_{\phantom{i}}^{\mu_2 \ldots\, \mu_r}$ in its standard local
 coordinate representation (\ref{eq:HAMMVF1}) satisfy the following conditions:
 \begin{enumerate}
  \item the coefficients $X_{\phantom{i}}^{\mu_1 \ldots\, \mu_r}$ depend
        only on the local coordinates $x^\rho$ for $M$ and, in the special
        case $N\!=\!1$, also on the local fiber coordinates $q^r$ for $E$,
  \item the coefficients $X_{\phantom{i}}^{i,\mu_2 \ldots\, \mu_r}$ are
        ``antisymmetric polynomials in the multimomentum variables'' of
        degree $r\!-\!1$, i.e., they can be written in the form
        \begin{equation} \label{eq:LHAMMVF01}
         X_{\phantom{i}}^{i,\mu_2 \ldots\, \mu_r}~
         =~\sum_{s=1}^r X_{s-1}^{i,\mu_2 \ldots\, \mu_r}~,
        \end{equation}
        with
        \begin{equation} \label{eq:LHAMMVF02}
         X_{s-1}^{i,\mu_2 \ldots\, \mu_r}~
         =~\frac{1}{(s\!-\!1)!} \frac{1}{(r\!-\!s)!}
           \sum_{\pi \ssmin S_{r-1}} \! (-1)^\pi \,
           p_{i_2}^{\mu_{\pi(2)}} \!\ldots\, p_{i_s}^{\mu_{\pi(s)}} \,
           Y_{s-1}^{i i_2 \ldots\, i_s,\mu_{\pi(s+1)} \ldots\, \mu_{\pi(r)}}~,
        \end{equation}
        where $S_{r-1}$ denotes the permutation group of $\{2,\ldots,r\}$ and
        the coefficients $Y_{s-1}^{i i_2 \ldots i_s,\mu_{s+1} \ldots \mu_r}$
        depend only on the local coordinates $x^\rho$ for~$M$ as well as the
        local fiber coordinates $q^r$ for~$E$ and are totally antisymmetric
        in $i,i_2,\ldots,i_s$ as well as in $\mu_{s+1},\ldots,\mu_r$.
  \item the remaining coefficients $X_i^{\mu_1 \ldots\, \mu_r}$ and
        $\tilde{X}_{\phantom{i}}^{\mu_2 \ldots\, \mu_r}$ can be
        expressed in terms of the previous ones and of new coefficients
        $X_-^{\mu_1 \ldots\, \mu_r}$ depending only on the local
        coordinates $x^\rho$ for $M$ as well as the local fiber
        coordinates $q^r$ for~$E$ and are totally antisymmetric
        in $\mu_1,\ldots,\mu_r$, according to
        \begin{eqnarray} \label{eq:LHAMMVF03}
          X_i^{\mu_1 \ldots\, \mu_r}  \!\!
          &=&\!\! \mbox{} - \, p~
                  \frac{\partial X_{\phantom{i}}^{\mu_1 \ldots\, \mu_r}}
                       {\partial q^i} \, + \,
                  p\>\!_i^\mu \;
                  \frac{\partial X_{\phantom{i}}^{\mu_1 \ldots\, \mu_r}}
                       {\partial x_{\phantom{i}}^\mu} \, - \,
                  \sum_{s=1}^r \, p\>\!_i^{\mu_s} \;
                  \frac{\partial X_{\phantom{i}}^{\mu_1 \ldots\, \mu_{s-1} \nu
                                                  \mu_{s+1} \ldots\, \mu_r}}
                       {\partial x_{\phantom{i}}^\nu}         \nonumber \\
          & &\!\! \mbox{} - \, \Sigma^{-1} \left(
                  \sum_{s=1}^r \, (-1)^{s-1} \, p\>\!_j^{\mu_s} \;
                  \frac{\partial X_{\phantom{i}}^{j,\mu_1 \ldots\, \mu_{s-1}
                                                    \mu_{s+1} \ldots\, \mu_r}}
                        {\partial q^i} \right)                          \\[1ex]
          & &\!\! \mbox{} + \,
                  \frac{\partial X_-^{\mu_1 \ldots\, \mu_r}}{\partial q^i}~,
                                                              \nonumber
        \end{eqnarray}
        (the first term being absent as soon as $\, N>1$) and
        \begin{eqnarray} \label{eq:LHAMMVF04}
         \tilde{X}_{\phantom{i}}^{\mu_2 \ldots\, \mu_r}  \!\!
         &=&\!\! (-1)^r \; p~
                 \frac{\partial X_{\phantom{i}}^{\mu_2 \ldots\, \mu_r \nu}}
                      {\partial x_{\phantom{i}}^\nu}          \nonumber \\
         & &\!\! \mbox{} - \, \Sigma^{-1} \left( p\>\!_i^\mu \;
                 \frac{\partial X_{\phantom{i}}^{i,\mu_2 \ldots\, \mu_r}}
                      {\partial x_{\phantom{i}}^\mu} \, - \,
                 \sum_{s=2}^r \, p\>\!_i^{\mu_s} \;
                 \frac{\partial X_{\phantom{i}}^{i,\mu_2 \ldots\, \mu_{s-1} \nu
                                                   \mu_{s+1} \ldots\, \mu_r}}
                      {\partial x_{\phantom{i}}^\nu} \right)            \\[1ex]
         & &\!\! \mbox{} - \, (-1)^r \,
                 \frac{\partial X_-^{\mu_2 \ldots\, \mu_r \nu}}
                      {\partial x_{\phantom{i}}^\nu}~.        \nonumber
        \end{eqnarray}
 \end{enumerate}
 It is exact Hamiltonian if and only if, in addition, the coefficents
 $X_{\phantom{i}}^{i,\mu_2 \ldots\, \mu_r}$ depend only on the local
 coordinates $x^\rho$ for $M$ as well as the local fiber coordinates
 $q^r$ for~$E$ and the coefficients $X_-^{\mu_1 \ldots\, \mu_r}$ vanish.
\end{thm}
\textbf{Proof.}~~The proof will be carried out by ``brute force'' computation.
First, we apply the exterior derivative to eq.~(\ref{eq:HAMMVF2}) and use
eq.~(\ref{eq:CMPDF6}) to simplify the expressions involving derivatives with
respect to the space-time variables. Collecting the terms, we get
\vspace{2mm}
\begin{eqnarray*}
 L_X^{} \omega \!\!
 &=&\!   \mbox{} - \; \frac{1}{(r\!-\!2)!} \;
         \frac{\partial \tilde{X}_{\phantom{i}}^{\mu_3 \ldots\, \mu_r \nu}}
              {\partial x_{\phantom{i}}^\nu}~
         d^{\,n} x_{\mu_3 \ldots\, \mu_r}^{}                            \\[1mm]
 & &\!   \mbox{} - \; \frac{1}{(r\!-\!1)!} \;
         \biggl( \frac{\partial \tilde{X}_{\phantom{i}}^{\mu_2 \ldots\, \mu_r}}
                      {\partial q^i} \, - \, (-1)^{r-1} \,
                 \frac{\partial X_i^{\mu_2 \ldots\, \mu_r \nu}}
                     {\partial x_{\phantom{i}}^\nu} \biggr) \;
         dq^i \,\smwedge\> d^{\,n} x_{\mu_2 \ldots\, \mu_r}^{}          \\[1mm]
 & &\!   \mbox{} - \; \frac{1}{(r\!-\!1)!} \;
         \biggl( \frac{\partial \tilde{X}_{\phantom{i}}^{\mu_2 \ldots\, \mu_r}}
                      {\partial p\>\!_i^\mu} \, + \,
                 \frac{\partial X_{\phantom{i}}^{i,\mu_2 \ldots\, \mu_r}}
                      {\partial x_{\phantom{i}}^\mu} \, - \,
                 \sum_{s=2}^r \, \delta_\mu^{\mu_s} \,
                 \frac{\partial X_{\phantom{i}}^{i,\mu_2 \ldots\, \mu_{s-1} \nu
                                                   \mu_{s+1} \ldots\, \mu_r}}
                      {\partial x_{\phantom{i}}^\nu} \biggr)            \\[1mm]
 & &     \hspace{1cm} \times \;
         dp\>\!_i^\mu \,\smwedge\> d^{\,n} x_{\mu_2 \ldots\, \mu_r}^{}  \\[3mm]
 & &\!   \mbox{} - \; \frac{1}{(r\!-\!1)!} \;
         \biggl( \frac{\partial \tilde{X}_{\phantom{i}}^{\mu_2 \ldots\, \mu_r}}
                      {\partial p} \, + \, (-1)^{r-1} \,
                 \frac{\partial X_{\phantom{i}}^{\mu_2 \ldots\, \mu_r \nu}}
                      {\partial x_{\phantom{i}}^\nu} \biggr) \;
         dp \,\,\smwedge\> d^{\,n} x_{\mu_2 \ldots\, \mu_r}^{}          \\[3mm]
 & &\!   \mbox{} - \; \frac{(-1)^r}{r!} \;
         \biggl( \frac{\partial X_{\phantom{i}}^{\mu_1 \ldots\, \mu_r}}
                      {\partial q^i} \, + \,
                 \frac{\partial X_i^{\mu_1 \ldots\, \mu_r}}
                      {\partial p} \biggr) \;
         dq^i \,\smwedge\, dp \,\,\smwedge\>
         d^{\,n} x_{\mu_1 \ldots\, \mu_r}^{}                            \\[1mm]
 & &\!   \mbox{} - \; \frac{(-1)^r}{r!} \;
         \biggl( \frac{\partial X_{\phantom{i}}^{\mu_1 \ldots\, \mu_r}}
                      {\partial p\>\!_i^\mu} \, - \,
                 \sum_{s=1}^r \, (-1)^{s-1} \, \delta_\mu^{\mu_s} \,
                 \frac{\partial X_{\phantom{i}}^{i,\mu_1 \ldots\, \mu_{s-1}
                                                   \mu_{s+1} \ldots\, \mu_r}}
                      {\partial p} \biggr)                              \\[1mm]
 & &     \hspace{1cm} \times \;
         dp\>\!_i^\mu \,\smwedge\, dp \,\,\smwedge\>
         d^{\,n} x_{\mu_1 \ldots\, \mu_r}^{}
\end{eqnarray*}
\begin{eqnarray*}
 \phantom{L_X^{} \omega \!\!}
 & &\!   \mbox{} + \; \frac{(-1)^r}{r!} \,
         \biggl( \delta_i^k \, \delta_\kappa^\mu \,
                 \frac{\partial X_{\phantom{i}}^{\mu_1 \ldots\, \mu_r}}
                      {\partial x_{\phantom{i}}^\mu} \, - \,
                 \sum_{s=1}^r \, \delta_i^k \, \delta_\kappa^{\mu_s} \,
                 \frac{\partial X_{\phantom{i}}^{\mu_1 \ldots\, \mu_{s-1} \nu
                                                 \mu_{s+1} \ldots\, \mu_r}}
                      {\partial x_{\phantom{i}}^\nu}                    \\
 & &     \hspace{2.2cm} \mbox{} - \,
                 \sum_{s=1}^r \, (-1)^{s-1} \, \delta_\kappa^{\mu_s} \,
                 \frac{\partial X_{\phantom{i}}^{k,\mu_1 \ldots\, \mu_{s-1}
                                                   \mu_{s+1} \ldots\, \mu_r}}
                      {\partial q^i} \, - \,
                 \frac{\partial X_i^{\mu_1 \ldots\, \mu_r}}
                      {\partial p\>\!_k^\kappa} \biggr)                 \\[1mm]
 & &     \hspace{1cm} \times \;
         dq^i \,\smwedge\, dp\>\!_k^\kappa \,\smwedge\>
         d^{\,n} x_{\mu_1 \ldots\, \mu_r}^{}                            \\[3mm]
 & &\!   \mbox{} - \; \frac{(-1)^r}{r!} \;
         \frac{\partial X_i^{\mu_1 \ldots\, \mu_r}}{\partial q^j}~
         dq^i \,\smwedge\, dq^j \,\smwedge\>
         d^{\,n} x_{\mu_1 \ldots\, \mu_r}^{}                            \\[1mm]
 & &\!   \mbox{} + \; \frac{(-1)^{r-1}}{(r\!-\!1)!} \;
         \frac{\partial X_{\phantom{i}}^{l,\mu_2 \ldots\, \mu_r}}
              {\partial p\>\!_k^\kappa}~
         dp\>\!_k^\kappa \,\smwedge\, dp\>\!_l^\lambda \,\smwedge\>
         d^{\,n} x_{\lambda \mu_2 \ldots\, \mu_r}^{}                    \\[1mm]
 & &\!   \mbox{} - \; \frac{1}{r!} \,
         \frac{\partial X_{\phantom{i}}^{\mu_1 \ldots\, \mu_r}}{\partial q^j}~
         dq^i \,\smwedge\, dq^j \,\smwedge\, dp\>\!_i^\mu \,\smwedge\>
         d^{\,n} x_{\mu \mu_1 \ldots\, \mu_r}^{}                        \\[1mm]
 & &\!   \mbox{} - \; \frac{1}{r!} \,
         \frac{\partial X_{\phantom{i}}^{\mu_1 \ldots\, \mu_r}}
              {\partial p\>\!_k^\kappa}~
         dq^l \,\smwedge\, dp\>\!_k^\kappa \,\smwedge\,
         dp\>\!_l^\lambda \,\smwedge\>
         d^{\,n} x_{\lambda \mu_1 \ldots\, \mu_r}^{}                    \\[1mm]
 & &\!   \mbox{} + \; \frac{1}{r!} \,
         \frac{\partial X_{\phantom{i}}^{\mu_1 \ldots\, \mu_r}}{\partial p}~
         dq^i \,\smwedge\, dp\>\!_i^\mu \,\smwedge\,
         dp \,\,\smwedge\> d^{\,n} x_{\mu \mu_1 \ldots\, \mu_r}^{}~.
         \rule[-6mm]{0mm}{7mm} \hspace{4.5cm}
\end{eqnarray*}
(Note that the last three terms would have to be omitted if $\, r=n$.)
Numbering the terms in this equation from $1$ to $12$, we begin by analyzing
terms no.\ 6, 10, 11 and 12.
\begin{itemize}
 \item Term No.\ 12: Given mutually different indices $\, \kappa_1,\ldots,
       \kappa_r$, we choose indices $k$ and $\, \kappa \,\nsmin \{\kappa_1,
       \ldots,\kappa_r\}$ (here we use the hypothesis that $\, r<n$) and,
       when $\, r<n-1$, a complementary set of indices $\, \nu_1,\ldots,
       \nu_{n-r-1} \,$ to contract this term with the multivector
       field $\; \partial_k^{} \,\smwedge\> \partial_\kappa^{\;\!k}
       \,\smwedge\> \partial_0^{} \,\smwedge\> \partial_{\nu_{1}}^{}
       \,\smwedge \ldots \smwedge\> \partial_{\nu_{n-r-1}}^{}$ (no sum
       over $k$), concluding that $X_{\phantom{i}}^{\kappa_1 \ldots\,
       \kappa_r}$ cannot depend on $p\>\!$.
 \item Term No.\ 11: Given indices $i$, $\mu$ and mutually different
       indices $\, \kappa_1,\ldots,\kappa_r$, we choose indices $j$
       and $\, \nu \,\nsmin \{\kappa_1,\ldots,\kappa_r\}$ (here we use
       the hypothesis that $\, r<n$) such that either $j \neq i$
       or $\nu \neq \mu$ and, when $\, r<n-1$, a complementary set
       of indices $\, \nu_1,\ldots,\nu_{n-r-1} \,$ to contract this
       term with the multivector field $\; \partial_j^{} \,\smwedge\>
       \partial_\mu^{\,i} \,\smwedge\> \partial_\nu^{\;\!j}
       \,\smwedge\> \partial_{\nu_{1}}^{} \,\smwedge \ldots \smwedge\>
       \partial_{\nu_{n-r-1}}^{}$ (no sum over~$j$), concluding that
       $X_{\phantom{i}}^{\kappa_1 \ldots\,\kappa_r}$ cannot depend on
       $p\>\!_i^\mu$. Obviously, there is one case where this argument
       does not work: namely when $N\!=\!1$, $r=n\!-\!1$ and
       $\, \mu \,\nsmin \{\kappa_1,\ldots,\kappa_r\}$. This
       situation will however be covered in the next item.
 \item Term No.\ 6 (first part): Given indices $k$, $\kappa$ and
       mutually different indices $\, \kappa_1,\ldots,\kappa_r$
       such that $\kappa \,\nsmin \{\kappa_1,\ldots,\kappa_r\}$,
       we choose a complementary set of indices $\, \nu_1,\ldots,
       \nu_{n-r-1} \,$ to contract this term with the multivector
       field $\; \partial_\kappa^{\;\!k} \,\smwedge\> \partial_0^{}
       \,\smwedge\> \partial_{\kappa}^{} \,\smwedge\> \partial_{\nu_{1}}^{}
       \,\smwedge \ldots \smwedge\> \partial_{\nu_{n-r-1}}^{}$, concluding
       that $X_{\phantom{i}}^{\kappa_1 \ldots\,\kappa_r}$ cannot depend on
       $p\;\!_k^\kappa$, since in this case the second term in the bracket
       gives no contribution. In particular, this settles the remaining case
       of the previous item.
 \item Term No.\ 10: Given an index $l$ and mutually different indices
       $\, \kappa_1,\ldots,\kappa_r$, we choose indices $k$ and
       $\, \kappa \,\nsmin \{\kappa_1,\ldots,\kappa_r\}$ (here we use
       the hypothesis that $\, r<n$) such that $k \neq l$ and, when
       $\, r<n-1$, a complementary set of indices $\, \nu_1,\ldots,
       \nu_{n-r-1} \,$ to contract this term with the multivector
       field $\; \partial_k^{} \,\smwedge\> \partial_l^{} \,\smwedge\>
       \partial_\kappa^{\;\!k} \,\smwedge\> \partial_{\nu_{1}}^{}
       \,\smwedge \ldots \smwedge\> \partial_{\nu_{n-r-1}}^{}$
       (no~sum over $k$), concluding that $X_{\phantom{i}}^{\kappa_1
       \ldots\, \kappa_r}$ cannot depend on $q^l$. Obviously, there
       is one case where this argument does not work: namely when
       $N\!=\!1$. In this situation, the whole term vanishes
       identically, and no conclusion can be drawn.
\end{itemize}
This proves the statements in item 1.\ of the theorem. Moreover, it allows to
simplify term no.~6, as follows:
\[
 \mbox{} - \; \frac{(-1)^{r-1}}{(r-1)!} \;
 \frac{\partial X_{\phantom{i}}^{i,\mu_2 \ldots\, \mu_r}}{\partial p}~
 dp\>\!_i^\mu \,\smwedge\, dp \,\,\smwedge\>
 d^{\,n} x_{\mu \mu_2 \ldots\, \mu_r}^{}~.
\]
Next we analyze terms no.\ 6 and 9.
\begin{itemize}
 \item Term No.\ 6 (second part): Given an index $k$ and mutually different
       indices $\, \kappa_2,\ldots,\kappa_r$, we choose an index $\, \kappa
       \,\nsmin \{\kappa_2,\ldots,\kappa_r\} \,$ and a complementary set of
       indices $\, \nu_1,\ldots,\nu_{n-r} \,$ to contract this term, in
       the simplified form given in the previous equation,
       with the multivector field $\; \partial_\kappa^{\;\!k} \,\smwedge\>
       \partial_0^{} \,\smwedge\> \partial_{\nu_{1}}^{} \,\smwedge \ldots
       \smwedge\> \partial_{\nu_{n-r}}^{}$, concluding that
       $X_{\phantom{i}}^{k,\kappa_2 \ldots\, \kappa_r}$
       cannot depend on $p\>\!$.
 \item Term No.\ 9: Given indices $i$, $j$, $\mu$, $\nu$ and mutually
       different indices $\, \kappa_2,\ldots,\kappa_r$, we choose a
       set of indices $\, \nu_1,\ldots,\nu_{n-r} \,$ such that
       $\, \{\kappa_2,\ldots,\kappa_r\} \smcap \{\nu_1,\ldots,
       \nu_{n-r}\} = \emptyset \,$ to contract this term with the
       multivector field $\;  \partial_\mu^{\,i} \,\smwedge\>
       \partial_\nu^{\;\!j} \,\smwedge\> \partial_{\nu_{1}}^{}
       \,\smwedge \ldots \smwedge\> \partial_{\nu_{n-r}}^{}$,
       obtaining
       \begin{equation} \label{eq:TERM9}
        \frac{\partial X_{\phantom{i}}^{j,\mu_2 \ldots\, \mu_r}}
             {\partial p\>\!_i^\mu} \;
        \epsilon_{\nu \mu_2 \ldots\, \mu_r \, \nu_1 \ldots\, \nu_{n-r}}~
        =~\frac{\partial X_{\phantom{i}}^{i,\mu_2 \ldots\, \mu_r}}
               {\partial p\>\!_j^\nu} \;
          \epsilon_{\mu \mu_2 \ldots\, \mu_r \, \nu_1 \ldots\, \nu_{n-r}}~.
       \end{equation}
       Now assume $\nu$ to be chosen so that $\, \nu \,\nsmin \{\kappa_2,
       \ldots,\kappa_r,\nu_1,\ldots,\nu_{n-r}\}$. Then if $\, \mu \,\nsmin
       \{\kappa_2,\ldots,\kappa_r\}$, we can take $\, \mu = \nu_1$, say,
       to conclude that $X_{\phantom{i}}^{j,\kappa_2 \ldots\, \kappa_r}$
       cannot depend on $p\>\!_i^\mu$:
       \begin{equation} \label{eq:GPOL1}
        \frac{\partial X_{\phantom{i}}^{j,\mu_2 \ldots\, \mu_r}}
             {\partial p\>\!_i^\mu}~=~0 \qquad
        \mbox{if $\, \mu \nsmin \{\mu_2,\ldots,\mu_r\}$}~.
       \end{equation}
       Moreover, if $\, \mu \,\smin \{\kappa_2,\ldots,\kappa_r\}$, this
       result implies that applying an operator $\partial_\mu^{\;\!i'}$
       (with arbitrary $i'$) to eq.~(\ref{eq:TERM9}) gives zero since on
       the rhs, the $\epsilon$-tensor kills all terms in the sum over the
       indices $\mu_2,\ldots,\mu_r$ in which the index $\mu$ appears among
       them:
       \begin{equation} \label{eq:GPOL2}
        \frac{\partial^{\>\!2} X_{\phantom{i}}^{j,\mu_2 \ldots\, \mu_r}}
             {\partial p\>\!_{i_1}^\mu \, \partial p\>\!_{i_2}^\mu}~=~0 \qquad
        \mbox{if $\, \mu \smin \{\mu_2,\ldots,\mu_r\} \,$
              (no sum over $\mu$)}~.
       \end{equation}
       The general solution to eqs~(\ref{eq:GPOL1}) and~(\ref{eq:GPOL2}) can
       be written in the form
       \[
        X_{\phantom{i}}^{j,\mu_2 \ldots\, \mu_r}~
        =~\sum_{s=1}^r \frac{1}{(s\!-\!1)!} \frac{1}{(r\!-\!s)!}
          \sum_{\pi \ssmin S_{r-1}} \! (-1)^\pi \,
          p_{j_2}^{\mu_{\pi(2)}} \!\ldots p_{j_s}^{\mu_{\pi(s)}} \,
          Y_{s-1}^{j,j_2 \ldots j_s,\mu_{\pi(s+1)} \ldots \mu_{\pi(r)}}~,
       \]
       where $S_{r-1}$ denotes the permutation group of $\{2,\ldots,r\}$ and
       the coefficients $Y_{s-1}^{i,j_2 \ldots j_s,\mu_{s+1} \ldots \mu_r}$
       are local functions on $E$: they do not depend on the multi\-momentum
       variables $p\>\!_k^\kappa$ or the energy variable $p\>\!$ and are
       totally antisymmetric both in $j_2,\ldots,j_s$ and in $\mu_{s+1},
       \ldots,\mu_r$. Differentiating this expression with respect to
       $p\>\!_i^\mu$ with $\, \mu = \mu_2 \,$ gives
       \begin{eqnarray*}
        \lefteqn{\frac{\partial X_{\phantom{i}}^{j,\mu \mu_3 \ldots\, \mu_r}}
                      {\partial p\>\!_i^\mu} \;
                 \epsilon_{\nu \mu \mu_3 \ldots\, \mu_r \,
                           \nu_1 \ldots\, \nu_{n-r}}
        \qquad \mbox{(no sum over $\mu$)}}                \hspace{5mm} \\
        &=&\!\! \sum_{s=2}^r \frac{1}{(s\!-\!2)!} \frac{1}{(r\!-\!s)!}
                \sum_{\pi \ssmin S_{r-2}} \! (-1)^\pi \,
                p_{j_3}^{\mu_{\pi(3)}} \!\ldots p_{j_s}^{\mu_{\pi(s)}} \,
                Y_{s-1}^{j,i j_3 \ldots j_s,
                         \mu_{\pi(s+1)} \ldots \mu_{\pi(r)}}           \\[-1ex]
        & &\hspace{4.5cm} \times \,
                \epsilon_{\nu \mu \mu_3 \ldots\, \mu_r \,
                          \nu_1 \ldots\, \nu_{n-r}}~,
       \end{eqnarray*}
       where $S_{r-2}$ denotes the permutation group of $\{3,\ldots,r\}$,
       which shows that eq.~(\ref{eq:TERM9}) will hold provided that
       \[
        Y_{s-1}^{j,i j_3 \ldots j_s,\mu_{\pi(s+1)} \ldots \mu_{\pi(r)}}~
        = \; - \, Y_{s-1}^{i,j j_3 \ldots j_s,
                           \mu_{\pi(s+1)} \ldots \mu_{\pi(r)}}~.
       \]
\end{itemize}
This proves the statements in item 2.\ of the theorem. To proceed further,
we write down the equations obtained from the remaining terms.
\begin{itemize}
 \item Term No.\ 1:
       \begin{equation} \label{eq:LHAMMVF11}
        \frac{\partial \tilde{X}_{\phantom{i}}^{\mu_3 \ldots\, \mu_r \nu}}
             {\partial x_{\phantom{i}}^\nu}~=~0~.
       \end{equation}
 \item Term No.\ 2:
       \begin{equation} \label{eq:LHAMMVF12}
        \frac{\partial \tilde{X}_{\phantom{i}}^{\mu_2 \ldots\, \mu_r}}
             {\partial q^i}~
        =~(-1)^{r-1} \, \frac{\partial X_i^{\mu_2 \ldots\, \mu_r \nu}}
                             {\partial x_{\phantom{i}}^\nu}~.
       \end{equation}
 \item Term No.\ 3:
       \begin{equation} \label{eq:LHAMMVF13}
        \frac{\partial \tilde{X}_{\phantom{i}}^{\mu_2 \ldots\, \mu_r}}
             {\partial p\>\!_i^\mu}~
        = \; - \; \frac{\partial X_{\phantom{i}}^{i,\mu_2 \ldots\, \mu_r}}
                       {\partial x_{\phantom{i}}^\mu} \, + \,
          \sum_{s=2}^r \, \delta_\mu^{\mu_s} \,
          \frac{\partial X_{\phantom{i}}^{i,\mu_2 \ldots\, \mu_{s-1} \nu
                                            \mu_{s+1} \ldots\, \mu_r}}
               {\partial x_{\phantom{i}}^\nu}~.
       \end{equation}
 \item Term No.\ 4:
       \begin{equation} \label{eq:LHAMMVF14}
        \frac{\partial \tilde{X}_{\phantom{i}}^{\mu_2 \ldots\, \mu_r}}
             {\partial p}~
        =~(-1)^r \, \frac{\partial X_{\phantom{i}}^{\mu_2 \ldots\, \mu_r \nu}}
                         {\partial x_{\phantom{i}}^\nu}~.
       \end{equation}
 \item Term No.\ 5:
       \begin{equation} \label{eq:LHAMMVF15}
        \frac{\partial X_i^{\mu_1 \ldots\, \mu_r}}{\partial p}~
        = \; - \, \frac{\partial X_{\phantom{i}}^{\mu_1 \ldots\, \mu_r}}
                       {\partial q^i}~.
       \end{equation}
 \item Term No.\ 7:
       \begin{equation} \label{eq:LHAMMVF17}
        \begin{array}{rcl}
         {\displaystyle
          \frac{\partial X_i^{\mu_1 \ldots\, \mu_r}}
               {\partial p\>\!_k^\kappa}} \!\!
         &=&\!\! {\displaystyle
                  \delta_i^k \, \delta_\kappa^\mu \,
                  \frac{\partial X_{\phantom{i}}^{\mu_1 \ldots\, \mu_r}}
                       {\partial x_{\phantom{i}}^\mu} \, - \,
                  \sum_{s=1}^r \, \delta_i^k \, \delta_\kappa^{\mu_s} \,
                  \frac{\partial X_{\phantom{i}}^{\mu_1 \ldots\, \mu_{s-1} \nu
                                                  \mu_{s+1} \ldots\, \mu_r}}
                       {\partial x_{\phantom{i}}^\nu}}                  \\[4ex]
         & &\!\! \mbox{} - \,
                 {\displaystyle
                  \sum_{s=1}^r \, (-1)^{s-1} \, \delta_\kappa^{\mu_s} \,
                  \frac{\partial X_{\phantom{i}}^{k,\mu_1 \ldots\, \mu_{s-1}
                                                    \mu_{s+1} \ldots\, \mu_r}}
                       {\partial q^i}}~.
        \end{array}
       \end{equation}
 \item Term No.\ 8:
       \begin{equation} \label{eq:LHAMMVF18}
         \frac{\partial X_i^{\mu_1 \ldots\, \mu_r}}{\partial q^j}~
         =~\frac{\partial X_j^{\mu_1 \ldots\, \mu_r}}{\partial q^i}~.
       \end{equation}
\end{itemize}
Beginning with eqs~(\ref{eq:LHAMMVF14}) and~(\ref{eq:LHAMMVF15}), we observe
first of all that the rhs of both equations does not depend on the energy
variable, so they can be immediately integrated with respect to $p\>\!$.
Moreover, the rhs of eq.~(\ref{eq:LHAMMVF13}) does not depend on the
$p\>\!_j^\mu$, not only when $\, \mu \,\nsmin \{\mu_2,\ldots,\mu_r\} \,$
but even when $\, \mu \,\smin \{\mu_2,\ldots,\mu_r\}$. (Of course, it also
does not depend on $p\>\!$.) Indeed, assuming that $\mu=\mu_2$, say, we have
\begin{eqnarray*}
\lefteqn{\frac{\partial}{\partial p\>\!_j^\mu}
         \left( - \; \frac{\partial X^{i,\mu_2 \ldots\, \mu_r}}
                          {\partial x^\mu} \, + \,
                \sum_{s=2}^r \, \delta_\mu^{\mu_s} \,
                \frac{\partial X_{\phantom{i}}^{i,\mu_2 \ldots\, \mu_{s-1} \nu
                                                  \mu_{s+1} \ldots\, \mu_r}}
                     {\partial x_{\phantom{i}}^\nu} \right)}
                                                          \hspace{1cm} \\[2ex]
 &=&\!\! \mbox{} - \,
         \frac{\partial^{\>\!2} X^{i,\mu_2 \ldots\, \mu_r}}
              {\partial x_{\phantom{k}}^\mu \, \partial p\>\!_j^\mu} \, + \,
         \frac{\partial^{\>\!2} X^{i,\nu\mu_3 \ldots\, \mu_r}}
              {\partial x_{\phantom{k}}^{\nu\vphantom{\mu}} \,
               \partial p\>\!_j^\mu}~,
\end{eqnarray*}
and in the sum over $\nu$, only the term with $\nu=\mu_2$ survives, as for all
other terms one has $\, \mu \nsmin \{\nu,\mu_3,\ldots,\mu_r\}$, but this term
cancels exactly the first summand. Thus according to the lemma formulated in
the appendix, we can integrate eq.~(\ref{eq:LHAMMVF13}) explicitly to
obtain\footnote{Recall that $\Sigma^{-1}$ is the operator that acts on
  polynomials in the multimomentum variables and the energy variable
  without constant term by multiplying the homogeneous component of
  degree $s$ by $1/s$.}
\begin{eqnarray} \label{eq:LHAMMVF16}
 \tilde{X}_{\phantom{i}}^{\mu_2 \ldots\, \mu_r}  \!\!
 &=&\!\! (-1)^r \; p~
         \frac{\partial X_{\phantom{i}}^{\mu_2 \ldots\, \mu_r \nu}}
              {\partial x_{\phantom{i}}^\nu} \, - \,
         \Sigma^{-1} \! \left( \! p\>\!_i^\mu \;
         \frac{\partial X_{\phantom{i}}^{i,\mu_2 \ldots\, \mu_r}}
              {\partial x_{\phantom{i}}^\mu} \, - \,
         \sum_{s=2}^r \, p\>\!_i^{\mu_s} \;
         \frac{\partial X_{\phantom{i}}^{i,\mu_2 \ldots\, \mu_{s-1} \nu
                                           \mu_{s+1} \ldots\, \mu_r}}
              {\partial x_{\phantom{i}}^\nu} \right)    \nonumber \\[2ex]
 & &\!\! \mbox{} + \, \tilde{Y}_{\phantom{i}}^{\mu_2 \ldots\, \mu_r}~,
\end{eqnarray}
where the $\tilde{Y}_{\phantom{i}}^{\mu_2 \ldots\, \mu_r}$ are local functions
on~$E$: they do not depend on the multimomentum variables or on the energy
variable. The same procedure works for eq.~(\ref{eq:LHAMMVF17}): its rhs
does not depend on the $p\;\!_l^\kappa$, not only when $\, \kappa \,\nsmin
\{\mu_1,\ldots,\mu_r\} \,$ but even when $\, \kappa \,\smin \{\mu_1,\ldots,
\mu_r\}$. (Of course, it also does not depend on $p\>\!$.) Indeed, assuming
that $\kappa=\mu_1$, say, we have
\begin{eqnarray*}
\lefteqn{\frac{\partial}{\partial p\>\!_l^\kappa}
         \left( \delta_i^k \,
                \frac{\partial X_{\phantom{i}}^{\mu_1 \ldots\, \mu_r}}
                     {\partial x_{\phantom{i}}^\kappa} \, - \,
                \sum_{s=1}^r \, \delta_i^k \, \delta_\kappa^{\mu_s} \,
                \frac{\partial X_{\phantom{i}}^{\mu_1 \ldots\, \mu_{s-1} \nu
                                                \mu_{s+1} \ldots\, \mu_r}}
                     {\partial x_{\phantom{i}}^\nu} \right.}
                                                          \hspace{1cm} \\
 & &\!\! \mbox{} - \left.
                \sum_{s=1}^r \, (-1)^{s-1} \, \delta_\kappa^{\mu_s} \,
                \frac{\partial X_{\phantom{i}}^{k,\mu_1 \ldots\, \mu_{s-1}
                                                  \mu_{s+1} \ldots\, \mu_r}}
                     {\partial q^i} \right)                             \\[2ex]
 &=&\!\! \mbox{} - \, \delta_i^k \,
         \frac{\partial^{\>\!2} X^{\nu \mu_2 \ldots\, \mu_r}}
              {\partial x_{\phantom{k}}^\nu \,
               \partial p\>\!_l^\kappa} \, - \,
         \frac{\partial^{\>\!2} X^{k,\mu_2 \ldots\, \mu_r}}
              {\partial q^i \, \partial p\>\!_l^\kappa}~,
\end{eqnarray*}
and both of these terms vanish. Thus according to the lemma formulated in the
\linebreak
appendix, we can integrate eq.~(\ref{eq:LHAMMVF13}) explicitly to
obtain\addtocounter{footnote}{-1}\footnotemark
\begin{eqnarray} \label{eq:LHAMMVF19}
 X_i^{\mu_1 \ldots\, \mu_r}  \!\!
 &=&\!\! \mbox{} - \, p~
         \frac{\partial X_{\phantom{i}}^{\mu_1 \ldots\, \mu_r}}
              {\partial q^i} \, + \,
         p\>\!_i^\mu \;
         \frac{\partial X_{\phantom{i}}^{\mu_1 \ldots\, \mu_r}}
              {\partial x_{\phantom{i}}^\mu} \, - \,
         \sum_{s=1}^r \, p\>\!_i^{\mu_s} \;
         \frac{\partial X_{\phantom{i}}^{\mu_1 \ldots\, \mu_{s-1} \nu
                                         \mu_{s+1} \ldots\, \mu_r}}
              {\partial x_{\phantom{i}}^\nu}                  \nonumber \\
 & &\!\! \hspace{2.7cm} \mbox{} - \, \Sigma^{-1} \left(
         \sum_{s=1}^r \, (-1)^{s-1} \, p\>\!_k^{\mu_s} \;
         \frac{\partial X_{\phantom{i}}^{k,\mu_1 \ldots\, \mu_{s-1}
                                           \mu_{s+1} \ldots\, \mu_r}}
              {\partial q^i} \right) \qquad                             \\[2ex]
 & &\!\! \mbox{} + \, Y_i^{\mu_1 \ldots\, \mu_r}~,           \nonumber
\end{eqnarray}
where the $Y_i^{\mu_1 \ldots\, \mu_r}$ are local functions on $E$: they do
not depend on the multimomentum variables or on the energy variable. Direct
calculation now shows that eqs~(\ref{eq:LHAMMVF11}), (\ref{eq:LHAMMVF12})
and (\ref{eq:LHAMMVF18}) reduce to
\begin{equation} \label{eq:LHAMMVF21}
 \frac{\partial \tilde{Y}_{\phantom{i}}^{\mu_3 \ldots\, \mu_r \nu}}
      {\partial x_{\phantom{i}}^\nu}~=~0~,
\end{equation}
\begin{equation} \label{eq:LHAMMVF22}
 \frac{\partial \tilde{Y}_{\phantom{i}}^{\mu_2 \ldots\, \mu_r}}{\partial q^i}~
 =~(-1)^{r-1} \, \frac{\partial Y_i^{\mu_2 \ldots\, \mu_r \nu}}
                      {\partial x_{\phantom{i}}^\nu}~,
\end{equation}
and
\begin{equation} \label{eq:LHAMMVF23}
 \frac{\partial Y_i^{\mu_1 \ldots\, \mu_r}}{\partial q^j}~
 =~\frac{\partial Y_j^{\mu_1 \ldots\, \mu_r}}{\partial q^i}~,
\end{equation}
respectively. This system of equations is easily solved by setting
\begin{equation} \label{eq:LHAMMVF24}
 \tilde{Y}_{\phantom{i}}^{\mu_2 \ldots\, \mu_r}
 =~(-1)^{r-1} \, \frac{\partial X_-^{\mu_2 \ldots\, \mu_r \mu}}
                      {\partial x_{\phantom{i}}^\mu} \quad , \quad
 Y_i^{\mu_1 \ldots\, \mu_r}
 =~\frac{\partial X_-^{\mu_1 \ldots\, \mu_r}}{\partial q^i}~,
\end{equation}
where the $X_-^{\mu_1 \ldots\, \mu_r}$ are local functions on $E$: they do not
depend on the multimomentum \linebreak variables or on the energy variable.
This completes the proof of the statements in item~3.\ of the theorem.
\\[1mm]
All that remains to be shown are the final statements concerning exact
Hamiltonian multivector fields. To this end, we apply the exterior
derivative to eq.~(\ref{eq:HAMMVF3}) and use eq.~(\ref{eq:CMPDF6})
to simplify the expressions involving derivatives with respect to
the space-time variables. Combining this with eq.~(\ref{eq:HAMMVF2})
and collecting the terms, we get
\vspace{2mm}
\begin{eqnarray*}
 L_X^{} \theta \!\!
 &=&\!\! \frac{1}{(r\!-\!1)!} \;
         \biggl( \frac{\partial X_{\phantom{i}}^{\mu_2 \ldots\, \mu_r \nu}}
                      {\partial x_{\phantom{i}}^\nu} \; p \; - \, (-1)^r \;
                 \frac{\partial X_{\phantom{i}}^{i,\mu_2 \ldots\, \mu_r}}
                      {\partial x_{\phantom{i}}^\mu} \; p\>\!_i^\mu     \\
 & &     \hspace{1.6cm} \mbox{} + \,
                 (-1)^r \, \sum_{s=2}^r \,
                 \frac{\partial X_{\phantom{i}}^{i,\mu_2 \ldots\, \mu_{s-1} \nu
                                                   \mu_{s+1} \ldots\, \mu_r}}
                      {\partial x_{\phantom{i}}^\nu} \; p\>\!_i^{\mu_s} \, - \,
                 (-1)^r \; \tilde{X}_{\phantom{i}}^{\mu_2 \ldots\, \mu_r}
                 \biggr) \;
         d^{\,n} x_{\mu_2 \ldots\, \mu_r}                              \\[1mm]
 & &     \mbox{} - \; \frac{1}{r!} \;
         \biggl( \frac{\partial X_{\phantom{i}}^{\mu_1 \ldots\, \mu_r}}
                      {\partial x_{\phantom{i}}^\mu} \; p\>\!_i^\mu \, - \,
                 \sum_{s=1}^r \,
                 \frac{\partial X_{\phantom{i}}^{\mu_1 \ldots\, \mu_{s-1} \nu
                                                 \mu_{s+1} \ldots\, \mu_r}}
                      {\partial x_{\phantom{i}}^\nu} \; p\>\!_i^{\mu_s} \, - \,
                 \frac{\partial X_{\phantom{i}}^{\mu_1 \ldots\, \mu_r}}
                      {\partial q^i} \; p                               \\
 & &     \hspace{1.4cm} \mbox{} - \,
                 \sum_{s=1}^r \, (-1)^{s-1} \,
                 \frac{\partial X_{\phantom{i}}^{j,\mu_1 \ldots\, \mu_{s-1}
                                                   \mu_{s+1} \ldots\, \mu_r}}
                      {\partial q^i} \; p\>\!_j^{\mu_s} \, - \,
                 X_i^{\mu_1 \ldots\, \mu_r} \biggr) \;
         dq^i \,\smwedge\> d^{\,n} x_{\mu_1 \ldots\, \mu_r}
\end{eqnarray*}
\begin{eqnarray*}
 \phantom{L_X^{} \theta \!\!}
 & &     \mbox{} + \; \frac{1}{r!} \;
         \biggl( \frac{\partial X_{\phantom{i}}^{\mu_1 \ldots\, \mu_r}}
                      {\partial p\>\!_j^\nu} \; p \; + \,
                 \sum_{s=1}^r \, (-1)^{s-1} \,
                 \frac{\partial X_{\phantom{i}}^{i,\mu_1 \ldots\, \mu_{s-1}
                                                   \mu_{s+1} \ldots\, \mu_r}}
                      {\partial p\>\!_j^\nu} \; p\>\!_i^{\mu_s} \biggr) \;
         dp\>\!_j^\nu \,\smwedge\> d^{\,n} x_{\mu_1 \ldots\, \mu_r}     \\[1mm]
 & &     \mbox{} + \; \frac{1}{r!} \;
         \biggl( \frac{\partial X_{\phantom{i}}^{\mu_1 \ldots\, \mu_r}}
                      {\partial p} \; p \; + \,
                 \sum_{s=1}^r \, (-1)^{s-1} \,
                 \frac{\partial X_{\phantom{i}}^{i,\mu_1 \ldots\, \mu_{s-1}
                                                   \mu_{s+1} \ldots\, \mu_r}}
                      {\partial p} \; p\>\!_i^{\mu_s} \biggr) \;
         dp \,\,\smwedge\> d^{\,n} x_{\mu_1 \ldots\, \mu_r}             \\[2mm]
 & &     \mbox{} - \; \frac{(-1)^r}{r} \;
         \frac{\partial X_{\phantom{i}}^{\mu_1 \ldots\, \mu_r}}
              {\partial q^j} \; p\>\!_i^\mu \;
         dq^i \,\smwedge\, dq^j \,\smwedge\>
         d^{\,n} x_{\mu \mu_1 \ldots\, \mu_r}                           \\[2mm]
 & &     \mbox{} - \, \frac{(-1)^r}{r!} \;
         \frac{\partial X_{\phantom{i}}^{\mu_1 \ldots\, \mu_r}}
              {\partial p\>\!_j^\nu} \; p\>\!_i^\mu \;
         dq^i \,\smwedge\, dp\>\!_j^\nu \,\smwedge\>
         d^{\,n} x_{\mu \mu_1 \ldots\, \mu_r}                           \\[2mm]
 & &     \mbox{} - \, \frac{(-1)^r}{r!} \;
         \frac{\partial X_{\phantom{i}}^{\mu_1 \ldots\, \mu_r}}
              {\partial p} \; p\>\!_i^\mu \;
         dq^i \,\smwedge\, dp \;\smwedge\>
         d^{\,n} x_{\mu \mu_1 \ldots\, \mu_r}~.
\vspace{2mm}
\end{eqnarray*}
(Note that the last three terms would have to be omitted if $\, r=n$.)
Numbering the terms in this equation from $1$ to $7$, we see that the
conditions imposed by the fact that $X$ should be locally Hamiltonian
are already sufficient to eliminate the last four terms and imply that
the first three terms will vanish as well if and only if we have
\[
 (\Sigma^{-1} - 1)
 \left( \sum_{s=1}^r \, (-1)^{s-1} \, p\>\!_j^{\mu_s} \;
        \frac{\partial X_{\phantom{i}}^{j,\mu_1 \ldots\, \mu_{s-1}
                                          \mu_{s+1} \ldots\, \mu_r}}
             {\partial q^i} \right)\!~=~0~,
\]
\[
 (\Sigma^{-1} - 1)
 \left( p\>\!_i^\mu \;
        \frac{\partial X_{\phantom{i}}^{i,\mu_2 \ldots\, \mu_r}}
             {\partial x_{\phantom{i}}^\mu} \, - \,
        \sum_{s=2}^r \, p\>\!_i^{\mu_s} \;
        \frac{\partial X_{\phantom{i}}^{i,\mu_2 \ldots\, \mu_{s-1} \nu
                                          \mu_{s+1} \ldots\, \mu_r}}
             {\partial x_{\phantom{i}}^\nu} \right)\!~=~0~,
\]
and
\[
 \frac{\partial X_-^{\mu_1 \ldots\, \mu_r}}{\partial q^i}~=~0 \quad , \quad
 \frac{\partial X_-^{\mu_2 \ldots\, \mu_r \nu}}
      {\partial x_{\phantom{i}}^\nu}~=~0~.
\]
But this means that the coefficients of the multimomentum variables in the
above expressions must be independent of the multimomentum variables and that
the coefficients $X_-^{\mu_1 \ldots\, \mu_r}$ can without loss of generality
be assumed to vanish, which completes the proof of the theorem. \\
\PCPqed

\noindent
\textbf{Proof of Theorem~\ref{thm:FIBPOL} and
Theorem~\ref{thm:DECSCD}, item 1.}~~Clearly, it suffices to prove the
statement of Theorem~\ref{thm:FIBPOL} , namely the possibility to
decompose an arbitrary locally Hamiltonian $r$-multivector field~$X$
into the sum of a fiberwise polynomial locally Hamiltonian
$r$-multivector field and an $r$-multivector field taking values in
the kernel of~$\,\omega$, locally and in coordinates, since both
properties~-- that of being fiberwise polynomial as well as that of
taking values in the kernel of $\,\omega$~-- are algebraic conditions
which hold in any coordinate system as soon as they hold in one and
which are preserved when such local decompositions are glued together
by means of a partition of unity. The same goes for the main statement
of Theorem~\ref{thm:DECSCD}, namely the fact that the homogeneous
components $X_s^{}$ of a fiberwise polynomial locally Hamiltonian
$r$-multivector field $X$ take values in the kernel of~$\,\omega$ as
soon as $s$ lies outside the range between $-1$ and $r-1$. But in
adapted local coordinates, all these statements follow directly from
Theorem~\ref{thm:HAMMVF}.  In~fact, note that
\[
 \Bigl[  \; \Sigma \,,
         \frac{\partial}{\partial x_{\phantom{i}}^{\mu_1}}
         \;\smwedge \ldots \smwedge\;
         \frac{\partial}{\partial x_{\phantom{i}}^{\mu_r}} \; \Bigr]~
 =~0 \quad , \quad
 \Bigl[  \; \Sigma\,,
         \frac{\partial}{\partial q^i} \;\smwedge\;
         \frac{\partial}{\partial x_{\phantom{i}}^{\mu_2}}
         \;\smwedge \ldots \smwedge\;
         \frac{\partial}{\partial x_{\phantom{i}}^{\mu_r}} \; \Bigr]~
 =~0~,
\]
\vspace*{-5mm}
\begin{eqnarray*}
 \Bigl[  \; \Sigma \,,
         \frac{\partial}{\partial p\>\!_i^{\mu_1}} \;\smwedge\;
         \frac{\partial}{\partial x_{\phantom{i}}^{\mu_2}}
         \;\smwedge \ldots \smwedge\;
         \frac{\partial}{\partial x_{\phantom{i}}^{\mu_r}} \; \Bigr] \!\!
 &=&\!\! \mbox{} - \;
         \frac{\partial}{\partial p\>\!_i^{\mu_1}} \;\smwedge\;
         \frac{\partial}{\partial x_{\phantom{i}}^{\mu_2}}
         \;\smwedge \ldots \smwedge\;
         \frac{\partial}{\partial x_{\phantom{i}}^{\mu_r}}~,            \\[3mm]
 \Bigl[  \; \Sigma \,,
         \frac{\partial}{\partial p} \;\smwedge\;
         \frac{\partial}{\partial x_{\phantom{i}}^{\mu_2}}
         \;\smwedge \ldots \smwedge\;
         \frac{\partial}{\partial x_{\phantom{i}}^{\mu_r}} \; \Big] \!\!
 &=&\!\! \mbox{} - \;
         \frac{\partial}{\partial p} \;\smwedge\;
         \frac{\partial}{\partial x_{\phantom{i}}^{\mu_2}}
         \;\smwedge \ldots \smwedge\;
         \frac{\partial}{\partial x_{\phantom{i}}^{\mu_r}}~,
\end{eqnarray*}
so if $X$ has the local coordinate expression~(\ref{eq:HAMMVF1}), its Lie
derivative along the scaling vector field $\Sigma$ will, according to the
Leibniz rule for the Lie derivative, have the local coordinate expression
\vspace{2mm}
\begin{eqnarray}
 L_\Sigma^{} X \!\!
 &=&\!\! \frac{1}{(r\!-\!1)!} \,
         \Bigl( \Sigma \cdot X_{\phantom{i}}^{i,\mu_2 \ldots\, \mu_r} \Bigr) \;
         \frac{\partial}{\partial q^i} \;\smwedge\;
         \frac{\partial}{\partial x_{\phantom{i}}^{\mu_2}}
         \;\smwedge \ldots \smwedge\;
         \frac{\partial}{\partial x_{\phantom{i}}^{\mu_r}}    \nonumber \\[1mm]
 & &\!\! \mbox{} + \, \frac{1}{r!} \,
         \Bigl( (\Sigma - 1) \cdot X_i^{\mu_1 \ldots\, \mu_r} \Bigr) \;
         \frac{\partial}{\partial p\>\!_i^{\mu_1}} \;\smwedge\;
         \frac{\partial}{\partial x_{\phantom{i}}^{\mu_2}}
         \;\smwedge \ldots \smwedge\;
         \frac{\partial}{\partial x_{\phantom{i}}^{\mu_r}}    \nonumber \\[1mm]
 & &\!\! \mbox{} + \, \frac{1}{(r\!-\!1)!} \,
         \Bigl( (\Sigma - 1) \cdot
                \tilde{X}_{\phantom{i}}^{\mu_2 \ldots\, \mu_r} \Bigr) \;
         \frac{\partial}{\partial p} \;\smwedge\;
         \frac{\partial}{\partial x_{\phantom{i}}^{\mu_2}}
         \;\smwedge \ldots \smwedge\;
         \frac{\partial}{\partial x_{\phantom{i}}^{\mu_r}}    \nonumber \\[3mm]
 & &\!\! \mbox{} + \, L_\Sigma^{} \xi~,                       \nonumber
\end{eqnarray}
and if $X$ is locally Hamiltonian, Theorem~\ref{thm:HAMMVF} forces
the coefficient functions $X_{\phantom{i}}^{i,\mu_2 \ldots\, \mu_r}$,
$X_i^{\mu_1 \ldots\, \mu_r}$ and $\tilde{X}_{\phantom{i}}^{\mu_2
\ldots\, \mu_r}$ to be polynomials of degree $r-1$, $r$ and $r$,
respectively, in the multi\-momentum variables and the energy variable.
Finally, Proposition~\ref{prp:INVLIESIGMVF} implies that if $X$ is
locally Hamiltonian or globally Hamiltonian or exact Hamiltonian or
takes values in the kernel of~$\,\omega$, the same is true for all
its homogeneous components~$X_s^{}$. \\
\PCPqed

The following proposition clarifies the interpretation of homogeneous
locally Hamiltonian multivector fields.
\begin{prp} \label{prp:HOMHAMMVF}
 Let $X$ be a locally Hamiltonian $r$-multivector field on $P$. Then
 \begin{enumerate}
  \item $X$ is exact Hamiltonian iff\/ $[\Sigma,X]$ takes values
        in the kernel of\/ $\omega$.
  \item If\/ $[\Sigma,X] - sX$ takes values in the kernel of\/ $\omega$,
        for some integer $s$ between $0$ and $r-1$, then $X$ is globally
        Hamiltonian with associated Poisson form
        \[
         \frac{(-1)^{r-1}}{s+1} \; i_X^{} \theta~.
        \]
  \item If\/ $[\Sigma,X] + X$ takes values in the kernel of\/ $\omega$,
        then $\, i_X^{} \theta = 0$.
 \end{enumerate}
\end{prp}

%\pagebreak

\noindent
\textbf{Proof.}~~The first statement follows immediately from
eq.~(\ref{eq:LTIO}). Similarly, the second claim can be proved by
multiplying eq.~(\ref{eq:LTIO}) by $(-1)^{r-1}/(s+1)$ and combining it
with eq.~(\ref{eq:EXMSM}) and eq.~(\ref{eq:LDFMVF}) to give
\[
 d \left( \frac{(-1)^{r-1}}{s+1} \; i_X^{} \theta \right)\!~
 =~\frac{(-1)^{r-1}}{s+1} \; L_X^{} \theta \, + \,
   \frac{1}{s+1} \; i_X^{} \omega~
 =~\frac{1}{s+1} \; i_{[\Sigma,X]+X}^{} \>\! \omega~,
\]
which equals $i_X^{} \omega$ since, by hypothesis,
$\, i_{[\Sigma,X] - sX}^{} \>\! \omega = 0$. Finally, the
third statement follows by observing that the kernel of~$\,\omega$
is contained in the kernel of $\theta$ and hence according to the
hypothesis made,
\begin{equation*}
 0~=~i_{[\Sigma,X]+X}^{} \theta~
   =~L_\Sigma^{} i_X^{} \theta - i_X^{} L_\Sigma^{} \theta + i_X^{} \theta~
   =~L_\Sigma^{} i_X^{} \theta~,
\end{equation*}
where we have used eq.~(\ref{eq:MCANF03}). Therefore, according to
Proposition~A.1, $i_X^{} \theta$ is the pull-back to $P$ of an $n$-form
on $E$ via the projection that defines $P$ as a vector bundle over~$E$,
which in turn can be obtained as the pull back to~$E$ of $i_X^{} \theta$
via the zero section of $P$ over~$E$. But this pull-back is zero, since
$\theta$ vanishes along the zero section of $P$ over~$E$. \\
\PCPqed

It may be instructive to spell all this out more explicitly for locally
Hamiltonian vector fields ($r\!=\!1$) and bivector fields ($r\!=\!2$).

We begin by writing down the general form of a locally Hamiltonian vector
field $X$: in adapted local coordinates, it has the representation
\begin{equation} \label{eq:HAMVF1}
 X~=~X_{}^\mu \, \frac{\partial}{\partial x_{\phantom{i}}^\mu} \, + \,
     X_{}^i \, \frac{\partial}{\partial q^i} \, + \,
     X_i^\mu \, \frac{\partial}{\partial p\>\!_i^\mu} \, + \,
     \tilde{X} \, \frac{\partial}{\partial p}~,
\end{equation}
where according to Theorem~\ref{thm:HAMMVF}, the coefficient functions
$X_{}^\mu$ and $X_{}^i$ depend only on the local coordinates $x^\rho$
for~$M$ and on the local fiber coordinates $q^r$ for~$E$ (the $X_{}^\mu$
being independent of the latter as soon as $N\!>\!1$), whereas the
coefficient functions $X_i^\mu$ and $\tilde{X}$ are explicitly given
by
\begin{equation} \label{eq:LHAMVF03}
 X_i^\mu~=~- \, p \; \frac{\partial X_{\phantom{i}}^\mu}{\partial q^i} \, + \,
           p\>\!_i^\nu \; \frac{\partial X_{\phantom{i}}^\mu}
                               {\partial x_{\phantom{i}}^\nu} \, - \,
           p\>\!_i^\mu \; \frac{\partial X_{\phantom{i}}^{\nu\vphantom{\mu}}}
                               {\partial x_{\phantom{i}}^\nu} \, - \,
           p\>\!_j^\mu \; \frac{\partial X_{\phantom{i}}^j}
                               {\partial q^i} \, + \,
           \frac{\partial X_-^\mu}{\partial q^i}
\end{equation}
(the first term being absent as soon as $\, N>1$) and
\begin{equation} \label{eq:LHAMVF04}
 \tilde{X}~=~- \, p \; \frac{\partial X_{\phantom{i}}^\nu}
                            {\partial x_{\phantom{i}}^\nu} \, - \,
             p\>\!_i^\mu \; \frac{\partial X_{\phantom{i}}^i}
                                 {\partial x_{\phantom{i}}^\mu} \, + \,
             \frac{\partial X_-^\nu}{\partial x_{\phantom{i}}^\nu}
\end{equation}
with coefficient functions $X_-^\mu$ that once again depend only on the
local coordinates~$x^\rho$ for~$M$ and on the local fiber coordinates~$q^r$
for~$E$. Regarding the decomposition (\ref{eq:HAMMVFSCD1}), the situation
here is particularly interesting and somewhat special since $\,\omega$
is nondegenerate on vector fields, so there are no nontrivial vector
fields taking values in the kernel of~$\,\omega$ and hence the
decomposition (\ref{eq:HAMMVFSCD1}) can be improved:
\begin{cor}
 Any locally Hamiltonian vector field $X$ on $P$ can be uniquely decomposed
 into the sum of two terms,
 \begin{equation} \label{eq:HAMVFCD}
  X~=~X_-^{} + X_+^{}~,
 \end{equation}
 where
 \begin{itemize}
  \item $X_-^{}$ has scaling degree $-1$, i.e., $[\Sigma,X_-^{}] = - X_-^{}$,
        and is vertical with respect to the projection onto $E$.
  \item $X_+^{}$ has scaling degree $0$, i.e., $[\Sigma,X_+^{}] = 0$, is
        exact Hamiltonian, is projectable onto $E$ and coincides with the
        canonical lift of its projection onto $E$.
 \end{itemize}
\end{cor}
\textbf{Proof.}~~In adapted local coordinates, the two contributions to $X$
are, according to eqs~(\ref{eq:LHAMVF03}) and~(\ref{eq:LHAMVF04}), given by
\begin{equation} \label{eq:LHAMVF05}
 X_-~=~\frac{\partial X_-^\mu}{\partial q^i} \;
       \frac{\partial}{\partial p\>\!_i^\mu} \, + \,
       \frac{\partial X_-^\nu}{\partial x_{\phantom{i}}^\nu} \;
       \frac{\partial}{\partial p}~,
\end{equation}
and
\begin{eqnarray} \label{eq:LHAMVF06}
 X_+^{} \!\!
 &=&\!\! X^\mu \, \frac{\partial}{\partial x_{\vphantom{i}}^\mu} \, + \,
         X^i \, \frac{\partial}{\partial q_{\vphantom{i}}^i}  \nonumber \\
 & &\!\! \mbox{} -
         \left(
          \frac{\partial X^j}{\partial q_{\vphantom{i}}^i} \,
          p\;\!_j^\mu \, - \,
          \frac{\partial X^\mu}{\partial x_{\vphantom{i}}^\nu} \,
          p\;\!_i^\nu \, + \,
          \frac{\partial X^\nu}{\partial x_{\vphantom{i}}^\nu} \,
          p\;\!_i^\mu \, + \,
          \frac{\partial X^\mu}{\partial q_{\vphantom{i}}^i} \, p
         \right)
         \frac{\partial}{\partial p\;\!_i^\mu}                          \\
 & &\!\! \mbox{} -
         \left(
          \frac{\partial X^i}{\partial x_{\vphantom{i}}^\mu} \,
          p\;\!_i^\mu \, + \,
          \frac{\partial X^\nu}{\partial x_{\vphantom{i}}^\nu} \, p
         \right)
         \frac{\partial}{\partial p_{\vphantom{i}}^{\vphantom{j}}}~.
                                                              \nonumber
\end{eqnarray}
Thus all statements of the corollary follow from what has already been shown,
except for the very last one, which is based on the following remark. \\
\PCPqed

\noindent
\textbf{Remark.}~~Every bundle automorphism of $E$ (as a fiber bundle over~$M$)
admits a canonical lift to a bundle automorphism of its first order jet bundle
$JE$ (as an affine bundle over $E$) and, by appropriate (twisted affine)
dualization, to the extended multiphase space $P$ (as a vector bundle
over~$E$). Similarly, passing to generators of one-parameter groups, one
sees that every vector field $X_E^{}$ on $E$ that is projectable to a vector
field $X_M^{}$ on $M$ admits a canonical lift to a vector field $X_{JE}^{}$
on $JE$ and, by appropriate (twisted affine) dualization, to a vector field
$X_P$ on $P$. (See, for example, \cite[\S 4B]{GIM}.) When $N\!=\!1$, lifting
to $P$ is even possible for arbitrary diffeomorphisms of $E$ and arbitrary
vector fields on $E$, since in this case $P$ can be identified with the
$n^{\mathrm{th}}$ exterior power of the cotangent bundle of $E$. Explicitly,
in terms of adapted local coordinates $(x_{\vphantom{i}}^\mu,
q_{\vphantom{i}}^i,p\>\!_i^\mu,p\>\!)$, we may write
\begin{equation} \label{eq:VFBASE1}
 X_M^{}~=~X_{}^\mu \, \frac{\partial}{\partial x_{\vphantom{i}}^\mu}~,
\end{equation}
and
\begin{equation} \label{eq:VFTOTS1}
 X_E^{}~=~X_{}^\mu \, \frac{\partial}{\partial x_{\vphantom{i}}^\mu} \, + \,
          X_{}^i \, \frac{\partial}{\partial q_{\vphantom{i}}^i}~,
\vspace{1mm}
\end{equation}
where, except for $N\!=\!1$, the $X_{}^\mu$ do not depend on the
$q_{\vphantom{r}}^r$; then
\begin{eqnarray} \label{eq:VFEMFS1}
 X_P^{} \!\!
 &=&\!\! X^\mu \, \frac{\partial}{\partial x_{\vphantom{i}}^\mu} \, + \,
         X^i \, \frac{\partial}{\partial q_{\vphantom{i}}^i}  \nonumber \\
 & &\!\! \mbox{} -
         \left(
          \frac{\partial X^j}{\partial q_{\vphantom{i}}^i} \,
          p\;\!_j^\mu \, - \,
          \frac{\partial X^\mu}{\partial x_{\vphantom{i}}^\nu} \,
          p\;\!_i^\nu \, + \,
          \frac{\partial X^\nu}{\partial x_{\vphantom{i}}^\nu} \,
          p\;\!_i^\mu \, + \,
          \frac{\partial X^\mu}{\partial q_{\vphantom{i}}^i} \, p
         \right)
         \frac{\partial}{\partial p\;\!_i^\mu}                          \\
 & &\!\! \mbox{} -
         \left(
          \frac{\partial X^i}{\partial x_{\vphantom{i}}^\mu} \,
          p\;\!_i^\mu \, + \,
          \frac{\partial X^\nu}{\partial x_{\vphantom{i}}^\nu} \, p
         \right)
         \frac{\partial}{\partial p_{\vphantom{i}}^{\vphantom{j}}}~.
                                                              \nonumber
\end{eqnarray}
Obviously, $X_P^{}$ has scaling degree $0$ and hence is not only
locally but even exact Hamiltonian. Conversely, since the expressions
in eqs~(\ref{eq:LHAMVF06}) and~(\ref{eq:VFEMFS1}) are identical, we
see that all exact Hamiltonian vector fields are obtained by this
lifting procedure. Similarly, one can show that all diffeomorphisms of
$P$ that preserve the multicanonical form $\theta$ are obtained by
lifting of automorphisms or, for $N\!=\!1$, diffeomorphisms of $E$:
this is the field theoretical analogue of a well-known theorem in
geometric mechanics, according to which all diffeomorphisms of a
cotangent bundle that preserve the canonical form $\theta$ are induced
by diffeomorphisms of its base manifold.

Similarly, we write down the general form of a locally Hamiltonian bivector
field $X$: in adapted local coordinates, it has the representation
\begin{equation} \label{eq:HAMBVF1}
 \begin{array}{rcl}
  X \!\!
  &=&\!\! \frac{1}{2} \;
          {\displaystyle
           X_{\phantom{i}}^{\mu\nu} \;
           \frac{\partial}{\partial x_{\phantom{i}}^\mu}
           \;\smwedge\;
           \frac{\partial}
                {\partial x_{\phantom{i}}^{\nu\vphantom{\mu}}} \, + \,
           X_{\phantom{i}}^{i,\mu} \;
           \frac{\partial}{\partial q^i} \;\smwedge\;
           \frac{\partial}{\partial x_{\phantom{i}}^\mu}} \\[2ex]
  & &\!   \mbox{} + \, \frac{1}{2} \;
          {\displaystyle
           X_i^{\mu\nu} \;
           \frac{\partial}{\partial p\>\!_i^\mu} \;\smwedge\;
           \frac{\partial}
                {\partial x_{\phantom{i}}^{\nu\vphantom{\mu}}} \, + \,
           \tilde{X}_{}^\mu \;
           \frac{\partial}{\partial p} \;\smwedge\;
           \frac{\partial}{\partial x_{\phantom{i}}^\mu}} \\[3ex]
  & &\!    \mbox{} + \, \xi~,
 \end{array}
\end{equation}
with
\begin{equation} \label{eq:LHAMBVF2}
 X_{\phantom{i}}^{i,\mu}~=~p\>\!_j^\mu \, Y_1^{ij} \, + \, Y_0^{i,\mu}~,
\end{equation}
where according to Theorem~\ref{thm:HAMMVF}, the coefficient functions
$X_{\phantom{i}}^{\mu\nu}$, $Y_1^{ij}$ and $Y_0^{i,\mu}$ depend only on
the local coordinates~$x^\rho$ for~$M$ and on the local fiber
coordinates~$q^r$ for~$E$ (the $X_{}^{\mu\nu}$ being independent of
the latter as soon as $N\!>\!1$), whereas the coefficient functions
$X_i^{\mu\nu}$ and $\tilde{X}_{}^\mu$ are explicitly given by
\begin{equation} \label{eq:LHAMBVF03}
 \begin{array}{rcl}
  X_i^{\mu\nu} \!\!
  &=&\!\! {\displaystyle
           \mbox{} - \, p \; \frac{\partial X_{\phantom{i}}^{\mu\nu}}
                                  {\partial q^i} \, + \,
           p\>\!_i^\kappa \; \frac{\partial X_{\phantom{i}}^{\mu\nu}}
                                  {\partial x_{\phantom{i}}^\kappa} \, - \,
           p\>\!_i^\mu \; \frac{\partial X_{\phantom{i}}^{\kappa\nu}}
                               {\partial x_{\phantom{i}}^\kappa} \, - \,
           p\>\!_i^\nu \; \frac{\partial X_{\phantom{i}}^{\mu\kappa}}
                               {\partial x_{\phantom{i}}^\kappa}} \\[2ex]
  & &\!\! {\displaystyle
           \mbox{} - \,
           {\textstyle \frac{1}{2}} \,
           p\>\!_j^\mu \, p\>\!_k^\nu \;
           \frac{\partial Y_1^{jk}}{\partial q^i} \, - \,
           p\>\!_j^\mu \, \frac{\partial Y_0^{j,\nu}}{\partial q^i} \, + \,
           {\textstyle \frac{1}{2}} \,
           p\>\!_j^\nu \, p\>\!_k^\mu \;
           \frac{\partial Y_1^{jk}}{\partial q^i} \, + \,
           p\>\!_j^\nu \, \frac{\partial Y_0^{j,\mu}}{\partial q^i} \, + \,
           \frac{\partial X_-^{\mu\nu}}{\partial q^i}}
 \end{array}
\end{equation}
(the first term being absent as soon as $N\!>\!1$) and
\begin{equation} \label{eq:LHAMBVF04}
 \tilde{X}_{}^\mu~
 =~p \; \frac{\partial X_{\phantom{i}}^{\mu\nu}}
             {\partial x_{\phantom{i}}^\nu} \, - \,
   {\textstyle \frac{1}{2}} \, p\;\!_i^\nu p\;\!_j^\mu \,
   \frac{\partial Y_1^{ij}}{\partial x_{\phantom{i}}^\nu} \, - \,
   p\;\!_i^\nu \,
   \frac{\partial Y_0^{i,\mu}}{\partial x_{\phantom{i}}^\nu} \, + \,
   {\textstyle \frac{1}{2}} \, p\;\!_i^\mu p\;\!_j^\nu \,
   \frac{\partial Y_1^{ij}}{\partial x_{\phantom{i}}^\nu} \, + \,
   p\;\!_i^\mu \,
   \frac{\partial Y_0^{i,\nu}}{\partial x_{\phantom{i}}^\nu} \, - \,
   \frac{\partial X_-^{\mu\nu}}{\partial x_{\phantom{i}}^\nu}
\end{equation}
with coefficient functions $X_-^{\mu\nu}$ that once again depend only on the
local coordinates $x^\rho$ for~$M$ and on the local fiber coordinates $q^r$
for~$E$. Note that now
\begin{equation} \label{eq:LHAMBVF05}
  X_-~=~\frac{1}{2} \,
        \frac{\partial X_-^{\mu\nu}}{\partial q^i} \;
        \frac{\partial}{\partial p\>\!_i^\mu} \;\smwedge\;
        \frac{\partial}{\partial x_{\phantom{i}}^{\nu\vphantom{\mu}}} \, - \,
        \frac{\partial X_-^{\mu\nu}}{\partial x_{\phantom{i}}^\nu} \;
        \frac{\partial}{\partial p} \;\smwedge\;
        \frac{\partial}{\partial x_{\phantom{i}}^\mu}~.
\end{equation}
Moreover, the operator $1+L_\Sigma^{}$ kills $X_-$ and removes the factors
$\frac{1}{2}$ in front of the quadratic terms in eqs~(\ref{eq:LHAMBVF03})
and~(\ref{eq:LHAMBVF04}).

To conclude this section, let us note that the definition of projectability
of vector fields can be immediately generalized to multivector fields: an
$r$-multivector field $X_E^{}$ on the total space~$E$ of a fiber bundle over
a manifold~$M$ with bundle projection $\; \pi: E \rightarrow M \;$ is called
\emph{projectable} if for any two points $e_1^{}$ and $e_2^{}$ in $E$,
\begin{equation} \label{eq:PRJMVF1}
 \bwedge^r T_{e_1}^{} \pi \cdot X_E^{}(e_1^{})~
 =~\bwedge^r T_{e_2}^{} \pi \cdot X_E^{}(e_2^{})
 \qquad \mbox{if} \qquad \pi(e_1^{}) = \pi(e_2^{})~,
\end{equation}
or in other words, if there exists an $r$-multivector field $X_M^{}$ on~$M$
such that
\begin{equation} \label{eq:PRJMVF2}
 \bwedge^r T \pi \smcirc X_E^{}~=~X_M^{} \circ \pi~.
\end{equation}
In adapted local coordinates, this amounts to requiring that if we
write
\begin{equation} \label{eq:PRJMVF3}
 X_E^{}~=~\frac{1}{r!} \, X_{}^{\mu_1 \ldots\, \mu_r} \;
          \frac{\partial}{\partial x^{\mu_1}}
          \;\smwedge \ldots \smwedge\;
          \frac{\partial}{\partial x^{\mu_r}} \; + \; \ldots~,
\end{equation}
where the dots denote $1$-vertical terms, the coefficients $X_{}^{\mu_1
\ldots\, \mu_r}$ should depend only on the local coordinates $x^\rho$
for~$M$ but not on the local fiber coordinates $q^r$ for~$E$. Now we
introduce the following terminology.
\begin{dfn} \label{def:PROJMVF}
 An $r$-multivector field on~$P$ is called \textbf{projectable} if it is
 projectable with respect to any one of the three projections from~$P$:
 to~$P_0$, to~$E$ and to~$M$.
\end{dfn}
With this terminology, Theorem~\ref{thm:HAMMVF} states that for $\, 0 < r < n$,
locally Hamiltonian $r$-multivector fields on~$P$ are projectable as soon as
$N\!>\!1$ and are projectable to~$E$ but not necessarily to~$P_0$ or to~$M$
when $N\!=\!1$. (Inspection of eq.~(\ref{eq:LHAMMVF03}) shows, however, that
they are projectable to~$P_0$ if and only if they are projectable to~$M$.)

Considering the special case of vector fields ($r\!=\!1$), we believe that
vector fields on the total space of a fiber bundle over space-time which
are not projectable should be regarded as pathological, since they generate
transformations which do not induce transformations of space-time. It is
hard to see how such transformations might be interpreted as candidates
for symmetries of a physical system. By analogy, we shall adopt the same
point of view regarding multivector fields of higher degree, since although
these do not generate diffeomorphisms of~$E$ as a manifold, they may perhaps
allow for an interpretation as generators of superdiffeomorphisms of an
appropriate supermanifold built over~$E$ as its even part.

%%%%%%%%%%%%%%%%%%%%%%%%%%%%%%%%%%%%%%%%%%%%%%%%%%%%%%%%%%%%%%%%%%%%%%%%%%%%%%%
\section{Poisson forms and Hamiltonian forms}

Our aim in this section is to give an explicit construction of Poisson
$(n-r)$-forms and, more generally, of Hamiltonian $(n-r)$-forms on the
extended multiphase space~$P$, where $\, 0 \leqslant r \leqslant n$.
(Note that eq.~(\ref{eq:HAMMVFF}) only makes sense for $r$ in this range.)
A special role is played by closed forms, since closed forms are always
Hamiltonian and closed forms that vanish on the kernel of $\,\omega$ are
always Poisson: these are in a sense the trivial examples. In other words,
the main task is to understand the extent to which general Hamiltonian
forms deviate from closed forms and general Poisson forms deviate from
closed forms that vanish on the kernel of $\,\omega$.

As a warm-up exercise, we shall settle the extreme cases of tensor degree $0$
and $n$. The case $\, r=n \,$ has already been analyzed in Ref.~\cite{FPR},
so we just quote the result.
\begin{prp} \label{prp:POISF0}~~
 A function $f$ on~$P$, regarded as a $0$-form, is always Hamiltonian and
 even Poisson. Moreover, its associated Hamiltonian $n$-multivector field
 $X$ is, in adapted local coordinates and modulo terms taking values in
 the kernel of~$\,\omega$, given by eq.~(\ref{eq:HAMNVF4}).
\end{prp}
The case $\, r=0 \,$ is equally easy.
\begin{prp} \label{prp:POISFN}~~
 An $n$-form $f$ on $P$ is Hamiltonian or Poisson if and only if it can be
 written as the sum of a constant multiple of $\,\theta$ with a closed form
 which is arbitrary if $f$ is Hamiltonian and vanishes on the kernel
 of~$\,\omega$ if $f$ is Poisson.
\end{prp}
Indeed, if $f$ is a Hamiltonian $n$-form, the multivector field $X$ that
appears in eq.~(\ref{eq:HAMMVFF}) will in fact be a function which has
to be locally Hamiltonian and hence, by Proposition~\ref{prp:HAMFUN},
constant. Thus $df$ must be proportional to $\,\omega$ and so $f$ must
be the sum of some constant multiple of $\,\theta$ and a closed form.

The intermediate cases ($0 < r < n$) are much more interesting. To handle
them, the first step is to identify the content of the kernel condition
(\ref{eq:KERN1}) in adapted local coordinates (for completeness, we also
include the two extreme cases):
\begin{prp} \label{prp:POISFLC}~~
 An $(n-r)$-form $f$ on~$P$, with $\, 0 \leqslant r \leqslant n$, vanishes
 on the kernel of~$\,\omega$ if and only if, in adapted local coordinates,
 it can be written in the form
 \begin{equation} \label{eq:POISF01}
  \begin{split}
   f~&=~\frac{1}{r!} \; f_{}^{\mu_1 \ldots\, \mu_r} \;
        d^{\,n} x_{\mu_1 \ldots\, \mu_r}^{} \; + \;
        \frac{1}{(r\!+\!1)!} \; f_i^{\mu_0 \ldots\, \mu_r} \;
        dq^i \,\smwedge\; d^{\,n} x_{\mu_0 \ldots\, \mu_r}^{}              \\
     &\quad + \;
        \frac{1}{r!} \; f_{}^{i,\mu_1 \ldots\, \mu_r} \;
        dp\>\!_i^\mu \,\smwedge\; d^{\,n} x_{\mu \mu_1 \ldots\, \mu_r}^{}  \\
     &\quad+ \;
        \frac{1}{(r\!+\!1)!} \; f_{}^{\prime \, \mu_0 \ldots\, \mu_r} \,
        \Bigl( dp \;\smwedge\; d^{\,n} x_{\mu_0 \ldots\, \mu_r}^{} \; - \;
               dq^i \,\smwedge\; dp\>\!_i^\mu \,\smwedge\;
               d^{\,n} x_{\mu_0 \ldots\, \mu_r \mu}^{} \Bigr)~,
  \end{split}
 \end{equation}
 where the second term in the last bracket is to be omitted if
 $\, r = n\!-\!1$ whereas only the first term remains if $\, r = n$.
\end{prp}
Note that for one-forms (just as for functions), the kernel
condition~(\ref{eq:KERN1}) is void, since $\,\omega$ is
non-degenerate. Also, it is in this case usually more convenient to
replace eq.~(\ref{eq:POISF01}) by the standard local coordinate
representation
\begin{equation} \label{eq:ONEFORM}
 f~=~f_\mu \, dx^\mu \, + \, f_i \, dq^i \, + \,
     f_\mu^i \, dp\>\!_i^\mu \, + \, f_0 \, dp~.
\end{equation}
\textbf{Proof.}~~Dualizing the statements of the proof of Proposition
\ref{prp:KERNOMEGA}, we see first of all that forms of degree $n\!-\!r$
vanishing on the kernel of $\omega$ must be $(n\!-\!r\!-\!2)$-horizontal
(since they vanish on $3$-vertical multivector fields) and that the only
term which is not $(n\!-\!r\!-\!1)$-horizontal is
\[
 dq^i \,\smwedge\; dp\>\!_k^\kappa \>\smwedge\;
 d^{\,n} x_{\mu_0 \ldots\, \mu_r \mu}^{}~.
\]
Thus we may write any such form as
\begin{eqnarray} \label{eq:POISF02}
 f \!\!
 &=&\!\! \frac{1}{r!} \; f_{}^{\mu_1 \ldots\, \mu_r} \;
         d^{\,n} x_{\mu_1 \ldots\, \mu_r}^{} \; + \;
         \frac{1}{(r\!+\!1)!} \; f_i^{\mu_0 \ldots\, \mu_r} \;
         dq^i \,\smwedge\; d^{\,n} x_{\mu_0 \ldots\, \mu_r}^{}     \nonumber \\
 & &\!   \mbox{} + \;
         \frac{1}{(r\!+\!1)!} \; f_\kappa^{k,\mu_0 \ldots\, \mu_r} \;
         dp\>\!_k^\kappa \>\smwedge\; d^{\,n} x_{\mu_0 \ldots\, \mu_r}^{} \;
         + \; \frac{1}{(r\!+\!1)!} \; f_{}^{\prime \, \mu_0 \ldots\, \mu_r} \;
         dp \;\smwedge\; d^{\,n} x_{\mu_0 \ldots\, \mu_r}^{} \qquad          \\
 & &\!   \mbox{} + \;
         \frac{1}{(r\!+\!1)!} \; f_{i,\kappa}^{k,\mu_0 \ldots\, \mu_r \mu} \;
         dq^i \,\smwedge\; dp\>\!_k^\kappa \>\smwedge\;
         d^{\,n} x_{\mu_0 \ldots\, \mu_r \mu}^{}                   \nonumber
\end{eqnarray}
and conclude from the requirement that $f$ should also vanish on
multivector fields $\xi$ of the type given in eqs~(\ref{eq:KERN5})
and~(\ref{eq:KERN6}) that the local coordinate represntation of a
general form of degree $n\!-\!r$ vanishing on the kernel of $\omega$
is the one given in eq.~(\ref{eq:POISF01}). Indeed, contracting
eq.~(\ref{eq:POISF02}) with the bivector
\[
 \frac{\partial}{\partial q^i} \;\smwedge\;
 \frac{\partial}{\partial p\>\!_k^\kappa} \; + \; \delta_i^k \;
 \frac{\partial}{\partial p} \;\smwedge\;
 \frac{\partial}{\partial x_{\phantom{i}}^\kappa}
\]
leads to the conclusion that the expression
\[
 \left( \delta_i^k \, \delta_\kappa^\mu \,
        f_{}^{\prime \, \mu_0 \ldots\, \mu_r} \; + \;
        f_{i,\kappa}^{k,\mu_0 \ldots\, \mu_r \mu} \right) \,
 d^{\,n} x_{\mu_0 \ldots\, \mu_r \mu}^{}
\]
must vanish, so $f$ takes the form
\begin{eqnarray} \label{eq:POISF03}
 f \!\!
 &=&\!\! \frac{1}{r!} \; f_{}^{\mu_1 \ldots\, \mu_r} \;
         d^{\,n} x_{\mu_1 \ldots\, \mu_r}^{} \; + \;
         \frac{1}{(r\!+\!1)!} \; f_i^{\mu_0 \ldots\, \mu_r} \;
         dq^i \,\smwedge\; d^{\,n} x_{\mu_0 \ldots\, \mu_r}^{}     \nonumber \\
 & &\!   \mbox{} + \;
         \frac{1}{(r\!+\!1)!} \; f_\kappa^{k,\mu_0 \ldots\, \mu_r} \;
         dp\>\!_k^\kappa \,\smwedge\; d^{\,n} x_{\mu_0 \ldots\, \mu_r}^{}    \\
 & &\!   \mbox{} + \;
         \frac{1}{(r\!+\!1)!} \; f_{}^{\prime \, \mu_0 \ldots\, \mu_r} \,
         \Bigl( dp \;\smwedge\; d^{\,n} x_{\mu_0 \ldots\, \mu_r}^{} \; - \;
                dq^i \,\smwedge\; dp\>\!_i^\mu \,\smwedge\;
                d^{\,n} x_{\mu_0 \ldots\, \mu_r \mu}^{} \Bigr)
                                                            \qquad \nonumber
\end{eqnarray}
Similarly, contracting eq.~(\ref{eq:POISF03}) with the bivector
\[
 \frac{\partial}{\partial p\>\!_i^\mu} \;\smwedge\;
 \frac{\partial}{\partial x_{\phantom{i}}^{\nu\vphantom{\mu}}} \; + \;
 \frac{\partial}{\partial p\>\!_i^{\nu\vphantom{\mu}}} \;\smwedge\;
 \frac{\partial}{\partial x_{\phantom{i}}^\mu}
\]
leads to the conclusion that the expression
\[
 f_\mu^{i,\mu_0 \ldots\, \mu_r} \, d^{\,n} x_{\mu_0 \ldots\, \mu_r \nu}^{}
 \; + \;
 f_\nu^{i,\mu_0 \ldots\, \mu_r} \, d^{\,n} x_{\mu_0 \ldots\, \mu_r \mu}^{}
\]
must vanish: setting $\, \mu = \nu$, it is easily seen that this forces the
coefficients $f_\mu^{i,\mu_0 \ldots\, \mu_r}$ to vanish when the indices
$\, \mu_0, \ldots , \mu_r \,$ are all different from $\mu$, and letting
$\, \mu \neq \nu$, we then conclude that
\[
 f_\mu^{i,\mu_0 \ldots\, \mu_{s-1} \mu \mu_{s+1} \ldots\, \mu_r}~
 =~f_\nu^{i,\mu_0 \ldots\, \mu_{s-1} \nu \mu_{s+1} \ldots\, \mu_r}
 \qquad \mbox{(no sum over $\mu$ or $\nu$)}~,
\]
so that we can write
\begin{equation} \label{eq:POISF04}
 f_\kappa^{k,\mu_0 \ldots\, \mu_r}~
 =~\sum_{s=0}^r \, (-1)^s \; \delta_\kappa^{\mu_s} \,
   f_{}^{k,\mu_0 \ldots\, \mu_{s-1} \mu_{s+1} \ldots\, \mu_r}
\end{equation}
Inserting this expression into eq.~(\ref{eq:POISF03}), we arrive at
eq.~(\ref{eq:POISF01}). \\
\PCPqed

\noindent
The proposition above can be used to prove the following interesting and
useful fact.
\begin{prp} \label{prp:POISFGL}~~
 An $(n-r)$-form $f$ on~$P$, with $\, 0 \leqslant r \leqslant n$, vanishes on
 the kernel of~$\,\omega$ if and only if there exists an $(r+1)$-multivector
 field $X$ on~$P$ such that
 \begin{equation} \label{eq:POISF05}
  f~=~i_X^{} \omega~.
 \end{equation}
 Then obviously,
 \begin{equation} \label{eq:POISF06}
  df~=~L_X^{} \omega~.
 \end{equation}
 In particular, $f$ is closed if and only if $X$ is locally Hamiltonian.
\end{prp}
\textbf{Proof.}~~The ``if'' part being obvious, observe that it suffices
to prove the ``only if'' part locally, in the domain of definition of
an arbitrary system of adapted local coordinates, by constructing the
coefficients of $X$ from those of $f$. (Indeed, since the relation
between $f$ and $X$ postulated in eq.~(\ref{eq:POISF05}) is purely
algebraic, i.e., it does not involve derivatives, we can construct
a global solution patching together local solutions with a partition
of unity.) But comparing eqs~(\ref{eq:HAMMVF1}), (\ref{eq:HAMMVF2})
and~(\ref{eq:POISF01}) shows that when $\, r < n$, this can be
achieved by setting
\phantom{this is just to adjust vertical spacing}
\begin{equation} \label{eq:POISF07}
 \begin{array}{rcl}
  X_{\phantom{i}}^{\mu_0 \ldots\, \mu_r} \!\!
  &=&\!\! (-1)^r \, f_{}^{\prime \, \mu_0 \ldots\, \mu_r}~, \\[2mm]
  X_{\phantom{i}}^{i,\mu_1 \ldots\, \mu_r} \!\!
  &=&\!\! (-1)^r \, f_{}^{i,\mu_1 \ldots\, \mu_r}~, \\[2mm]
  X_i^{\mu_0 \ldots\, \mu_r} \!\!
  &=&\!\! (-1)^{r+1} \, f_i^{\mu_0 \ldots\, \mu_r}~, \\[2mm]
  \tilde{X}_{\phantom{i}}^{\mu_1 \ldots\, \mu_r} \!\!
  &=&\!\! \mbox{} - \, f_{}^{\mu_1 \ldots\, \mu_r}~,
 \end{array}
\end{equation}
while for $\, r=n$, only the last equation is pertinent (for $\, r = n-1$,
the same conclusion can also be reached by comparing eqs~(\ref{eq:HAMNVF1}),
(\ref{eq:HAMNVF2}) and~(\ref{eq:ONEFORM})). \\
\PCPqed
\begin{cor}
 An $(n-r)$-form $f$ on~$P$, with $\, 0 \leqslant r \leqslant n$, is a
 Hamiltonian form if and only if\/ $df$ vanishes on the kernel of~$\,\omega$
 and is a Poisson form if and only if both\/ $df$ and $f$ vanish on the kernel
 of~$\,\omega$.
\end{cor}

With these preliminaries out of the way, we can proceed to the construction
of Poisson forms which are not closed. As we shall see, there are two such
constructions which, taken together, will be sufficient to handle the
general case.

The first construction is a generalization of the universal multimomentum map
of Ref.~\cite{FPR}, which to each exact Hamiltonian $r$-multivector field $F$
on~$P$ associates a Poisson $(n-r)$-form $J(F)$ on~$P$ defined by
eq.~(\ref{eq:UNMMM1}) below. What remained unnoticed in
Ref.~\cite{FPR} is that this construction works even when $X$ is only
locally Hamiltonian. In~fact, we have the following generalization of
Proposition~4.3 of Ref.~\cite{FPR}:
\begin{prp} \label{prp:UNMMM}~~
 For every locally Hamiltonian $r$-multivector field $F$ on~$P$, with
 $\, 0 \leqslant r \leqslant n$, the formula
 \begin{equation} \label{eq:UNMMM1}
  J(F)~=~(-1)^{r-1} \, i_F^{} \theta
 \end{equation}
 defines a Poisson $(n-r)$-form $J(F)$ on~$P$ whose associated Hamiltonian
 multivector field is $\, F + [\Sigma,F]$, that is, we have
 \begin{equation} \label{eq:UNMMM2}
  d \left( J(F) \right)\!~=~i_{F+[\Sigma,F]}^{\vphantom{\mu}} \omega~.
 \end{equation}
\end{prp}
\textbf{Proof.}~~Obviously, $J(F)$ vanishes on the kernel of $\,\omega$ since
this is contained in the kernel of $\,\theta$. Moreover, since $L_F^{} \omega$
is supposed to vanish, we can use eqs~(\ref{eq:LDFMVF}), (\ref{eq:MSYMF04})
and~(\ref{eq:LXiY-iYLX}) to compute
\begin{eqnarray*}
 d \left( J(F) \right)\!\!
 &=&\!\! (-1)^{r-1} \, d \left( i_F^{} \theta \right)\!~
  =~     (-1)^{r-1} \, L_F^{} \theta \, - \, i_F^{} \, d \>\! \theta    \\[1mm]
 &=&\!\! (-1)^r \, L_F^{} i_\Sigma^{} \omega \, + \, i_F^{} \omega      \\[1mm]
 &=&\!\! (-1)^r \, L_F^{} i_\Sigma^{} \omega \, + \,
         i_\Sigma^{} L_F^{} \omega \, + \, i_F^{} \omega                \\[1mm]
 &=&\!\! \mbox{} - \, i_{[F,\Sigma]}^{} \omega \, + \, i_F^{} \omega~.
\end{eqnarray*}
\PCPqed

The second construction uses differential forms on~$E$, pulled back
to differential forms on~$P$ via the target projection $\, \tau :
P \rightarrow E$. Characterizing which of these are Hamiltonian
forms and which are Poisson forms is a simple exercise.
\begin{prp}
 Let $f_0^{}$ be an $(n-r)$-form on~$E$, with $\, 0 < r < n$. Then
 \begin{itemize}
  \item $\tau^* f_0$ is a Hamiltonian form on~$P$ if and only if
        $df_0$ is $(n-r)$-horizontal.
  \item $\tau^* f_0$ is a Poisson form on~$P$ if and only if $f_0$
        is $(n-r-1)$-horizontal and $df_0$ is $(n-r)$-horizontal.
 \end{itemize}
\end{prp}
\textbf{Proof.}~~In adapted local coordinates $(x_{\phantom{i}}^\mu,q^i)$
for~$E$ and $(x_{\phantom{i}}^\mu,q^i,p\;\!_i^\mu,p\;\!)$ for~$P$, we can
write
\begin{equation} \label{eq:POISF08}
 f_0~=~\frac{1}{r!} \; f_0^{\mu_1 \ldots\, \mu_r} \;
       d^{\,n} x_{\mu_1 \ldots\, \mu_r}^{} \; + \;
       \frac{1}{(r\!+\!1)!} \; (f_0)_i^{\mu_0 \ldots\, \mu_r} \;
       dq^i \,\smwedge\; d^{\,n} x_{\mu_0 \ldots\, \mu_r}^{} \; + \;
       \ldots~,
\end{equation}
where the dots denote higher order terms containing at least two $dq$'s.
Now applying Proposition~\ref{prp:POISFLC} to $\tau^* f_0$, we see that
$\tau^* f_0$ will vanish on the kernel of $\,\omega$ if and only the
terms denoted by the dots all vanish, i.e., if $f_0$ can be written
in the form
\begin{equation} \label{eq:POISF09}
 f_0~=~\frac{1}{r!} \; f_0^{\mu_1 \ldots\, \mu_r} \;
       d^{\,n} x_{\mu_1 \ldots\, \mu_r}^{} \; + \;
       \frac{1}{(r\!+\!1)!} \; (f_0)_i^{\mu_0 \ldots\, \mu_r} \;
       dq^i \,\smwedge\; d^{\,n} x_{\mu_0 \ldots\, \mu_r}^{}~.
\end{equation}
But this is precisely the condition for the $(n-r)$-form $f_0$ to be
$(n-r-1)$-horizontal. (Note that this equivalence holds even when
$\, r = n-1$, provided we understand the condition of being
$0$-horizontal to be empty.) Similarly, since
Proposition~\ref{prp:POISFGL} implies that a form on~$P$ is
Hamiltonian if and only if its exterior derivative vanishes on the
kernel of $\,\omega$, the same argument applied to $\, d(\tau^* f_0) =
\tau^* df_0 \,$ shows that, irrespectively of whether $\tau^* f_0$
itself vanishes on the kernel of $\,\omega$ or not and hence whether
we use eq.~(\ref{eq:POISF08}) or eq.~(\ref{eq:POISF09}) as our
starting point, $\tau^* f_0$ will be Hamiltonian if and only if
\begin{equation} \label{eq:POISF10}
 \begin{split}
  df_0~&=~\frac{1}{(r\!-\!1)!} \;
          \frac{\partial f_0^{\mu_2 \ldots\, \mu_r \nu}}
               {\partial x_{\phantom{i}}^\nu} \;
          d^{\,n} x_{\mu_2 \ldots\, \mu_r}^{}                           \\[1mm]
       &\quad + \;
          \frac{1}{r!}
          \left( \frac{\partial f_0^{\mu_1 \ldots\, \mu_r}}
                      {\partial q^i} \, - \,
                 \frac{\partial (f_0)_i^{\mu_1 \ldots\, \mu_r \nu}}
                      {\partial x_{\phantom{i}}^\nu} \right)
          dq^i \,\smwedge\; d^{\,n} x_{\mu_1 \ldots\, \mu_r}^{}~.
 \end{split}
\end{equation}
But this is precisely the condition for the $(n-r+1)$-form $df_0$ to be
$(n-r)$-horizontal. Moreover, it is easy to write down an associated
Hamiltonian $r$-multivector field $X_0$:
\begin{equation} \label{eq:HAMMVF5}
 \begin{array}{rcl}
  X_0 \!\!
  &=&\!\! {\displaystyle
           \frac{(-1)^r}{r!}
           \left( \frac{\partial f_0^{\mu_1 \ldots\, \mu_r}}
                       {\partial q^i} \, - \,
                 \frac{\partial (f_0)_i^{\mu_1 \ldots\, \mu_r \nu}}
                      {\partial x_{\phantom{i}}^\nu} \right) \,
           \frac{\partial}{\partial p\>\!_i^{\mu_1}} \;\smwedge\;
           \frac{\partial}{\partial x_{\phantom{i}}^{\mu_2}}
           \;\smwedge \ldots \smwedge\;
           \frac{\partial}{\partial x_{\phantom{i}}^{\mu_r}}} \\[3ex]
  & &\!   {\displaystyle
           \mbox{} - \; \frac{1}{(r\!-\!1)!} \;
           \frac{\partial f_0^{\mu_2 \ldots\, \mu_r \nu}}
                {\partial x_{\phantom{i}}^{\nu\vphantom{\mu}}} \;
           \frac{\partial}{\partial p} \;\smwedge\;
           \frac{\partial}{\partial x_{\phantom{i}}^{\mu_2}}
           \;\smwedge \ldots \smwedge\;
           \frac{\partial}{\partial x_{\phantom{i}}^{\mu_r}}}~.
 \end{array}
\end{equation}
\PCPqed

\vspace{3mm}

\noindent
Note also that if $f_0$ is $(n-r-1)$-horizontal and thus has the form stated
in eq.~(\ref{eq:POISF09}), $df_0$ would contain just one additional higher
order term, namely
\[
 \frac{1}{(r\!+\!1)!} \,
 \frac{\partial (f_0)_j^{\mu_0 \ldots\, \mu_r}}{\partial q^i}~
 dq^i \,\smwedge\, dq^j \,\smwedge\; d^{\,n} x_{\mu_0 \ldots\, \mu_r}^{}~.
\]

%\pagebreak

\noindent
Its absence means that
\vspace*{5mm}
\[
 \frac{\partial (f_0)_j^{\mu_0 \ldots\, \mu_r}}{\partial q^i}~
 =~\frac{\partial (f_0)_i^{\mu_0 \ldots\, \mu_r}}{\partial q^j}~,
\vspace*{2mm}
\]
so there exist local functions $f_0^{\mu_0 \ldots\, \mu_r}$ on~$E$ such that
\[
 (f_0)_i^{\mu_0 \ldots\, \mu_r}~
 =~\frac{\partial f_0^{\mu_0 \ldots\, \mu_r}}{\partial q^i}~.
\]
This implies that $f_0$ can be written as the sum
\begin{equation} \label{eq:POISF11}
 f_0~=~f_h + f_c
\end{equation}
of a horizontal form $f_h$ and a closed form $f_c$, defined by setting
\begin{equation} \label{eq:POISF12}
 f_h~=~\frac{1}{r!} \left( f_0^{\mu_1 \ldots\, \mu_r} \, - \,
                           \frac{\partial f_0^{\mu_1 \ldots\, \mu_r \nu}}
                                {\partial x_{\phantom{i}}^\nu} \right) \,
       d^{\,n} x_{\mu_1 \ldots\, \mu_r}^{}~,
\end{equation}
and
\begin{equation} \label{eq:POISF13}
 f_c~=~\frac{1}{r!} \;
       \frac{\partial f_0^{\mu_1 \ldots\, \mu_r \nu}}
            {\partial x_{\phantom{i}}^\nu} \;
       d^{\,n} x_{\mu_1 \ldots\, \mu_r}^{} \; + \;
       \frac{1}{(r\!+\!1)!} \;
       \frac{\partial f_0^{\mu_0 \ldots\, \mu_r}}{\partial q^i} \;
       dq^i \,\smwedge\; d^{\,n} x_{\mu_0 \ldots\, \mu_r}^{}~.
\end{equation}
The same kind of local decomposition into the sum of a horizontal form
and a closed form can also be derived if $f_0$ is arbitrary and thus
has the form stated in eq.~(\ref{eq:POISF08}); this case can be
handled by decreasing induction on the number of $dq$'s that appear
in the higher order terms denoted by the dots in eq.~(\ref{eq:POISF08}).
We shall refrain from working this out in detail, since unfortunately
the decomposition~(\ref{eq:POISF11}) depends on the system of adapted
local coordinates used in its construction: under coordinate
transformations, the terms $f_h$ and $f_c$ mix. Therefore,
this decomposition has no coordinate independent meaning
and is in general valid only locally.

Finally, we note that in the above discussion, we have deliberately excluded
the extreme cases $\, r=0$ ($n$-forms) and $\, r=n$ (functions). For $n$-forms,
the equivalences stated above would be incorrect since if $f_0$ has tensor
degree $n$ and hence $X_0$ has tensor degree $0$, $i_{X_0}^{} \omega$ would
by Proposition~\ref{prp:HAMFUN} be a constant multiple of $\omega$ whereas
$d(\tau^* f_0)$ would be reduced to a linear combination of terms of the
form $\, dq^i \,\smwedge\; d^{\,n} x$, implying that $\tau^* f_0$ can
only be Hamiltonian if it is closed. For functions, the construction
is uninteresting since according to Proposition~\ref{prp:POISF0}, all
functions on~$P$ are Poisson, and not just the ones lifted from~$E$.

Now we are ready to state our main decomposition theorem. (In what follows,
we shall simply write $f_0$ instead of $\tau^* f_0$ when there is no danger
of confusion, the main exception being the proof of Theorem~\ref{thm:POISFCD}
below).

%\pagebreak

\begin{thm} \label{thm:POISFCD}~~
 Any Hamiltonian $(n-r)$-form and, in particular, any Poisson $(n-r)$-form~$f$
 on~$P$, with $\, 0 < r < n$, admits a unique decomposition
 \begin{equation} \label{eq:POISFCD1}
  f~=~f_0^{} + f_+^{} + f_c^{}
  \qquad \mbox{with} \qquad
  f_+^{}~=~\sum_{s=1}^r f_s^{}~,
 \end{equation}
 where
 \begin{enumerate}
  \item $f_0^{}$ is (the pull-back to~$P$ of) an $(n-r)$-form on~$E$ whose
        exterior derivative is $(n-r)$-horizontal and which is otherwise
        arbitrary if $f$ is Hamiltonian whereas it is restricted to be
        $(n-r-1)$-horizontal iff $f$ is Poisson.
  \item $f_+^{}$ is of the form
        \begin{equation}
         f_+^{}~=~J(F)~=~(-1)^{r-1} \, i_F^{} \theta
         \qquad \mbox{with} \qquad
         F~=~\big( 1 + L_\Sigma^{} \big)^{-1} X_+^{}~,
        \end{equation}
        and correspondingly, for $\, s = 1,\ldots,r$, $f_s^{}$ is of the form
        \begin{equation}
         f_s^{}~=~\frac{(-1)^{r-1}}{s} \; i_{X_{s-1}^{}}^{} \theta~,
        \end{equation}
        where $X$ is any fiberwise polynomial Hamiltonian $r$-multivector field
        associated with~$f$, decomposed according to eq.~(\ref{eq:HAMMVFSCD1}).
  \item $f_c^{}$ is a closed $(n-r)$-form on~$P$ which vanishes on the zero
        section of~$P$ (as a vector bundle over~$E$) and which is otherwise
        arbitrary if $f$ is Hamiltonian whereas it is restricted to vanish
        on the kernel of~$\,\omega$ iff $f$ is Poisson.
 \end{enumerate}
 We shall refer to eq.~(\ref{eq:POISFCD1}) and to eq.~(\ref{eq:POISFCD2})
 below as the \textbf{canonical decomposition} of Hamiltonian forms or
 Poisson forms on~$P$.
\end{thm}
\textbf{Proof.}~~Let $f$ be a Poisson $(n-r)$-form and $X$ be a
Hamiltonian $r$-multivector field associated with $f$. As already
mentioned in the introduction, we may without loss of generality
assume $X$ to be fiberwise polynomial and decompose it into
homogeneous components with respect to scaling degree, according to
eq.~(\ref{eq:HAMMVFSCD1}):
\[
 X~=~X_-^{} + X_+^{} + \, \xi \qquad \mbox{with} \qquad
 X_+^{}~=~\sum_{s=1}^r X_{s-1}^{}~.
\]
Then defining $F$ as in the theorem, or equivalently, by
\[
 F~=~\sum_{s=1}^r F_{s-1}^{}
 \qquad \mbox{with} \qquad
 F_{s-1}^{}~=~\frac{1}{s} \, X_{s-1}^{}~,
\]
we obtain
\[
 F + [\Sigma,F]~=~X_+^{},
\]
and hence according to eq.~(\ref{eq:UNMMM2}), the exterior derivative
of the difference $\, f - J(F) \,$ is given by
\[
 d \big( f - J(F) \big)~=~df \, - \, d \big( J(F) \big)~
 =~i_X^{} \omega \, - \, i_{X_+^{}}^{} \omega~=~i_{X_-^{}}^{} \omega~.
\]
Applying the equivalence stated in eq.~(\ref{eq:HOMHAM1}), we see that
since $X_-^{}$ has scaling degree~$-1$, $i_{X_-^{}}^{} \omega$ must have
scaling degree $0$ and hence, according to Proposition~\ref{prp:Pullback},
is the pull-back to~$P$ of some $(n-r)$-form $f_0^{\,\prime}$ on~$E$:
\[
 d \big( f - J(F) \big)~=~i_{X_-^{}}^{} \omega~=~\tau^* f_0^{\,\prime}~.
\]
Next, we define $f_0^{}$ to be the restriction of $\, f - J(F) \,$
to the zero section of~$P$, or more precisely, its pull-back to~$E$
with the zero section $\, s_0^{} : E \rightarrow P$,
\begin{equation}
 f_0^{}~=~s_0^* \big( f - J(F) \big)~,
\end{equation}
and set
\begin{equation}
 f_c^{}~=~f - \tau^* f_0^{} - J(F)~.
\end{equation}
Then
\begin{eqnarray*}
 d f_c^{} \!\!
 &=&\!\! d \big( f - J(F) \big) \, - \,
         d \Big( \tau^* s_0^* \big( f - J(F) \big) \Big)                \\
 &=&\!\! d \big( f - J(F) \big) \, - \,
         \tau^* s_0^* \; d \big( f - J(F) \big)                         \\[1mm]
 &=&\!\! \tau^* f_0^{\,\prime} \, - \,
         \tau^* s_0^* \, \tau^* f_0^{\,\prime}                          \\[1mm]
 &=&\!\! 0~,
\end{eqnarray*}
and
\[
 s_0^* f_c^{}~
 =~s_0^* \big( f - J(F) \big) \, - \, s_0^* \, \tau^* f_0^{}~
 =~f_0^{} \, - \, s_0^* \, \tau^* f_0^{}~=~0~,
\vspace{1mm}
\]
showing that indeed, $f_c$ is closed and vanishes on the zero section
of~$P$. \\
\PCPqed

\noindent
\textbf{Proof of Theorem~\ref{thm:FIBPOL} and Theorem~\ref{thm:DECSCD},
item 2.}~~These statements are immediate consequences of
Theorem~\ref{thm:POISFCD}.\\
\PCPqed

\noindent
\textbf{Remark.}~~It should be noted that despite appearances, the
decompositions~(\ref{eq:POISFCD1}) of Theorem~\ref{thm:POISFCD}
and~(\ref{eq:HAMFORSCD1}) of Theorem~\ref{thm:DECSCD} are not
necessarily identical: for $s=1,\ldots,r$, the $f_s$ of
eq.~(\ref{eq:POISFCD1}) and the $f_s$ of eq.~(\ref{eq:HAMFORSCD1}) may
differ by homogeneous closed $(n-r)$-forms of scaling degree $s$. But
the decomposition~(\ref{eq:POISFCD1}) of Theorem~\ref{thm:POISFCD}
seems to be the more natural one.

%\pagebreak

Theorem~\ref{thm:POISFCD} implies that Poisson forms have a rather
intricate local coordinate re\-pre\-sen\-ta\-tion, involving two
locally Hamiltonian multivector fields. Indeed, if we take $f$ to be a
general Poisson $(n-r)$-form on~$P$, with $\, 0 < r < n$, we can apply
Propositions~\ref{prp:POISFGL} and~\ref{prp:UNMMM} to rewrite
eq.~(\ref{eq:POISFCD1}) in the form
\begin{equation} \label{eq:POISFCD2}
 f~=~f_0^{} \, + \, (-1)^{r-1} i_F^{} \theta \, + \, (-1)^r i_{F_c}^{} \omega~,
\end{equation}
where $f_0^{}$ is as before while $F$ and $F_c$ are two locally
Hamiltonian multivector fields on~$P$ of tensor degree $r$ and $r+1$,
respectively, satisfying $\, F_-^{} = 0 \,$ and $\, (F_c^{})_-^{} =
0$.\footnote{The condition $\, (F_c^{})_-^{} = 0 \,$ will guarantee
that $i_{F_c}^{} \omega$ vanishes on the zero section of~$P$.}
\linebreak In terms of the standard local coordinate
representations~(\ref{eq:POISF01}) for~$f$, (\ref{eq:POISF09}) for
$f_0^{}$ and~(\ref{eq:HAMMVF1}) for $F$ and for $F_c$, we obtain,
according to eqs~(\ref{eq:HAMMVF2}) and~(\ref{eq:HAMMVF3}),
\vspace{-2mm}
\begin{equation} \label{eq:POISF14}
 \begin{array}{rcl}
  f_{\phantom{i}}^{\mu_1 \ldots\, \mu_r} \!\!
  &=&\!\! {\displaystyle
           (-1)^{r-1} \, p \, F_{\phantom{i}}^{\mu_1 \ldots\, \mu_r} \, + \,
           \sum_{s=1}^r \, (-1)^{r-s} \, p\>\!_i^{\mu_s} \,
           F_{\phantom{i}}^{i,\mu_1 \ldots\, \mu_{s-1}
                              \mu_{s+1} \ldots\, \mu_r}} \\[3ex]
  & &\!\! \mbox{} + \, f_0^{\mu_1 \ldots\, \mu_r} \, + \,
          (-1)^{r-1} (\tilde{F}_c^{})_{\phantom{i}}^{\mu_1 \ldots\, \mu_r}~,
 \vspace{-1mm}
 \end{array}
\end{equation}
\begin{equation} \label{eq:POISF15}
  f_i^{\mu_0 \ldots\, \mu_r}~
  =~{\displaystyle
     \mbox{} - \, \sum_{s=0}^r \, (-1)^s \, p\>\!_i^{\mu_s} \,
     F_{\phantom{i}}^{\mu_0 \ldots\, \mu_{s-1}
                      \mu_{s+1} \ldots\, \mu_r}} \, + \,
     (f_0^{})_i^{\mu_0 \ldots\, \mu_r} \, - \,
     (F_c^{})_i^{\mu_0 \ldots\, \mu_r}~,
\vspace{-1mm}
\end{equation}
\begin{equation} \label{eq:POISF16}
 f_{\phantom{i}}^{i,\mu_1 \ldots\, \mu_r}~
 =~(F_c)_{\phantom{i}}^{i,\mu_1 \ldots\, \mu_r}~,
\vspace{1mm}
\end{equation}
\begin{equation} \label{eq:POISF17}
  f_{\phantom{i}}^{\prime \, \mu_0 \ldots\, \mu_r}~
  =~(F_c)_{\phantom{i}}^{\mu_0 \ldots\, \mu_r}~,
\vspace{2mm}
\end{equation}
where the coefficients of $F$ and of $F_c$ are subject to the
constraints listed in Theorem~\ref{thm:HAMMVF}; in particular,
the coefficients $(F_c^{})_i^{\mu_0 \ldots\, \mu_r}$ and
$(\tilde{F}_c^{})_{\phantom{i}}^{\mu_1 \ldots\, \mu_r}$
can be completely expressed in terms of the coefficients
$(F_c)_{\phantom{i}}^{\mu_0 \ldots\, \mu_r}$ and
$(F_c)_{\phantom{i}}^{i,\mu_1 \ldots\, \mu_r}$, according to
eqs~(\ref{eq:LHAMMVF03}) and~(\ref{eq:LHAMMVF04}) (with $r$ replaced
by $r+1$, $X$ replaced by $F_c^{}$ and $X_-^{}$ replaced by $0$).
In particular, we see that the coefficients $f_{\phantom{i}}^{\mu_1
\ldots\, \mu_r}$ are ``antisymmetric polynomials in the multimomentum
variables'' of degree~$r$. More explicitly, we can rewrite
eq.~(\ref{eq:POISF14}) in the form
\begin{equation} \label{eq:POISF18}
 f_{\phantom{i}}^{\mu_1 \ldots\, \mu_r}~
 =~(-1)^{r-1} \, p \, F_{\phantom{i}}^{\mu_1 \ldots\, \mu_r} \, + \,
   \sum_{s=1}^r \, f_s^{\mu_1 \ldots\, \mu_r} \, + \,
   f_0^{\mu_1 \ldots\, \mu_r} \, + \,
   (-1)^{r-1} (\tilde{F}_c^{})_{\phantom{i}}^{\mu_1 \ldots\, \mu_r}~,
\end{equation}
where inserting the expansion~(\ref{eq:LHAMMVF02}) (with $X$ replaced
by~$F$, $X_{s-1}$ replaced by $F_{s-1}$ and $Y_{s-1}$ replaced by
$\, G_{s-1} = \frac{1}{s} \, g_s$) gives, after a short calculation,
\begin{equation}
 f_s^{\mu_1 \ldots\, \mu_r}~
 =~(-1)^{r-1} \; \frac{1}{s!} \frac{1}{(r\!-\!s)!}
   \sum_{\pi \ssmin S_r} (-1)^\pi \,
   p\>\!_{i_1}^{\mu_{\pi(1)}} \!\ldots\, p\>\!_{i_s}^{\mu_{\pi(s)}} \,
   g_s^{i_1 \ldots\, i_s,\mu_{\pi(s+1)} \ldots\, \mu_{\pi(r)}}~.
\end{equation}

%\pagebreak

It is an instructive exercise to spell this out more explicitly for the case
of Poisson forms $f$ of degree $n-1$ ($r=1$), whose standard local coordinate
representation~(\ref{eq:POISF01}) reads
\begin{equation} \label{eq:POISF19}
 \begin{array}{rcl}
  f\!\!
  &=&\!\! f_{\phantom{i}}^\mu \; d^{\,n} x_\mu \, + \,
          \frac{1}{2} \, f_i^{\mu\nu} \;
          dq^i \,\smwedge\; d^{\,n} x_{\mu\nu}^{} \, + \,
          f_{\phantom{i}}^{i,\mu} \;
          dp\>\!_i^\mu \,\smwedge\; d^{\,n} x_\mu^{} \\[1ex]
  & &\!   \mbox{} + \, \frac{1}{2} \,
          f_{\phantom{i}}^{\prime \, \kappa\lambda\vphantom{\mu}} \,
          \Bigl( dp \;\smwedge\; d^{\,n} x_{\kappa\lambda}^{} \; - \;
                 dq^i \,\smwedge\; dp\>\!_i^\mu \,\smwedge\;
                 d^{\,n} x_{\kappa\lambda\mu}^{} \Bigr)~,
 \end{array}
\end{equation}
with coefficient functions $f_{\phantom{i}}^\mu$, $f_i^{\mu\nu}$,
$f_{\phantom{i}}^{i,\mu}$ and $f_{\phantom{i}}^{\prime \, \kappa
\lambda\phantom{\mu}}$ which, according to
eqs~(\ref{eq:HAMVF1})-(\ref{eq:LHAMVF04}),
(\ref{eq:HAMBVF1})-(\ref{eq:LHAMBVF04})
and~(\ref{eq:POISF14})-(\ref{eq:POISF17}), are given by
\begin{equation} \label{eq:POISF20}
 f_{\phantom{i}}^\mu~
 =~p \, F_{\phantom{i}}^\mu \, + \, p\>\!_i^\mu F_{\phantom{i}}^i \, + \,
   f_0^\mu \, + \, (\tilde{F}_c^{})_{\phantom{i}}^\mu~,
\vspace*{-1mm}
\end{equation}
\begin{equation} \label{eq:POISF21}
 f_i^{\mu\nu}~
 =~\mbox{} - \left( p\>\!_i^\mu F_{\phantom{i}}^{\nu\vphantom{\mu}} \, - \,
                    p\>\!_i^\nu F_{\phantom{i}}^\mu \right) + \,
   (f_0^{})_i^{\mu\nu} \, - \, (F_c^{})_i^{\mu\nu}~,
\end{equation}
\begin{equation} \label{eq:POISF22}
 f_{\phantom{i}}^{i,\mu}~=~(F_c)_{\phantom{i}}^{i,\mu}~,
\end{equation}
\begin{equation} \label{eq:POISF23}
 f_{\phantom{i}}^{\prime \, \mu\nu}~
 =~\!\! (F_c)_{\phantom{i}}^{\mu\nu}~,
\end{equation}
with
\begin{equation} \label{eq:HAMBVF2}
 (F_c)_{}^{i,\mu}~=~p\>\!_j^\mu \, (G_c)_{}^{ij}~,
\end{equation}
where the coefficient functions $F_{\vphantom{i}}^\mu$,
$F_{\vphantom{i}}^{i\vphantom{\mu}}$, $f_0^\mu$, $(f_0^{})_i^{\mu\nu}$,
$(F_c^{})_{\phantom{i}}^{\mu\nu}$ and $(G_c^{})_{}^{ij}$ all depend only on
the local coordinates~$x^\rho$ for~$M$ and on the local fiber coordinates
$q^r$ for~$E$ (the $F_{}^\mu$ and $(F_c^{})_{\phantom{i}}^{\mu\nu}$ being
independent of the  latter as soon as $\, N>1$), whereas the coefficients
$(F_c^{})_i^{\mu\nu}$ and $(\tilde{F}_c^{})_{\phantom{i}}^\mu$
can be completely expressed in terms of the
coefficients~$(F_c^{})_{\phantom{i}}^{\mu\nu}$ and~$(G_c^{})_{}^{ij}$,
according to eqs~(\ref{eq:LHAMBVF03}) and~(\ref{eq:LHAMBVF04}) (with $X$
replaced by $F_c^{}$, $Y$ replaced by $G_c^{}$ and $X_-^{}$ replaced by $0$).
Obviously, little structural insight can be gained from such an explicit
representation: the canonical decomposition~(\ref{eq:POISFCD1})
or~(\ref{eq:POISFCD2}) as such is much more instructive.

Finally, we want to clarify the relation between Poisson forms and
Hamiltonian multivector fields in terms of their standard local
coordinate representations.
\begin{thm} \label{thm:HAMMVFFOR}~~
 Let $f$ be a Poisson $(n-r)$-form and $X$ be a Hamiltonian $r$-multivector
 field on~$P$ associated with $f$. Assume that, in adapted local coordinates,
 $f$ and~$X$ are given by eqs~(\ref{eq:POISF01}) and~(\ref{eq:HAMMVF1}),
 respectively. Then
 \vspace{3mm}
 \begin{equation} \label{eq:HAMMVFFOR1}
  X_{\phantom{i}}^{\mu_1 \ldots\, \mu_r}~
  =~(-1)^{r-1}
    \left( \frac{\partial f_{\phantom{i}}^{\mu_1 \ldots\, \mu_r}}
                {\partial p} \, - \,
           \frac{\partial f_{\phantom{i}}^{\prime \, \mu_1 \ldots\, \mu_r \nu}}
                {\partial x_{\phantom{i}}^\nu} \right)\!~,
 \end{equation}
 \begin{equation} \label{eq:HAMMVFFOR2}
  X_{\phantom{i}}^{i,\mu_2 \ldots\, \mu_r}~
  =~\frac{1}{n-r+1} \,
    \frac{\partial f_{\phantom{i}}^{\mu_2 \ldots\, \mu_r \mu}}
                     {\partial p\>\!_i^\mu}~,
 \vspace{1mm}
 \end{equation}
 \begin{equation} \label{eq:HAMMVFFOR3}
  X_i^{\mu_1 \ldots\, \mu_r}~
  =~(-1)^r \,
    \left( \frac{\partial f_{\phantom{i}}^{\mu_1 \ldots\, \mu_r}}
                {\partial q^i} \, - \,
           \frac{\partial f_i^{\mu_1 \ldots\, \mu_r \nu}}
                {\partial x_{\phantom{i}}^\nu} \right)\!~,
 \vspace{1mm}
 \end{equation}
 \begin{equation} \label{eq:HAMMVFFOR4}
  \tilde{X}_{\phantom{i}}^{\mu_2 \ldots\, \mu_r}~
  = \; \mbox{} - \, \frac{\partial f_{\phantom{i}}^{\mu_2 \ldots\, \mu_r \nu}}
                         {\partial x_{\phantom{i}}^\nu}~,
 \end{equation}
 that is, locally and modulo terms taking values in the kernel of $\,\omega$,
 $X$ is given by
 \begin{equation} \label{eq:HAMMVFFOR5}
  \begin{array}{rcl}
   X &=& {\displaystyle - \;
          \frac{1}{(r\!-\!1)!}
          \left( \frac{\partial f_{\phantom{i}}^{\mu_2 \ldots\, \mu_r \mu}}
                      {\partial x_{\phantom{i}}^\mu} \,
                 \frac{\partial}{\partial p} \; - \;
                 \frac{1}{r} \,
                 \frac{\partial f_{\phantom{i}}^{\mu_2 \ldots\, \mu_r \mu}}
                      {\partial p}
                 \frac{\partial}{\partial x_{\phantom{i}}^\mu} \right.} \\[5mm]
     & & {\displaystyle \hspace{3cm} \left. + \;
                 \frac{1}{r} \,
                 \frac{\partial f_{\phantom{i}}^
                                  {\prime \, \mu_2 \ldots\, \mu_r \mu \nu}}
                      {\partial x_{\phantom{i}}^\nu} \,
                 \frac{\partial}{\partial x_{\phantom{i}}^\mu} \right)
          \,\smwedge\;
          \frac{\partial}{\partial x_{\phantom{i}}^{\mu_2}}
          \;\smwedge \ldots \smwedge\;
          \frac{\partial}{\partial x_{\phantom{i}}^{\mu_r}}}            \\[7mm]
     & & {\displaystyle + \;
          \frac{1}{(r\!-\!1)!}
          \left( \frac{1}{n-r+1} \,
                 \frac{\partial f_{\phantom{i}}^{\mu_2 \ldots\, \mu_r \mu}}
                      {\partial p\>\!_i^\mu} \,
                 \frac{\partial}{\partial q^i} \; - \;
                 \frac{1}{r} \,
                 \frac{\partial f_{\phantom{i}}^{\mu_2 \ldots\, \mu_r \mu}}
                      {\partial q^i} \,
                 \frac{\partial}{\partial p\>\!_i^\mu} \right.}         \\[5mm]
     & & {\displaystyle \hspace{3cm} \left. + \;
                 \frac{1}{r} \,
                 \frac{\partial f_i^{\mu_2 \ldots\, \mu_r \mu \nu}}
                      {\partial x_{\phantom{i}}^\nu} \,
                 \frac{\partial}{\partial p\>\!_i^\mu} \right)
          \,\smwedge\;
          \frac{\partial}{\partial x_{\phantom{i}}^{\mu_2}}
          \;\smwedge \ldots \smwedge\;
          \frac{\partial}{\partial x_{\phantom{i}}^{\mu_r}}}~.
  \end{array}
 \end{equation}
 If, in the canonical decomposition  (\ref{eq:POISFCD1}) or~
 (\ref{eq:POISFCD2}) of~$f$, the closed term $\, f_c^{} = (-1)^r
 \, i_{F_c}^{} \omega \,$ is absent, then  $\, f_{}^{\prime \,
 \mu_0 \ldots\, \mu_r} = 0$. If $f$ is horizontal with respect to
 the projection onto~$M$, then $\, f_i^{\mu_0 \ldots\, \mu_r} = 0$.
 In these cases, the above formulas simplify accordingly.
\end{thm}
\textbf{Proof.}~~There are several methods for proving this, with
certain overlaps. Let us begin with the ``trivial'' case of closed
forms $f$, for which we must have $\, X = 0$. Assuming $f$ to be
of the form $\, f_c^{} = (-1)^r \, i_{F_c}^{} \omega \,$ and using
eqs~(\ref{eq:POISF14})-(\ref{eq:POISF17}) to rewrite the expressions
on the rhs of the above equations in terms of the components of
$F_c^{}$, we must show that
\[
 \frac{\partial (\tilde{F}_c^{})_{\phantom{i}}^{\mu_1 \ldots\, \mu_r}}
      {\partial p} \, + \,
 (-1)^r \, \frac{\partial (F_c)_{\phantom{i}}^{\mu_1 \ldots\, \mu_r \nu}}
                {\partial x^\nu}~=~0~,
\]
\[
 \frac{\partial (\tilde{F}_c^{})_{\phantom{i}}^{\mu_2 \ldots\, \mu_r \mu}}
      {\partial p\>\!_i^\mu}~=~0~,
\]
\[
 \frac{\partial (\tilde{F}_c^{})_{\phantom{i}}^{\mu_1 \ldots\, \mu_r}}
      {\partial q^i} \, - \,
 (-1)^r \, \frac{\partial (F_c^{})_i^{\mu_1 \ldots\, \mu_r \nu}}
                {\partial x^\nu}~=~0~,
\]
\[
 \frac{\partial (\tilde{F}_c^{})_{\phantom{i}}^{\mu_2 \ldots\, \mu_r \nu}}
      {\partial x^\nu}~=~0~.
\vspace{2mm}
\]
But this follows directly from the analogues of eqs~(\ref{eq:LHAMMVF04}),
(\ref{eq:LHAMMVF13}), (\ref{eq:LHAMMVF12}) and~(\ref{eq:LHAMMVF11}),
respectively, which hold since $F_c^{}$ is locally Hamiltonian. To handle
the remaining cases where $f$ is of the form $\, f = f_0^{} \, + \,
(-1)^{r-1} \, i_F^{} \theta \,$, it is easier to proceed by direct
inspection of eq.~(\ref{eq:HAMMVFF}). Indeed, we may for a general
Poisson form $f$ apply the exterior derivative to eq.~(\ref{eq:POISF01})
and compare the result with eq.~(\ref{eq:HAMMVF2}). In this way,
eqs~(\ref{eq:HAMMVFFOR4}), (\ref{eq:HAMMVFFOR3})
and~(\ref{eq:HAMMVFFOR1}) can be obtained directly
by equating the coefficients of $\, d^{\,n} x_{\mu_2 \ldots\, \mu_r}$,
of $\, dq^i \,\smwedge\; d^{\,n} x_{\mu_1 \ldots\, \mu_r} \,$ and of
$\, dp \;\smwedge\; d^{\,n} x_{\mu_1 \ldots\, \mu_r}$, respectively.
The only case which requires an additional argument is
eq.~(\ref{eq:HAMMVFFOR2}), since collecting terms proportional to
$\, dp\;\!_i^\mu \,\smwedge\; d^{\,n} x_{\mu_1 \ldots\, \mu_r}$ leads to
\begin{eqnarray} \label{eq:HAMMVFFOR6}
\lefteqn{\frac{(-1)^{r-1}}{(r\!-\!1)!} \, X_{}^{i,\mu_2 \ldots\, \mu_r} \;
         dp\>\!_i^\mu \,\smwedge\; d^{\,n} x_{\mu \mu_2 \ldots\, \mu_r}}
                                                              \nonumber \\[1mm]
 &=&\!\! \frac{1}{r!} \,
         \frac{\partial f^{\mu_1 \ldots\, \mu_r}}{\partial p\>\!_i^\mu}~
         dp\>\!_i^\mu \,\smwedge\; d^{\,n} x_{\mu_1 \ldots\, \mu_r}     \\
 & &\!\! \mbox{} - \; \frac{1}{(r\!-\!1)!} \,
         \frac{\partial f_{}^{i,\mu_2 \ldots\, \mu_r \nu}}{\partial x^\nu}~
         dp\>\!_i^\mu \,\smwedge\; d^{\,n} x_{\mu \mu_2 \ldots\, \mu_r}^{} \;
         - \; \frac{(-1)^r}{r!} \,
         \frac{\partial f_{}^{i,\mu_1 \ldots\, \mu_r}}{\partial x^\mu}~
         dp\>\!_i^\mu \,\smwedge\; d^{\,n} x_{\mu_1 \ldots\, \mu_r}^{}~.
                                                              \nonumber
\end{eqnarray}
But when $f$ is of the form $\, f = f_0^{} \, + \, (-1)^{r-1} \, i_F^{}
\theta \,$, eq.~(\ref{eq:POISF16}) implies that the last two terms on the
rhs of eq.~(\ref{eq:HAMMVFFOR6}) vanish. Moreover, since $F$ is Hamiltonian,
we know from Theorem~\ref{thm:HAMMVF} that the $F_{}^{\mu_1 \ldots\, \mu_r}$
depend on the $p\>\!_i^\mu$ only if $\, \mu \smin \{\mu_1,\ldots\,,\mu_r\}$,
and hence according to eq.~(\ref{eq:POISF14}), the same is true for the
$f_{}^{\mu_1 \ldots\, \mu_r}$. This reduces the first term on the rhs of
eq.~(\ref{eq:HAMMVFFOR6}) to an expression which, when compared with the
lhs, leads to the conclusion that for any choice of mutually different
indices $\mu$ and $\, \mu_2,\ldots,\mu_r$, we have
\begin{equation} \label{eq:HAMMVFFOR7}
 X_{\phantom{i}}^{i,\mu_2 \ldots\, \mu_r}~
 =~\frac{\partial f_{\phantom{i}}^{\mu_2 \ldots\, \mu_r \mu}}
                    {\partial p\>\!_i^\mu} \qquad
 \mbox{if $\, \mu \nsmin \{\mu_2,\ldots,\mu_r\} \,$ (no sum over $\mu$)}~.
\end{equation}
Summing over $\mu$ gives eq.~(\ref{eq:HAMMVFFOR2}). \\
\PCPqed

%%%%%%%%%%%%%%%%%%%%%%%%%%%%%%%%%%%%%%%%%%%%%%%%%%%%%%%%%%%%%%%%%%%%%%%%%%%%%%%
\section{Poisson brackets}

In the characterization of locally Hamiltonian multivector fields and of
Poisson forms derived in the previous two sections, the decomposition into
homogeneous terms with respect to scaling degree plays a central role.
It is therefore natural to ask how this decomposition complies with the
Schouten bracket of Hamiltonian multivector fields and with the Poisson
bracket of Poisson forms. To this end, let us first recall the definition
of the Poisson bracket between Poisson forms given in \cite{FR1} for
$(n-1)$-forms and in \cite{FPR} for forms of arbitrary degree.
\begin{dfn}~~
 Let $f$ and $g$ be Poisson forms of tensor degree $n-r$ and $n-s$ on~$P$,
 respectively. Their Poisson bracket is the Poisson form of tensor degree
 $n-r-s+1$ on~$P$ defined by
 \begin{equation} \label{eq:POISBR1}
  \{f,g\}~=~(-1)^{r(s-1)} \, i_Y^{} i_X^{} \omega \, + \, d \,
            \Bigl( \! (-1)^{(r-1)(s-1)} \, i_Y^{} f \, - \, i_X^{} g \, - \,
                      (-1)^{(r-1)s} \, i_Y^{} i_X^{} \theta \Bigr)~,
 \end{equation}
 where $X$ and $Y$ are Hamiltonian multivector fields associated with $f$ and
 $g$, respectively.
\end{dfn}
We find the following properties of the two mentioned bracket operations
with respect to scaling degree.
\begin{prp}~~
 Let $X$ and $Y$ be homogeneous multivector fields on~$P$ of scaling
 degree~ $k$ and~$l$, respectively. Then their Schouten bracket
 $[X,Y]$ is of scaling degree $k+l$:
 \begin{equation}
  L_\Sigma^{} X~=~k X~~,~~L_\Sigma^{} Y~=~l \, Y
  \quad \Longrightarrow \quad
  L_\Sigma^{} [X,Y]~=~(k+l) \, [X,Y]~.
 \end{equation}
\end{prp}
\textbf{Proof.}~~The proposition is a consequence of the graded Jacobi
identity for multivector fields \cite{Tul}, which can be rewritten as
the statement that the Schouten bracket with a given multivector field
$Z$ of odd/even tensor degree acts as an even/odd superderivation:
\[
 [Z,[X,Y]]~=~[[Z,X],Y] \, + \, (-1)^{(t-1)(r-1)} [X,[Z,Y]]~.
\]
In particular, since $\Sigma$ has tensor degree $1$,
\[
 [\Sigma,[X,Y]]~=~[[\Sigma,X],Y] \, + \, [X,[\Sigma,Y]]~,
\]
from which the proposition follows immediately. \\
\PCPqed
\begin{cor}~~
 Let $X$ and $Y$ be locally Hamiltonian multivector fields on~$P$ of scaling
 degree  $-1$. Then their Schouten bracket $[X,Y]$ takes values in the kernel
 of~$\,\omega$.
\end{cor}
\textbf{Proof.}~~From the preceding proposition, $[X,Y]$ is a locally
 Hamiltonian multivector field of scaling degree $-2$. and hence, by
 Theorems~\ref{thm:FIBPOL} and~\ref{thm:DECSCD}, must take values in
 the kernel of~$\,\omega$. \\
\PCPqed

\noindent
For the Poisson bracket of Poisson forms, we have the following property.
\begin{prp}~~
 Let $f$ and $g$ be homogeneous Poisson forms on~$P$ of scaling degree~$k$
 and~$l$, respectively. Then their Poisson bracket $\{f,g\}$ is of scaling
 degree $k+l-1$:
 \begin{equation}
  L_\Sigma^{} f~=~k f~~,~~L_\Sigma^{} g~=~l \, g
  \quad \Longrightarrow \quad
  L_\Sigma^{} \{f,g\}~=~(k+l-1) \, \{f,g\}~.
 \end{equation}
\end{prp}
\textbf{Proof.}~~As explained in the last paragraph of Section~1
(see, in particular, eq.~(\ref{eq:HOMHAM1})), we can find homogeneous
Hamiltonian multivector fields~$X$ of scaling degree~$k-1$ and~$Y$
of scaling degree~$l-1$ such that $\, i_X^{} \omega = df \,$ and
$\, i_Y^{} \omega = dg \,$. We shall consider each of the terms
in the definition of the Poisson bracket separately. We find
\begin{eqnarray*}
 L_\Sigma^{} \left( i_Y^{} i_X^{} \omega \right)\!\!
 &=&\!\! i_Y^{} L_\Sigma^{} i_X^{} \omega \, + \,
         i_{[\Sigma,Y]}^{} i_X^{} \omega \\
 &=&\!\! i_Y^{} i_X^{} L_\Sigma^{} \omega \, + \,
         i_Y^{} i_{[\Sigma,X]}^{} \omega \, + \,
         i_{[\Sigma,Y]}^{} i_X^{} \omega \\
 &=&\!\! i_Y^{} i_X^{} \omega \, + \,
         (k-1) \, i_Y^{} i_X^{} \omega \, + \,
         (l-1) \, i_Y^{} i_X^{} \omega \\
 &=&\!\! (k+l-1) \, i_Y^{} i_X^{} \omega~.
\end{eqnarray*}
The same calculation works with $\,\omega$ replaced by $\theta$, so that,
since $L_\Sigma$ commutes with $d$,
\[
 L_\Sigma^{} \left( d \left( i_Y^{} i_X^{} \theta \right) \right)\!~
 =~(k+l-1) \, d \left( i_Y^{} i_X^{} \theta \right).
\]
Moreover,
\begin{eqnarray*}
 L_\Sigma^{} \left( d \left( i_Y^{} f \right) \right)\!
 &=& d \left( L_\Sigma^{} i_Y^{} f \right) \\
 &=& d \left( i_Y^{} L_\Sigma^{} f \, + \,
              i_{[\Sigma,Y]}^{} f \right) \\
 &=& d \left( k \, i_Y^{} f \, + \,
              (l-1) \, i_Y^{} f \right) \\
 &=& (k+l-1) \, d \left( i_Y^{} f \right).
\end{eqnarray*}
and similarly,
\[
 L_\Sigma^{} \left( d \left( i_X^{} g \right) \right)\!~
 =~(k+l-1) \, d \left( i_X^{} g \right).
\]
Putting the pieces together, the proposition follows. \\
\PCPqed

Having shown in what sense both the Schouten bracket and the Poisson bracket
respect scaling degree, let us use the canonical decomposition of Poisson
forms to express their Poisson bracket in terms of known operations on the
simpler objects from which they can be constructed. To start with, we settle
the case of homogeneous Poisson forms of positive scaling degree.
\begin{prp} \label{prp:JX}~~
 Let $X_{k-1}^{}$ be a homogeneous locally Hamiltonian $r$-multivector field
 on~$P$ of scaling degree $k-1$ and $Y_{l-1}^{}$ be a homogeneous locally
 Hamiltonian $s$-multivector field on~$P$ of scaling degree $l-1$, with
 $\, 1 \leqslant k,l \leqslant r$. Set
 \begin{equation}
  f_k^{}~=~\frac{(-1)^{r-1}}{k} \; i_{X_{k-1}^{}}^{} \theta~~~,~~~
  g_l^{}~=~\frac{(-1)^{s-1}}{l} \; i_{Y_{l-1}^{}}^{} \theta~.
 \end{equation}
 Then
 \begin{equation} \label{eq:POISBR2}
  \begin{split}
   \{f_k^{},g_l^{}\}~=~
   &\frac{(-1)^{r+s}}{k+l-1} \; i_{[Y_{l-1}^{},X_{k-1}^{}]}^{} \theta \\[1mm]
   &\mbox{} - \, (-1)^{(r-1)s} \, \frac{(k-1)(l-1)(k+l)}{kl(k+l-1)} \;
    d \Big( i_{X_{k-1}^{}}^{} i_{Y_{l-1}^{}}^{} \theta \Big)~.
  \end{split}
 \end{equation}
\end{prp}
\textbf{Proof.}~~From the defining equation~(\ref{eq:POISBR1}) for the Poisson
bracket, we find
\[
 \begin{split}
  \{f_k^{},g_l^{}\}~
  &=~(-1)^{r(s-1)} \, i_{Y_{l-1}^{}}^{} i_{X_{k-1}^{}}^{} \omega \\[1mm]
  & \hspace*{2em}  + \,
    d \, \Big( \frac{(-1)^{(r-1)s}}{k} \;
               i_{Y_{l-1}^{}}^{} i_{X_{k-1}^{}}^{} \theta \, - \,
               \frac{(-1)^{(s-1)}}{l} \;
               i_{X_{k-1}^{}}^{} i_{Y_{l-1}^{}}^{} \theta \\
  & \hspace*{10em} - \, (-1)^{(r-1)s} \,
               i_{Y_{l-1}^{}}^{} i_{X_{k-1}^{}}^{} \theta \Big) \\[2mm]
  &=~(-1)^{r(s-1)} \, i_{Y_{l-1}^{}}^{} i_{X_{k-1}^{}}^{} \omega \, + \,
     (-1)^{(r-1)s} \, \Big( \frac{1}{k} + \frac{1}{l} - 1 \Big) \;
     d \, \Big( i_{Y_{l-1}^{}}^{} i_{X_{k-1}^{}}^{} \theta \Big)~.
 \end{split}
\]
On the other hand, we compute
\[
 \begin{split}
  i_{[Y_{l-1}^{},X_{k-1}^{}]}^{} \, \theta~
  &=~(-1)^{(s-1)r} L_{Y_{l-1}^{}}^{} i_{X_{k-1}^{}}^{} \theta \, - \,
     i_{X_{k-1}^{}}^{} L_{Y_{l-1}^{}}^{} \theta \\[2mm]
  &=~(-1)^{(s-1)r} \, d \, i_{Y_{l-1}^{}}^{} i_{X_{k-1}^{}}^{} \theta \, + \,
     (-1)^{(s-1)(r-1)} \, i_{Y_{l-1}^{}}^{} \, d \, i_{X_{k-1}^{}}^{} \theta \\
  &\qquad - \,
     i_{X_{k-1}^{}}^{} \, d \, i_{Y_{l-1}^{}}^{} \theta \, - \,
     (-1)^{s-1} \, i_{X_{k-1}^{}}^{} i_{Y_{l-1}^{}}^{} \, d \>\! \theta \\[2mm]
  &=~(-1)^{(s-1)r} \, d \, i_{Y_{l-1}^{}}^{} i_{X_{k-1}^{}}^{} \theta \, + \,
     (-1)^{s(r-1)} k \, i_{Y_{l-1}^{}}^{} i_{X_{k-1}^{}}^{} \omega \\
  &\qquad + \,
     (-1)^{s(r-1)} l \, i_{Y_{l-1}^{}}^{} i_{X_{k-1}^{}}^{} \omega \, - \,
     (-1)^{s(r-1)} i_{Y_{l-1}^{}}^{} i_{X_{k-1}^{}}^{} \omega \\[2mm]
  &=~(-1)^{(s-1)r} \, d \, i_{Y_{l-1}^{}}^{} i_{X_{k-1}^{}}^{} \theta \, + \,
     (-1)^{s(r-1)} (k+l-1) \, i_{Y_{l-1}^{}}^{} i_{X_{k-1}^{}}^{} \omega~.
 \end{split}
\]
Thus
\[
 \begin{split}
  \{f_k^{},g_l^{}\}~
  &=~\frac{(-1)^{r+s}}{k+l-1} \; i_{[Y_{l-1},X_{k-1}]^{}}^{} \theta \, - \,
     \frac{(-1)^{(r-1)s}}{k+l-1} \;
     d \, \Big( i_{Y_{l-1}^{}}^{} i_{X_{k-1}^{}}^{} \theta \Bigr) \\
  &\qquad + \,
     (-1)^{(r-1)s} \, \Big( \frac{1}{k} + \frac{1}{l} - 1 \Big) \;
     d \Big( i_{Y_{l-1}^{}}^{} i_{X_{k-1}^{}}^{} \theta \Big)~.
   \end{split}
\]
Now the claim follows because
\begin{equation*}
 \frac{1}{k} + \frac{1}{l} - 1 - \frac{1}{k+l-1}~
 = \; \mbox{} - \, \frac{(k-1)(l-1)(k+l)}{kl(k+l-1)}~.
\end{equation*}
\PCPqed

\noindent
As a special case, consider homogeneous Poisson forms of scaling degree~$1$,
which arise by contracting $\theta$ with a Hamiltonian multivector field of
scaling degree $0$, that is, with an exact Hamiltonian multivector field (see
the first statement in Proposition~\ref{prp:HOMHAMMVF}). These Poisson forms
have been studied in \cite{FPR} under the name ``universal multimomentum map''.
\begin{cor}~~
 The space of\/ homogeneous Poisson forms on~$P$ of\/ scaling degree~$1$
 closes under the Poisson bracket.
\end{cor}
Obviously, it also follows from the proposition that no such statement holds
for homogeneous Poisson forms of scaling degree $> 1$, since the second term
in eq.~(\ref{eq:POISBR2}) vanishes only for $k=1$ or $l=1$.

Turning to homogeneous Poisson forms on~$P$ of scaling degree~$0$, which come
from forms on~$E$ by pull-back, we have
\begin{prp}~~
 The space of\/ homogeneous Poisson forms on~$P$ of\/ scaling degree~$0$
 is abelian under the Poisson bracket:
 \begin{equation}
  \{f_0^{},g_0^{}\}~=~0~.
 \end{equation}
\end{prp}
\textbf{Proof.}~~Without loss of generality, we may assume the Hamiltonian
multivector fields $X_-^{}$ and $Y_-^{}$ associated with $f_0^{}$ and with
$g_0^{}$, respectively, to be homogeneous of scaling degree $-1$. Therefore,
using the fact that if a multivector field $X$ is homogeneous of scaling
degree $k$ and a differential form $\alpha$ is homogeneous of scaling
degree $l$, then the differential form $i_X^{} \alpha$ is homogeneous
of scaling degree $k+l$,
\[
 L_\Sigma^{} X~=~k X~~,~~L_\Sigma^{} \alpha~=~l \>\! \alpha
 \quad \Longrightarrow \quad
 L_\Sigma^{} \, i_X^{} \alpha~=~(k+l) \>\! i_X^{} \alpha~,
\]
which follows immediately from the formula $\; L_\Sigma^{} i_X^{} \alpha
= i_X^{} L_\Sigma^{} \alpha + i_{[\Sigma,X]}^{} \alpha \,$, we see that
all four terms in the definition~(\ref{eq:POISBR1}) of the Poisson bracket
between $f_0^{}$ and $g_0^{}$ are differential forms of scaling degree $-1$
and hence must vanish. \\
\PCPqed

\noindent
For the mixed case of the Poisson bracket between a homogeneous Poisson form
of strictly positive scaling degree with one of scaling degree zero, we find
the following result.
\begin{prp}~~
 Let $X_{k-1}^{}$ be a homogeneous locally Hamiltonian $r$-multivector field
 on~$P$ of scaling degree $k-1$, with $\, 1 \leqslant k \leqslant r$, and let
 $g_0^{}$ be a homogeneous Poisson $(n-s)$-form on~$P$ of scaling degree zero,
 with associated Hamiltonian $s$-multivector field~$Y_-^{}$. Set
 \begin{equation}
  f_k^{}~=~\frac{(-1)^{r-1}}{k} \; i_{X_{k-1}^{}}^{} \theta~.
 \end{equation}
 Then
 \begin{equation}
  \{f_k^{},g_0^{}\}~=~- L_{X_{k-1}^{}}^{} g_0^{}~.
 \end{equation}
\end{prp}
\textbf{Proof.}~~By Proposition~\ref{prp:HOMHAMMVF}, $i_{Y_-}^{} \theta$
vanishes. Hence only two of the four terms in the defining
equation~(\ref{eq:POISBR1}) for the Poisson bracket survive:
\[
 \begin{split}
  \{f_k^{},g_0^{}\}~
  &=~(-1)^{r(s-1)} \, i_{Y_-^{}}^{} i_{X_{k-1}^{}}^{} \omega \, - \,
     d i_{X_{k-1}^{}}^{} g_0^{} \\
  &=~- \big( d i_{X_{k-1}^{}}^{} g_0^{} \, - \,
             (-1)^r\,i_{X_{k-1}^{}}^{} \, d g_0^{} \big)~
   =~- L_{X_{k-1}^{}}^{} g_0^{}~.
 \end{split}
\]
\PCPqed

Finally, let us consider closed Poisson forms, whose associated Hamiltonian
multivector fields vanish. Still, the Poisson bracket of a closed Poisson
form with an arbitrary Poisson form does not vanish, but it is once again
a closed Poisson form.
\begin{prp}~~
 Let $f$ be a Poisson $(n-r)$-form on~$P$, with associated Hamiltonian
 $r$-multivector field $X$, and let $g$ be a closed Poisson $(n-s)$-form
 on~$P$. Set
 \begin{equation}
  g~=~(-1)^s i_{G_c}^{} \omega~.
 \end{equation}
 Then
 \begin{equation}
  \{f,g\}~=~(-1)^{r+s-1} i_{[G_c,X]}^{} \omega~.
 \end{equation}
\end{prp}
\textbf{Proof.}~~As the Hamiltonian multivector field associated
with~$g$ vanishes, only one of the four terms in the defining
equation~(\ref{eq:POISBR1}) for the Poisson bracket survives:
\[
 \{f,g\}~= \; - \, d \left( i_X^{} g \right)\!~
         =~(-1)^{s-1} \, d \left( i_X^{} i_{G_c}^{} \omega \right)\!~
         =~(-1)^{rs-1} i_{[X,G_c]}^{} \omega~
         =~(-1)^{r+s-1} i_{[G_c,X]}^{} \omega~.
\]
(For the penultimate equation, see, e.g., Proposition~3.3 of
Ref.\ \cite{FPR}.) \\
\PCPqed

In view of the canonical decomposition for Poisson forms stated in
Theorem~\ref{thm:POISFCD}, the above propositions exhaust the possible
combinations for the computation of Poisson brackets.

%%%%%%%%%%%%%%%%%%%%%%%%%%%%%%%%%%%%%%%%%%%%%%%%%%%%%%%%%%%%%%%%%%%%%%%%%
\section{Conclusions and Outlook}

In this paper, we have achieved three goals. First, we have determined the
general structure of locally Hamiltonian multivector fields on the extended
multiphase space of classical first order field theories. According to
Theorem~\ref{thm:HAMMVF}, the basic structure that arises from explicit
calculations in adapted local coordinates is the decomposition of any
such multivector field $X$, of tensor degree $r$ ($0 < r < n$), into a
sum of terms of homogeneous scaling degree plus a remainder $\xi$ which
is a multivector field taking values in the kernel of~$\,\omega$:
\begin{equation}
 X~=~X_{-1} + X_0 + \ldots + X_{r-1} + \xi
 \qquad \mbox{with} \qquad
 L_\Sigma^{} X_k~=~k X_k~.
\end{equation}
Moreover, according to Proposition~\ref{prp:HOMHAMMVF}, all homogeneous locally
Hamiltonian multi\-vector fields of nonnegative scaling degree are in fact
globally Hamiltonian, and they are exact Hamiltonian if and only if they
have zero scaling degree. At the level of local coefficient functions,
this decomposition arises because the coefficient functions have to be
antisymmetric polynomials in the multimomentum variables; see
eqs~(\ref{eq:LHAMMVF01}) and~(\ref{eq:LHAMMVF02}).

Second, we have extended the scaling degree analysis to the study of
Hamiltonian forms by means of the formula
\[
 L_\Sigma^{} i_X^{} \omega~=~i_{X+[\Sigma,X]}^{} \omega~.
\]
As shown in Theorem~\ref{thm:POISFCD}, this leads to a canonical decomposition
of any Hamiltonian $(n-r)$-form $f$ ($0 < r < n$) into a sum of terms
of homogeneous scaling degree plus a remainder $f_c$ which is a closed
form:
\begin{equation}
 f~=~f_0 + f_1 + \ldots + f_r + f_c
 \qquad \mbox{with} \qquad
 L_\Sigma^{} f_s~=~s f_s~.
\end{equation}
Moreover, if $X$ is a Hamiltonian multivector field associated with $f$, then
\begin{equation} \label{eq:FFROMX}
 f_s^{}~=~\frac{(-1)^{r-1}}{s} \; i_{X_{s-1}^{}}^{} \theta
 \qquad \mbox{for $\, s>0$}~,
\end{equation}
where the $X_{s-1}$ are the homogeneous components of $X$ of nonnegative
scaling degree as described before, whereas $f_0$ arises by pull-back from
a form on the total space of the configuration bundle of the theory.
Locally, this form can be decomposed into the sum of a horizontal form
and a closed form (we prove this explicitly only for Poisson forms),
but this decomposition has no global, coordinate invariant meaning.
The canonical decomposition of Poisson forms is also useful for
deriving local formulas for $X$ in terms of $f$; these are given
in Theorem~\ref{thm:HAMMVFFOR}. They clearly show that the situation
in multisymplectic geometry resembles that encountered in symplectic
geometry but exhibits a significantly richer structure. In particular,
the notion of conjugate variables requires a conceptual extension.

Third, we have used the canonical decomposition of Poisson forms to
derive explicit formulas for the Poisson bracket between Poisson forms.
The resulting Lie algebra shows an interesting and nontrivial structure.
It has a trivial part, namely the space of closed Poisson forms, which
constitutes an ideal that one might wish to divide out: this ideal
is abelian but not central. It commutes with the most interesting
and useful part, namely the subalgebra of homogeneous Poisson forms
of scaling degree $1$, which by means of eq.~(\ref{eq:FFROMX}),
specialized to the case $s\!=\!1$, correspond to the exact
Hamiltonian multivector fields, and in such a way that the
Poisson bracket on this subalgebra corresponds to the Schouten
bracket for exact Hamiltonian multivector fields (up to signs).
The nontrivial mixing occurs through the spaces of homogeneous
Poisson forms of scaling degree $0$ and of scaling degree $>1$:
they close under the operation of taking the Poisson bracket with
a homogeneous Poisson forms of scaling degree $1$ but not under the
operation of taking mutual Poisson brackets, since these contain
contributions lying in the ideal of closed Poisson forms.

An important aspect of our results is that they confirm, once again,
the apparently unavoidable appearance of strong constraints on the
dependence of Hamiltonian multi\-vector fields and Hamiltonian forms
on the multimomentum variables and the energy variable in extended
multiphase space, expressed through the ``antisymmetric poly\-nomial''
structure of their coefficient functions. This strongly suggests
that there should be some product structure complementing the
Poisson bracket operation. So far, such a structure seems to
exist only for a very restricted class of Poisson forms, namely
the horizontal forms studied by Kanatchikov \cite{Ka1}. Also,
one might wonder whether the structural properties derived here
still hold in the multisymplectic formulation of higher order
field theories \cite{Got}.

Finally, a central question that remains is how the various proposals
of Poisson brackets in the multisymplectic formalism that can be found
in the literature, including the one proposed in Refs~\cite{FR1}
and~\cite{FPR}, relates to the Peierls\,-\,DeWitt bracket that comes
from the functional approach based on the concept of covariant phase
space.  Briefly, covariant phase space is defined as the space
$\mathscr{S}$ of solutions of the equations of motion and, formally
viewed as an infinite-dimensional manifold, carries a naturally
defined symplectic form~$\Omega$~\cite{CW,Cr,Zu}. A systematic general
investigation of the Peierls\,-\,DeWitt bracket in the multisymplectic
framework, including a proof of the fact that it is precisely the
canonical Poisson bracket for functionals on~$\mathscr{S}$ derived
from the symplectic form $\Omega$ on~$\mathscr{S}$, has been carried
out recently~\cite{Ro,FSR}. In order to establish the desired
relation, we must restrict this bracket to a certain class of
functionals, namely functionals $\mathslf{F}$ obtained by using fields
to pull Hamiltonian forms or Poisson forms $f$ on extended multiphase
space back to space-time and then integrate over submanifolds $\Sigma$
of the corresponding dimension. Explicitly, using the notation of
Ref.~\cite{FSR}, we have
\begin{equation} \label{eq:FLAG1}
 \mathslf{F}\,[\phi]~
 =~\int_\Sigma \big( \mathbb{F} \mathscr{L} \smcirc
                     (\varphi,\partial\varphi) \big)^\ast f
\end{equation}
in the Lagrangian framework and
\begin{equation} \label{eq:FHAM1}
 \mathslf{F}\,[\phi]~
 =~\int_\Sigma \big( \mathscr{H} \smcirc (\varphi,\pi) \big)^\ast f
\end{equation}
in the Hamiltonian framework. Now using the classification of Hamiltonian
vector fields and Hamiltonian $(n-1)$-forms obtained in this paper, it has
been shown recently that the Peierls\,-\,DeWitt bracket $\{\mathslf{F},
\mathslf{G}\,\}$ between two functionals $\mathslf{F}$ and $\mathslf{G}\,$
derived from Hamiltonian $(n-1)$-forms $f$ and~$g$, respectively, is the
functional derived from the Hamiltonian $(n-1)$-form $\{f,g\}$~\cite{Sa};
details will be published elsewhere. The question of how to extend this
result to Poisson forms of other degree is currently under investigation.

%%%%%%%%%%%%%%%%%%%%%%%%%%%%%%%%%%%%%%%%%%%%%%%%%%%%%%%%%%%%%%%%%%%%%%%%%%%%%%%
\section{Acknowledgements}

Part of this work has been financially supported by CNPq (``Conselho Nacional
de Desenvolvimento Cient\'{\i}fico e Tecnol\'ogico''), Brazil (M.F.), by DFG
(``Deutsche Forschungsgemeinschaft''), Germany, under the Emmy Noether
Programme (C.P.), and by FAPESP (``Fundação de Amparo \`a Pesquisa do
Estado de S\~ao Paulo'', Brazil (H.R.).
%%%%%%%%%%%%%%%%%%%%%%%%%%%%%%%%%%%%%%%%%%%%%%%%%%%%%%%%%%%%%%%%%%%%%%%%%%%%%%%
\begin{appendix}

\section{Appendix}

Let $V$ be a vector bundle over a manifold $M$ with projection $\pi$ and let
$\, i_0^{} : M \rightarrow V \,$ be its zero section. For any point $v$ in $V$,
we shall denote the zero vector in its fiber by~$v_0^{}$; thus $\, v_0^{} =
(i_0^{} \smcirc \pi) \, v$. Next, let $\Sigma$ the scaling or Euler vector
field on $V$ and denote its flux by $F$; thus
\[
 \Sigma(v)~=~v~~,~~F_\lambda^{}(v)~=~\mathrm{e}^\lambda v \qquad
 \mbox{for $\, \lambda \smin \mathbb{R} \,$ and $\, v \smin V$}~.
\]
Then
\[
 \lim_{\lambda \rightarrow -\infty} F_\lambda(v)~=~v_0^{}~.
\]
Next, consider the tangent bundle $TV$ of the total space $V$, together with
the vertical bundle which is defined to be the kernel of the tangent map
$T\pi$ to the projection $\pi$. Given any point $\, v \smin V \,$ and any
tangent vector $\, w \smin\, T_v V \,$ at this point, we define a new tangent
vector $\, w_0^{} \smin\, T_{v_0^{}} V \,$ at the corresponding zero vector by
\[
 w_0^{}~=~T_v \left( i_0^{} \smcirc \pi \right) \cdot w~.
\]
Since $i_0^{}$ is an immersion, $w_0^{}$ will vanish if and only if $w$ is
vertical. With this tool at hand, we can investigate the properties of the
tangent map $\; T_v F_\lambda : T_v V \rightarrow T_{F_\lambda(v)} V \,$:
the idea is that it should rescale vertical vectors by a factor
$\mathrm{e}^\lambda$ but leave horizontal vectors invariant:
\begin{itemize}
 \item Under the standard identification of the vertical tangent spaces of
       a vector bundle with the fibers of that vector bundle, the restriction
       of $T_v F_\lambda$ to the vertical space at $v$ is identified with
       $F_\lambda$ itself, since this is a fiberwise linear map.
 \item $T_v F_\lambda$ satisfies
       \[
        T_{F_\lambda(v)} \pi \smcirc T_v F_\lambda~=~T_v \pi~.
       \]
\end{itemize}
This implies that
\[
 \lim_{\lambda \rightarrow -\infty} T_v F_\lambda \cdot w~=~w_0^{}~,
\]
a relation that can be checked most easily by employing an arbitrary local
trivialization: denoting the typical fiber of $V$ by $\tilde{V}$ and choosing
a trivialization $\, V\vert_{\,U} \cong U \times \tilde{V} \,$ of $V$ over
some open subset $U$ of $M$, we have the following correspondences:
\[
 \begin{array}{ccccccc}
           v            &\leftrightarrow&       (x,\tilde{v})   &\quad , \quad&
           w            &\leftrightarrow&       (u,\tilde{w})           \\[1mm]
         v_0^{}         &\leftrightarrow&           (x,0)       &\quad , \quad&
         w_0^{}         &\leftrightarrow&           (u,0)               \\[1mm]
      F_\lambda(v)      &\leftrightarrow& (x,\mathrm{e}^\lambda \tilde{v})
  &\quad , \quad&
  T_v F_\lambda \cdot w &\leftrightarrow& (u,\mathrm{e}^\lambda \tilde{w})
 \end{array}
\]
where $\, x \smin\, U$, $\tilde{v} \smin \tilde{V}$, $u \smin\, T_x M$,
$\tilde{w} \smin \tilde{V}$.

Now we are ready to prove the following
\begin{prp}\label{prp:Pullback}
 Let $V$ be a vector bundle over a manifold $M$ with projection $\pi$ and let
 $\Sigma$ be the scaling or Euler vector field on $V$. A differential form
 $\alpha$ on the total space~$V$ will be the pull-back of a differential form
 $\alpha_0^{}$ on the base space $M$ to $V$ via $\pi$ if and only if it is
 scale invariant:
 \[
  \alpha~=~\pi^* \alpha_0^{} \quad \Longleftrightarrow \quad
  L_\Sigma^{} \alpha~=~0~.
 \]
\end{prp}
\textbf{Proof.}~~Assume first that the form $\alpha$ on $V$ is the pull-back
of a form $\alpha_0^{}$ on $M$; then $\, \alpha = \pi^* \alpha_0^{} \,$ and
hence $\, d \alpha = \pi^* d \alpha_0^{}$. Therefore, $\alpha$ and $d \alpha$
are both horizontal. This means that for any vertical vector field $X$ on $V$,
including $\Sigma$, we have
\[
 i_X^{} \alpha~=~0~,
\]
as well as $\, i_X^{} d \alpha = 0$, so
\[
 L_X^{} \alpha~=~0~.
\]
Conversely, assume that the form $\alpha$ on $V$, of degree $r$, say,
satisfies $L_\Sigma^{} \alpha = 0$, so $\alpha$ is invariant under the
flow $F$ of $\Sigma\,$:
\[
 \frac{d}{d\lambda} \, F_\lambda^* \alpha~=~0~.
\]
This means that given $\, v \smin V \,$ and $\; w_1,\ldots,w_r \smin\,
T_v V$, the expression
\[
 (F_\lambda^* \alpha)_v^{}(w_1,\ldots,w_r)~
 =~\alpha_{F_\lambda(v)}^{}(T_v F_\lambda \cdot w_1 \,,\, \ldots \,,\,
                            T_v F_\lambda \cdot w_r)
\]
does not depend on $\lambda$, so its value
\[
 \alpha_v^{}(w_1,\ldots,w_r)
\]
at $\, \lambda = 0 \,$ is equal to its value
\[
 \alpha_{v_0^{}}^{}((w_1)_0^{},\ldots,(w_r)_0^{})
\]
obtained in the limit $\, \lambda \rightarrow -\infty$. But this
means that $\alpha$ is equal to $\pi^* \alpha_0^{}$ where
$\alpha_0^{}$ is defined as $\, \alpha_0^{} = i_0^* \alpha$.
\PCPqed

\end{appendix}
%%%%%%%%%%%%%%%%%%%%%%%%%%%%%%%%%%%%%%%%%%%%%%%%%%%%%%%%%%%%%%%%%%%%%%%%%%%%%%%

\end{document}